\documentclass[twocolumn,10pt,nofootinbib,floatfix,raggedbottom]{revtex4-2}

\usepackage[T1]{fontenc}
\usepackage[utf8]{inputenc}
\usepackage{amsmath,amssymb,bm}
\usepackage{graphicx}
\usepackage{booktabs}
\usepackage{multirow}
\usepackage{array}[=2016-10-06]
\usepackage{xcolor}
\usepackage[colorlinks=true,allcolors=blue]{hyperref}
\usepackage{natbib}
\usepackage{placeins}
\usepackage{comment}
\usepackage{wrapfig}
\usepackage{orcidlink}
\usepackage{fontawesome5}

\newcommand{\mrs}{\affiliation{Aix-Marseille Universit\'e, Universit\'e de Toulon, CNRS, CPT, Marseille, France}}

\newcommand{\plandscape}{%
  \texttt{posterior-landscape}\,%
  \href{https://github.com/Mik3M4n/posterior-landscape}{\faGithub}%
}

\graphicspath{{figures/}}

\begin{document}

\title{\texttt{posterior-landscape}: morphological feature finding in two-dimensional probability densities}

\author{Michele Mancarella\texorpdfstring{$\,$}{ }\orcidlink{0000-0002-0675-508X}}
\email{michele.MANCARELLA@univ-amu.fr}
\mrs 
 
\author{Viola De Renzis\texorpdfstring{$\,$}{ }\orcidlink{0000-0001-7038-735X}}
\mrs
 
\author{Alessandro Agapito$\,$\orcidlink{0009-0005-9004-3163}
}
\mrs

\begin{abstract}
We present \plandscape{}, a Python framework to identify and characterize morphological features in two-dimensional probability densities. 
The framework combines concepts from topological data analysis,
statistical density-ridge estimation, and scale-space image analysis. 
%
It extracts localized concentrations, valleys, ridges, and changes of slope; for Bayesian reconstructions, it propagates uncertainty in feature recovery, geometry, and correspondence across posteriors. The framework also connects two-dimensional structures to features in one-dimensional marginals, and quantifies their correlation with other inferred parameters. We validate the method on controlled densities, and demonstrate its use on gravitational-wave population reconstructions, where it provides a systematic way to identify and interpret population features and assess which ones carry information relevant for cosmology.
The resulting catalogues also offer a common morphological
representation for future comparisons between  predictions and population reconstructions.
\end{abstract}

\maketitle

\section{Introduction}
\label{sec:introduction}

When looking at a two-dimensional map, one immediately recognizes structures:
isolated concentrations, extended ridges, valleys, and boundaries. This is
especially true for probability densities, although related questions arise
for other positive fields, including geographical maps. Turning this visual
notion of ``structure'' into a reproducible quantitative description is less
straightforward. A local maximum, for example, identifies a point but not
necessarily the finite region associated with a feature; extended structures
may not be characterized by a single critical point at all.

The problem can be stated more quantitatively: given a normalized
two-dimensional density $p(x,y)$, how can one identify its morphological
features and assign to them well-defined locations, regions, and summary
properties? In Bayesian applications, a second layer becomes necessary because
one typically has an ensemble of posterior realizations,
${p^{(s)}(x,y)}$, rather than a unique density. Features can then move,
change shape, appear, disappear, split, or merge across realizations.
Uncertainty in their recovery, location, and geometry must therefore be
propagated together with uncertainty in the underlying density.

Here, we introduce \plandscape{}, a Python framework that performs feature identification directly on
the density. 
The package is available open--source at Ref.~\cite{Mancarella_posterior-landscape_morphological_feature_2026}
The method identifies localized concentrations, valleys, ridges,
and changes of slope, and provides a feature catalogue for each realization of
the normalized input map. For Bayesian reconstructions, uncertainty is
propagated by identifying features independently in each realization and
matching compatible structures across the ensemble.

The framework brings together ideas developed in several fields. Topological data analysis describes how concentrations and depressions are separated and connected as the density level of a field is varied \citep{Edelsbrunner2002Persistence,Carr2003ContourTrees,Edelsbrunner2003MorseSmale}. Statistical methods for locating elongated concentrations use the local slope and curvature of a density to follow their shape \citep{Ozertem2011PrincipalCurves,Genovese2014Ridges}. Methods from computer vision examine a map at different smoothing resolutions to assess which structures remain recognisable \citep{Witkin1983ScaleSpace,Lindeberg1994ScaleSpace}. Research on visualising uncertain maps addresses how these structures vary across an ensemble of realisations \citep{Favelier2019PersistenceAtlas,Athawale2022MorseEnsembles}. Our contribution is to combine these complementary ideas into a framework that identifies features, assigns them finite measurement regions, and follows their properties and uncertainties across posterior draws.

The framework also tracks how two-dimensional structures appear after
marginalization. Features are identified independently in both one-dimensional
marginals and compared with the projections of the two-dimensional feature
regions. This makes it possible to determine whether a structure visible in a
marginal has a unique origin in the joint density, whether several joint
structures contribute to it, or whether part of the two-dimensional morphology
is lost under projection. Global tail scales can be extracted from both
marginals and treated separately from localized morphological features. When
the reconstructed density is sampled jointly with other inferred parameters,
the feature summaries and tail scales can additionally be compared with those
parameters through correlation- and information-based diagnostics.

Our immediate motivation comes from gravitational-wave population inference.
As compact-binary catalogues grow~\cite{LIGOScientific:2025hdt,LIGOScientific:2025slb,
LIGOScientific:2026wfs}, increasingly flexible methods are being
used to reconstruct joint distributions of masses, spins, and redshift~\cite{LIGOScientific:2025pvj,LIGOScientific:2026ctl}. 
In such analyses, physically interesting structures are often not explicit model
parameters and must instead be extracted from the reconstructed density. The
two-dimensional component-mass distribution provides a particularly useful
example: one may ask which structures in the joint mass plane are present, which generate
features observed in the one-dimensional mass spectra and, in spectral-siren
applications~\citep{
LIGOScientific:2025jau,LIGOScientific:2026uyd}, which of those structures are associated with cosmological
parameters such as the Hubble constant $H_0$.

A common morphological description could also connect observational
reconstructions to physical population predictions. For example,
a population-synthesis prediction and a weakly modelled reconstruction
could each supply a joint component-mass density to the same feature
analysis. Comparing feature locations, extents, strengths, and
connections, with their uncertainties and under consistent analysis
choices, would provide quantitative tests of predicted population
structure beyond visual comparisons or manually selected features.
We regard this as a promising application of the framework.

In this paper we present the feature-finding framework for a single density,
its extension to posterior ensembles, its treatment of one-dimensional
marginals and projections, and optional tools for association with external
inferred quantities. We validate the method on a controlled synthetic
two-dimensional density, first as a single map and then as an ensemble of
uncertain realizations constructed from the same underlying density. We
conclude with an application to publicly available LIGO--Virgo--KAGRA
population reconstructions of the joint component-mass distribution.

\section{Method}
\label{sec:method}

\plandscape{} identifies morphology directly from a gridded
probability density, independently of the model used to reconstruct it.
The analysis separates three tasks: identifying a point or curve, attaching
an appropriate finite measurement region, and following the resulting
feature through posterior realisations. This distinction allows a feature
to be located even when its full extent cannot be measured, and separates
uncertainty in its recovery from uncertainty in its properties. We describe
the principal construction here; numerical prescriptions, independent
one-dimensional detection, and supplementary analyses are collected in
Appendix~\ref{app:technical_methods}.
Appendix~\ref{app:software_io} contains a description of the software input/ouput; extensive documentation is provided on the software repository.

The input is a non-negative density on a tensor-product grid
$(x_{1,i},x_{2,j})$, restricted to a domain $\Omega$ by a mask $M$.
Density measure and spatial geometry are separate choices. For density
$p_x$ per $dx_1\,dx_2$, the density per $d\log_a x_1\,d\log_a x_2$ is
\begin{equation}
 F_{\log}=x_1x_2(\ln a)^2p_x.
 \label{eq:method_jacobian}
\end{equation}
We denote the selected feature density and measure by $F$ and
$d\nu=d\nu_1d\nu_2$, with $\int_\Omega F\,d\nu=1$.
Distances, smoothing, and spatial derivatives instead use analysis
coordinates $g_i=x_i$ or $g_i=\log_a x_i$. 

The implementation described here is version 0.8.8, using NumPy and SciPy
\citep{Harris2020NumPy,Virtanen2020SciPy}, with optional Numba compilation
~\citep{Lam2015Numba}.

\subsection{Morphological feature identification}
\label{sec:feature_identification}

\paragraph{Resolution and boundaries.}
We analyse the density at several supplied Gaussian smoothing widths $s$ to test whether the same morphological structure remains
recognisable when the map is viewed at nearby resolutions.
This follows the scale-space approach: a structure recognised at nearby
resolutions is less dependent on the finest details of a single map
\citep{Witkin1983ScaleSpace,Lindeberg1994ScaleSpace}. 
For a density $F$, we define
\begin{equation}
 F_s=\frac{1}{\mathcal Z_s}
       \frac{G_{\boldsymbol\sigma}*(MF)}{G_{\boldsymbol\sigma}*M}.
 \label{eq:method_smoothing}
\end{equation}
Here $G_{\boldsymbol\sigma}$ is a Gaussian kernel with standard deviations $\sigma_i=s/\operatorname{med}(\Delta g_i)$, $*$ denotes discrete convolution and $\mathcal Z_s$ is the
normalisation. Mask-normalised convolution avoids treating the excluded
domain as zero density \citep{Knutsson1993NormalizedConvolution}.
This definition averages
density amplitudes, compensates for the fraction of the kernel inside the
mask, and restores unit integral. 
On a grid approximately uniform in the analysis coordinates, $s$ is a
spatial Gaussian standard deviation; $s=0$ leaves the normalised field
unchanged. The default scales are $(0,0.025,0.05)$.\\

\paragraph{Feature classes.}
The density can be viewed as the height of a landscape. Peaks and pits
locate concentrations and depressions; saddles are passes connecting them.
The corresponding feature classes are constructed as follows.
\begin{itemize}
\item \emph{Peaks.}
Starting from the highest values
of $F_s$, the algorithm activates grid nodes one by one. A newly activated
node with no active neighbour starts a new high-density region and marks a
local maximum.
As the threshold is lowered, these regions expand. When two regions first meet,
the joining level defines their saddle, or pass. The region containing the higher
maximum survives the merger, while the lower maximum is paired with the
saddle at which it loses its independent identity. 

\item \emph{Pits.}
The complementary construction starts from minima of the density, then raises the density threshold and grows
connected low-density regions from local minima. At a merger, the region
containing the deeper minimum survives; the shallower minimum is paired
with the joining saddle. A pit is therefore the low-density counterpart
of a peak.

\item \emph{Ridges and valleys.}
At a merger, uphill paths from the saddle to the associated peaks define
ridge branches; downhill paths to the corresponding pits define valley
branches. Each branch is identified by its saddle and extremum endpoints.
Two branches can share a saddle while terminating at different extrema,
so a branch and the full two-sided structure are distinct objects.

\item \emph{Shoulders.}
A shoulder is a flattening followed by renewed steepening along an
otherwise monotonic density profile, and need not contain an extremum.
For positive coordinates, define
\begin{equation}
\begin{split}
 t&=\frac{\ln x_1+\ln x_2}{2},\qquad
 r=\frac{\ln x_1-\ln x_2}{2},\\
 \alpha&=\left.\frac{\partial\ln F_s}{\partial t}\right|_r,
 \qquad \kappa=\left.\frac{\partial\alpha}{\partial t}\right|_r.
\end{split}
\label{eq:method_shoulder_coordinates}
\end{equation}
Increasing $t$ at fixed $r$ rescales both coordinates while preserving their ratio.
A local maximum of $\alpha$ marks the flattening. A bounded
transition has four ordered landmarks: a preceding positive curvature
maximum $t_+$, the slope maximum $t_\alpha$, a following negative curvature
minimum $t_-$, and the next negative-to-nonnegative curvature crossing
$t_0$. A bounded
transition therefore satisfies $t_+<t_\alpha<t_-<t_0$. The reported position
of the shoulder on each profile is $t_-$, where the renewed steepening after
the flattening is strongest. Compatible detections on neighbouring $r$ profiles are linked
into an extended front.
Shoulder detection and the measured portion of each valley are
restricted to a highest-density support $\mathcal H_\gamma(F_s)$
containing at least probability $\gamma$, with $\gamma=0.995$ by
default. 

\end{itemize}

The peak--pit construction connects the landscape picture to topological
data analysis: one follows the appearance, growth, and merger of connected
regions as the density threshold changes. The connectivity of the
high- and low-density regions is recorded by join and split trees,
respectively~\citep{Carr2003ContourTrees}.
The idea of representing a scalar field through its extrema, saddles, and the
ascending or descending paths that connect them comes from the Morse--Smale
description of scalar fields \citep{Edelsbrunner2003MorseSmale}.
The discrete paths found this way have the required topological connections, but their
shape can reflect finite  grid resolution. The result is refined using the local shape of
the density field, with Gradient- and Hessian-based corrections. This links the topological construction to
differential density-ridge methods
\citep{Ozertem2011PrincipalCurves,Genovese2014Ridges}. 
Endpoint positions remain
fixed, so refinement does not redefine the branch identity.
All details are given in Appendix~\ref{app:technical_topology}.\\

\paragraph{Catalogue selection.}
The feature classes defined above use different geometric constructions, but
their inclusion in the catalogue follows a common two-stage logic. First, each
candidate must satisfy a class-specific criterion on a field analysed at a
fixed smoothing width $s$. Second, where applicable, the same morphology must
remain recognisable under a change of smoothing resolution.

For both peaks and pits, the saddle pairing supplies a non-local measure of
feature strength. 
For a non-essential extremum $e$, whose component disappears at a
merger with a stronger one, paired with saddle $z_e$, we define the density-level
persistence~\citep{Edelsbrunner2002Persistence}
\begin{equation}
 \Pi_e=\begin{cases}
 F_s(e)-F_s(z_e),&\text{peak},\\
 F_s(z_e)-F_s(e),&\text{pit}.
 \end{cases}
 \label{eq:method_persistence}
\end{equation}
It is the density change required before the component born at $e$ joins one
associated with a stronger extremum. Small values therefore describe shallow
local structure, while large values indicate a feature separated from its
surroundings by a substantial density contrast.
A a mathematical
motivation for using this quantity is provided by the stability theorem for persistence diagrams: under a uniformly bounded perturbation of size $\varepsilon$,
matched birth and merger levels change by at most $\varepsilon$, and their
difference by at most $2\varepsilon$
\citep{CohenSteiner2007Stability}. A sufficiently persistent structure
cannot consequently be erased by an arbitrarily small change in the field. 

At each scale, a single persistence threshold $T_s$ is calibrated from
the combined finite peak and pit persistences. The algorithm searches
for the largest eligible logarithmic gap $G_*$ between adjacent sorted
values, excluding only the gap between the two largest values. If
$G_*\geq g_{\min}$, where $g_{\min}$ is a user-specified minimum
separation with default value $1$, then $T_s$ is set to the geometric
mean of the two bounding persistence values. Numerical floors and
fallback rules are specified in
Appendix~\ref{app:technical_persistence}.
A user-supplied numerical threshold can instead impose one fixed
cutoff across all smoothing scales.

The fixed-scale inclusion rule is then applied separately to each feature
class. Non-essential peaks and pits require
$\Pi_e\geq T_s$; the global maximum and minimum remain eligible
independently. Ridge and valley branches require a retained extremum
endpoint and a sufficiently persistent merger, as specified in
Appendix~\ref{app:technical_persistence}. Shoulders do not enter this
persistence calibration;  their inclusion instead requires resolved
slope and curvature excursions, including a pre- to post-transition
slope change, and a sufficiently extended front
(Appendix~\ref{app:technical_shoulders}).

Topological candidates at the first requested scale define initial
identities. Same-type counterparts are matched across scales using point
positions or branch endpoints; a feature must recur at at least
$\lceil2N_{\rm scale}/3\rceil$ scales, and a retained branch must end at a
retained extremum. With the default three scales, two recoveries are
required. Shoulder fronts are compared separately and require compatible
detections at at least two scales when several widths are supplied.

\subsection{Feature regions and strength}
\label{sec:feature_measurements}
\label{sec:posterior_measurements}

\begin{figure*}[t]
    \centering
    \includegraphics[width=0.8\textwidth]{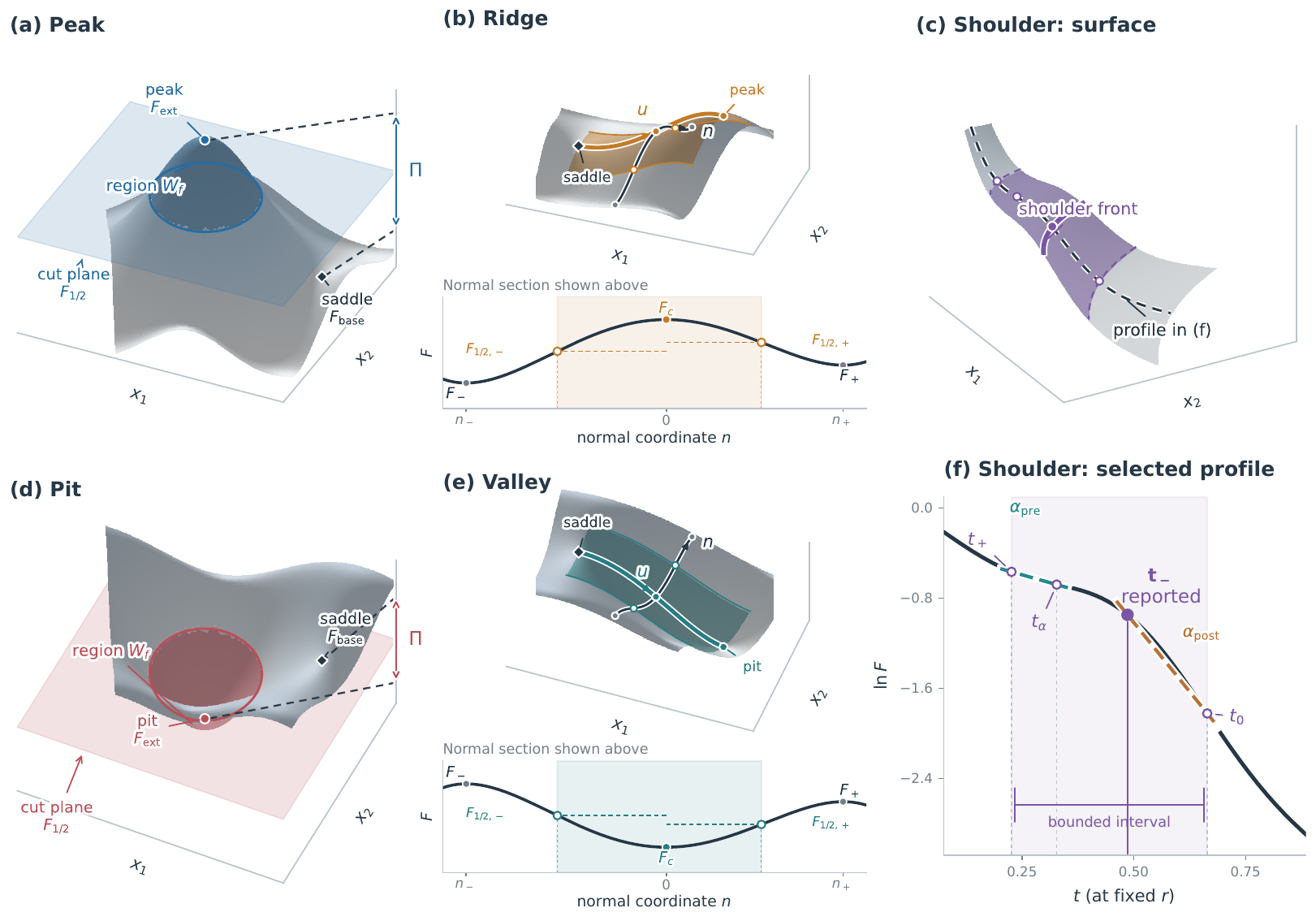}
    \caption{
    Illustration of the feature definitions and measurement
    regions. 
    \textbf{(a,d) Peaks and pits.}
    The horizontal plane at
    $F_{1/2}=(F_{\rm ext}+F_{\rm base})/2$
    cuts the surface around the selected extremum.
    The coloured surface patches project onto the connected measurement
    regions $W_f$.
    The paired saddle supplies $F_{\rm base}$. 
    The vertical bracket spans the full persistence
    $\Pi=|F_{\rm ext}-F_{\rm base}|$.
    \textbf{(b,e) Ridges and valleys.}
    Each surface shows one saddle--extremum branch and a selected normal
    section, reproduced below. Arrows indicate the along-curve coordinate
    $u$ and the normal coordinate $n$.
    Starting from the central density $F_c$, the first side minima
    for a ridge, or maxima for a valley, supply $F_-$ and $F_+$.
    Open markers indicate the crossings of
    $F_{1/2,\pm}=(F_c+F_\pm)/2$.
    Their separation defines the local strip width, shaded in the
    lower profile.
    \textbf{(c,f) Shoulders.}
    The dashed curve on the surface selects the fixed-$r$ profile
    whose logarithmic density is shown in~(f).
    Its four landmarks are the preceding positive curvature maximum
    $t_+$, the logarithmic-slope maximum $t_\alpha$, the following
    negative curvature minimum $t_-$, and the next
    negative-to-nonnegative curvature crossing $t_0$.
    The reported position is $t_-$.
    The coloured secants define $\alpha_{\rm pre}$ and
    $\alpha_{\rm post}$ over $[t_+,t_\alpha]$ and $[t_-,t_0]$,
    respectively.
    On the surface, the reported positions form the shoulder front,
    bounded by the onset and terminal curves at $t_+$ and $t_0$.
    }
    \label{fig:algorithmic_feature_definitions}
\end{figure*}

We attach a measurement region $\mathcal W_f$ to each measurable
feature using its local geometry.\footnote{Note that this is distinct from a posterior
credible region for the feature's position.}

The regions follow the different geometries of the feature classes.
\begin{itemize}
\item \emph{Peaks and interior pits.}
For extremal and base densities $F_{\rm ext}$ and $F_{\rm base}$, define the half-level
\begin{equation}
 F_{1/2}=\frac{F_{\rm ext}+F_{\rm base}}2.
 \label{eq:method_half_level}
\end{equation}
For a peak, $\mathcal W_f$ is the connected component containing the maximum above this
level; for a pit, it is the corresponding connected
component containing the minimum below it.
For an extremum paired with a saddle, $F_{\rm base}$ is the density
at that saddle.
The dimensionless relative peak prominence and
pit depth are
\begin{equation}
 A_{\rm P}=1-\frac{F_{\rm base}}{F_{\rm peak}},\qquad
 A_{\rm D}=1-\frac{F_{\rm pit}}{F_{\rm base}}.
 \label{eq:method_point_strength}
\end{equation}
\item \emph{Ridges and valleys.}
A ridge or valley branch is intrinsically a one-dimensional curve. To measure
its transverse width and contrast, we construct a finite strip around the
curve from profiles taken normal to it in the selected analysis geometry. Along the branch, the algorithm searches outward from the
curve for the first resolved minimum on either side of a ridge, or the first
resolved maximum on either side of a valley. If
$\overline c$ denotes the physical-arclength-weighted median
transverse logarithmic contrast over the valid sections of that event,
its relative strength is
\begin{equation}
 A_{\rm curve}=1-\exp(-\overline c).
 \label{eq:method_curve_strength}
\end{equation}
This contrast measures the density variation across the crest or
trough.
Two same-type branches sharing a saddle form a paired curve event,
whose measurement region is the union of their strips.

\item \emph{Shoulders.}
The onset and terminal curves defined by $t_+$ and $t_0$ bound the
transition around the front. Its longest sufficiently extended bounded
segment defines $\mathcal W_f$ within $\mathcal H_\gamma(F_s)$.
On each bounded profile, the slope change is
\begin{equation}
 \Delta\alpha=\alpha_{\rm pre}-\alpha_{\rm post},
 \label{eq:method_shoulder_contrast}
\end{equation}
where the two slopes are secants of $\ln F_s$ over
$[t_+,t_\alpha]$ and $[t_-,t_0]$, respectively.
We define the bounded slope-change strength as
\begin{equation}
 \begin{split}
 A_{{\rm S},f}
   &=1-\exp\!\left(-\overline{\Delta\alpha}_f\right) \, , \\
   \overline{\Delta\alpha}_f
   &=\operatorname{median}_{j\,\mathrm{bounded}}\Delta\alpha_j
 \end{split}
 \label{eq:method_shoulder_strength}
\end{equation}
for $\overline{\Delta\alpha}_f\geq0$, giving
$0\leq A_{{\rm S},f}<1$. 
\end{itemize}

Appendix~\ref{app:technical_regions} gives the transverse searches,
boundary and fallback rules, and detailed morphology-specific strengths.

Each measurable feature can be characterised by a region centroid,
defined as a physical-coordinate first moment with
morphology-specific weighting.
Concentrations and shoulders are weighted by density; depressions are
weighted by missing density relative to their local background. Writing
$(v)_+=\max(v,0)$, an interior pit uses
$\Delta F=(F_{\rm base}-F_s)_+$; a valley uses
$\Delta F=(F_{\rm bg}-F_s)_+$, where $\ln F_{\rm bg}$ interpolates between
the two transverse side maxima.\footnote{The weighting specifies the quantity being localized. For peaks and ridges, density weighting gives the mean population location within the measured feature region. For pits and valleys, deficit weighting localizes the depletion relative to the adopted reference background. Excess weighting for positive features would also be a valid choice, yielding a centroid of the enhancement above a specified background. We adopt density weighting for concentrations to preserve their direct interpretation in terms of population probability.} The common weighting field is
\begin{equation}
 \rho_f(\boldsymbol x)=
 \begin{cases}
 F_s(\boldsymbol x),&\text{peak, ridge event, shoulder},\\
 \Delta F(\boldsymbol x),&\text{interior pit, valley event}.
 \end{cases}
 \label{eq:method_region_weight}
\end{equation}
The centroid is defined as
\begin{equation}
 \boldsymbol\mu_f=
 \frac{\int_{\mathcal W_f}\boldsymbol x \rho_f(\boldsymbol x)\,d\nu}
      {\int_{\mathcal W_f}\rho_f(\boldsymbol x)\,d\nu}.
 \label{eq:method_region_centroid}
\end{equation}
All regions and their measurements are reconstructed separately on each
posterior draw.

Figure~\ref{fig:algorithmic_feature_definitions} provides an intuitive illustration of the
different feature classes.
The surface views and their
selected sections show how the local constructions define both
the reported feature geometry and its finite measurement region.

Table~\ref{tab:method_catalogue_selection} summarises the selection
criteria, finite measurement regions, and reported strengths of the
principal two-dimensional feature classes. Selection determines whether
an object is retained; its region and strength characterise the structure
when the required geometry is measurable. Optionally, one can search for plateaus and
depression floors; these are treated separately in
Appendix~\ref{app:technical_optional}.

\begin{table*}[t]
\centering
\caption{
Mathematical selection criteria, finite measurement regions, and
reported strengths of the principal two-dimensional feature classes.
The inclusion column applies at one smoothing scale; global extrema
remain eligible independently of $T_s$.
Reference-catalogue selection additionally requires compatible
same-type detections at
$N_{\rm rec}\geq\lceil2N_{\rm scale}/3\rceil$ scales for topological
features, with retained extremum endpoints for branches.
Shoulders require compatible fronts at two scales when several
widths are supplied.
}
\label{tab:method_catalogue_selection}
\small
\setlength{\tabcolsep}{3pt}
\renewcommand{\arraystretch}{1.15}
\begin{tabular}{@{}
p{0.10\textwidth}
p{0.35\textwidth}
p{0.30\textwidth}
p{0.20\textwidth}
@{}}
\toprule
\raggedright Feature
&
\raggedright Fixed-scale inclusion
&
\raggedright Finite measurement region
&
\raggedright Reported strength
\tabularnewline
\midrule

\raggedright Peak
&
\raggedright
$\Pi_{\rm peak}\geq T_s$,
with $\Pi_{\rm peak}=F_{\rm peak}-F_s(z_e)$;
Eq.~\eqref{eq:method_persistence}.
&
\raggedright
Connected component containing the maximum in
$F_s\geq F_{1/2}$;
Eq.~\eqref{eq:method_half_level}.
Density weighting.
&
\raggedright
Relative prominence:
\[
 A_{\rm P}
 =1-\frac{F_{\rm base}}{F_{\rm peak}}.
\]
\tabularnewline[2mm]

\raggedright Pit
&
\raggedright
$\Pi_{\rm pit}\geq T_s$,
with $\Pi_{\rm pit}=F_s(z_e)-F_{\rm pit}$;
Eq.~\eqref{eq:method_persistence}.
&
\raggedright
For an interior pit, the connected component containing the minimum
in $F_s\leq F_{1/2}$.
Weighting by
$\Delta F=(F_{\rm base}-F_s)_+$.
&
\raggedright
Relative depth:
\[
 A_{\rm D}
 =1-\frac{F_{\rm pit}}{F_{\rm base}}.
\]
\tabularnewline[2mm]

\raggedright Ridge
&
\raggedright
$\Pi_{\rm event}\geq T_s$ and a retained peak endpoint, where
$\Pi_{\rm event}=\max_{c\in\mathcal D}\Pi_c$. $\mathcal D$ contains the components disappearing at a merger;
Eq.~\eqref{eq:method_event_persistence}.
&
\raggedright
Transverse half-level strips,
$F_{1/2,\pm}=(F_c+F_\pm)/2$.
Two branches sharing a saddle form a paired-event region.
Density weighting.
&
\raggedright
Transverse event contrast:
\[
 A_{\rm curve}=1-e^{-\overline c}.
\]
$\overline c$ is the
physical-arclength-weighted median transverse logarithmic contrast.
Local $c_{\rm R}$:
Eq.~\eqref{eq:method_transverse_contrast}.
\tabularnewline[2mm]

\raggedright Valley
&
\raggedright
$\Pi_{\rm event}\geq T_s$ and a retained pit endpoint, where
$\Pi_{\rm event}=\max_{c\in\mathcal D}\Pi_c$;
Eq.~\eqref{eq:method_event_persistence}.
&
\raggedright
Paired union of valid transverse strips requiring two resolved
side maxima.
Weighting by
$\Delta F=(F_{\rm bg}-F_s)_+$;
Appendix~\ref{app:technical_regions}.
&
\raggedright
Transverse event contrast:
\[
 A_{\rm curve}=1-e^{-\overline c}.
\]
Local $c_{\rm V}$:
Eq.~\eqref{eq:method_transverse_contrast}.
\tabularnewline[2mm]

\raggedright Shoulder
&
\raggedright
Derivative tests and minimum front extent, Eqs.~\eqref{eq:method_shoulder_selection}, \eqref{eq:method_shoulder_extent}
&
\raggedright
Transition between the onset and terminal curves defined by
$t_+$ and $t_0$, restricted to $\mathcal H_\gamma(F_s)$.
The bounded segment must independently satisfy the minimum-extent
requirement.
Density weighting.
&
\raggedright
Bounded slope-change strength:
\[
 A_{\rm S}
 =1-e^{-\overline{\Delta\alpha}_f}.
\]
 $\overline{\Delta\alpha}_f$ is the median slope change over bounded
cross-sections;
Eq.~\eqref{eq:method_shoulder_strength}.
\tabularnewline

\bottomrule
\end{tabular}

\medskip
\end{table*}

\subsection{Posterior tracking and uncertainty}
\label{sec:posterior_matching}
\label{sec:posterior_geometry}

For a single density field, the preceding sections identify morphological
objects and, where available, their finite measurement regions. Across a
posterior ensemble, these objects can move, change shape, or fail to be
recovered. We first establish correspondence between draws, then distinguish
the probability of recovering a feature from the uncertainty in its
measurable properties.\\

\paragraph{Reference identities and correspondence.}
For density draws $F_b$ with normalised weights $\omega_b$, the
single-field multiscale analysis is applied to the normalised pointwise
weighted median,
\begin{equation}
 F_{\rm ref}(\boldsymbol x)=
 \frac{\operatorname{Med}_{\omega}[F_b(\boldsymbol x)]}
 {\int_\Omega\operatorname{Med}_{\omega}[F_b(\boldsymbol x)]\,d\nu}.
 \label{eq:method_reference}
\end{equation}
This defines reference identities and representative geometry. 

Each posterior draw is then analysed independently.\footnote{Draw-level
topological extraction uses the first reference smoothing scale $s_0$
and its fixed numerical cutoff $T_{s_0}$. Shoulders repeat their
multiscale search in every draw.}
Compatible candidates are matched one-to-one to reference identities
by minimising spatial mismatch, using linear assignment
\citep{Crouse2016Assignment}. Matching compares point positions,
saddle--extremum landmarks, or representative front points, as appropriate.
Unmatched candidates do not create new reference identities; their counts
and the frequency of ambiguous matches are recorded separately.
The resulting posterior summaries therefore describe the tracked reference
catalogue, rather than an exhaustive union of everything found in individual
draws. Matching details are given in
Appendix~\ref{app:technical_posterior}.\\

\paragraph{Recovery and measurement availability.}
For feature $f$
in draw $b$, let $M_{fb}$ indicate a morphological match, $L_{fb}$
usable point or curve geometry, and $R_{fb}$ a valid finite measurement
region. These indicators satisfy $R_{fb}\leq L_{fb}\leq M_{fb}$, with
posterior supports
\begin{equation}
\begin{split}
 P_{{\rm morph},f}&=\sum_b\omega_bM_{fb},\\
 P_{{\rm loc},f}&=\sum_b\omega_bL_{fb},\\
 P_{{\rm reg},f}&=\sum_b\omega_bR_{fb}.
\end{split}
\label{eq:method_supports}
\end{equation}
These probabilities describe recovery under the supplied posterior and
the adopted identification rules. 
They answer, respectively, whether the reference morphology is recovered,
whether it can be localised, and whether its finite-region properties
can be measured. For example, a valley may be located without two
resolved transverse sides, and a shoulder front may be recovered without
a sufficiently extended bounded transition. 
The default requirement to be reported in the catalogue is
$P_{\rm morph}\geq0.05$.
\\

\paragraph{Location and centroid uncertainty.}
Having established availability, the interpretation of a spatial summary
depends on what is being located. 

For point features, the full posterior distribution of the location is
constructed directly from the matched draw-level positions.
This produces an
estimated posterior density for the feature position. A highest-density
region is then the set of locations with the largest posterior location
density that together contain a specified posterior mass. 
This follows from the highest-density-region construction for summarizing possibly
non-Gaussian multivariate distributions~\citep{Hyndman1996HDR}.

For an extended feature, the corresponding object is a curve rather than
a single point. Its \emph{location corridor} is constructed from positions
sampled along the available posterior curves, with each draw contributing
its posterior weight. Therefore, a $90\%$ corridor contains $90\%$ of this spatial
probability, but not necessarily $90\%$ of complete curves. Its spatial extent
reflects both the length of the structure and its variation across draws.
Related spatial summaries of scalar-field ensembles have been developed
in uncertainty visualisation
\citep{Favelier2019PersistenceAtlas,Athawale2022MorseEnsembles}.

Finally, a finite-region centroid is a different location observable.
Each $\boldsymbol\mu_{f,b}$ is calculated from that draw's own
$\mathcal W_{f,b}$ and density or deficit weight, using
Eq.~\eqref{eq:method_region_centroid}. Its posterior therefore includes
changes in the region and in the distribution of weight within it,
conditional on a valid region measurement. Centroid intervals describe
where the feature's density or deficit is concentrated.\\ 

\paragraph{Uncertainty in the measurement region.}
The preceding summaries describe positions or curves. To describe the
varying finite region itself, we can also ask how often each position
belongs to it. The unconditional inclusion map is
\begin{equation}
 P_{{\rm incl},f}(\boldsymbol x)=
 \sum_b\omega_b R_{fb}\,
 \mathbf1\{\boldsymbol x\in\mathcal W_{f,b}\},
 \label{eq:method_region_inclusion}
\end{equation}
where an unavailable region contributes zero to the indicator.
For $P_{{\rm reg},f}>0$, division by $P_{{\rm reg},f}$ gives the
inclusion probability conditional on region availability.
For example, an inclusion value of $0.9$ in the conditional map means that the
position belongs to the measurement region in $90\%$ of the
region-valid posterior weight.

\subsection{Joint--marginal correspondence}
\label{sec:marginal_features}

We identify marginal features
independently and then compare them with the joint catalogue.
In particular, a feature visible in a marginal need not have a unique joint counterpart; 
several joint regions can contribute at the same coordinate. 
To obtain the projected profile $q_{f\to i}$ along coordinate $x_i$,
we integrate the density $F_s$ for concentrations and shoulders, or
the missing density $\Delta F$ for depressions, over the other
coordinate within the measurement region $\mathcal W_f$.
Positive marginal features are compared
with peaks, paired ridges, and shoulders; dips with interior pits and paired
valleys.
The marginalisation, one-dimensional detection and posterior
tracking are described in Appendix~\ref{app:technical_marginals}.

Let $I_1$ be a marginal feature interval and $I_2$ the projected extent of
a joint measurement region. Two complementary comparisons are the overlap $J$ and the capture fraction $C$, defined as
\begin{align}
 J&=\frac{|\ln I_1\cap\ln I_2|}{|\ln I_1\cup\ln I_2|},
 \label{eq:method_projection_comparison}\\
 C&=\frac{\int_{I_1}q_{f\to i}\,d\nu_i^{\rm 1D}}
          {\int q_{f\to i}\,d\nu_i^{\rm 1D}}.
 \label{eq:method_capture_fraction}
\end{align}
Here $|\ln I|$ is interval length in logarithmic coordinates and
$d\nu_i^{\rm 1D}$ is $dx_i$ or $d\ln x_i$. $J$ measures geometric overlap;
$C$ measures the fraction of projected feature weight captured by the
independent marginal interval. High capture does not require identical
boundaries. All compatible pairs are retained, and the largest positive
reference overlap selects a primary counterpart. Correspondence is not
one-to-one: several joint structures may contribute to one marginal
feature, or one joint structure to several marginal features. Posterior
comparisons use the common draws where both measurements are valid.

\subsection{Association with an external parameter}
\label{sec:parameter_association}
\label{sec:methods_information}

When each density realisation is accompanied by an aligned parameter sample
$\lambda_b$, a feature measurement $Q_{fb}$ defines the paired posterior sample
$(\lambda_b,Q_{fb})$.
A feature may be recovered more often at some values of the external parameter,
or its measured properties may vary with that parameter even when it is
consistently recovered. We distinguish these two questions: association with
feature recovery or measurement availability, and association with a
continuous feature property.\\

\paragraph{Discrete associations.}
Binary recovery or availability states compare the parameter distributions
between presence and absence using the rank-biserial effect derived from
the Mann--Whitney statistic \citep{Mann1947U,Cureton1956RankBiserial}.
Positive values mean larger $\lambda$ when the state is present.\\

\paragraph{Continuous associations.}
For
continuous measurements, Spearman's rank correlation describes monotonic
association \citep{Spearman1904Association}, while mutual information $I$ also
captures non-monotonic dependence~\citep{Shannon1948CommunicationI}. On the common valid subset $\mathcal V$, this is
\begin{equation}
\begin{split}
 I(\lambda;Q)
 &=\int\pi_{\mathcal V}(\lambda,Q)
 \log_2\frac{\pi_{\mathcal V}(\lambda,Q)}
 {\pi_{\mathcal V}(\lambda)\pi_{\mathcal V}(Q)}\,d\lambda\,dQ\\
 &=\mathbb E_{Q\mid\mathcal V}\!\left[
 D_{\rm KL}\{\pi_{\mathcal V}(\lambda\mid Q)
 \Vert\pi_{\mathcal V}(\lambda)\}\right].
\end{split}
\label{eq:method_mutual_information}
\end{equation}
The expected-divergence interpretation links information to the change in
the parameter distribution when the feature measurement is specified
\citep{Shannon1948CommunicationI,Kullback1951Information,Lindley1956Information}.
Information is symmetric and does not establish a causal direction.
Both individual centroid coordinates and their vector are analysed;
the latter can retain dependence not represented by either coordinate
alone. We then use conditional information $I(\lambda;\mu_2\mid\mu_1)$~\citep{WYNER197851} to ask what
dependence the second coordinate retains once the first is known. The same
question can compare a joint feature location with an independently
measured marginal scale.

We use the nearest-neighbour estimators of
\citet{Kraskov2004MutualInformation} and
\citet{Frenzel2007ConditionalMI} for the mutual and conditional information respectively. Their full equations and numerical
conventions are in Appendix~\ref{app:technical_association}. Finite-sample
offsets are assessed using within-chain circular shifts of $\lambda$, in
the spirit of surrogate-data diagnostics \citep{Theiler1992Surrogate}.
The reported two-dimensional information subtracts the surrogate median as an empirical correction
and truncates negative differences at zero.

\section{Controlled validation}
\label{sec:controlled_validation}

\subsection{Deterministic recovery}
\label{sec:controlled_fixed}

We first test recovery from a density with known morphology, defined on a
uniform grid in $u_i=\ln x_i\in[-3.5,0]$ and normalised with respect to
$d\ln x_1\,d\ln x_2$. Its principal structures are two peaks $P_1,P_2$ joined by a ridge
$R$, two pits $D_1,D_2$ joined by a transverse valley $V$, and a common
saddle at the ridge--valley intersection. A broad high-coordinate shelf
also produces a bounded slope-change front $E$. %
The complete analytic construction is specified in Appendix~\ref{app:synthetic_density}.

\begin{figure}[!t]
\centering
\includegraphics[width=0.48\textwidth]
{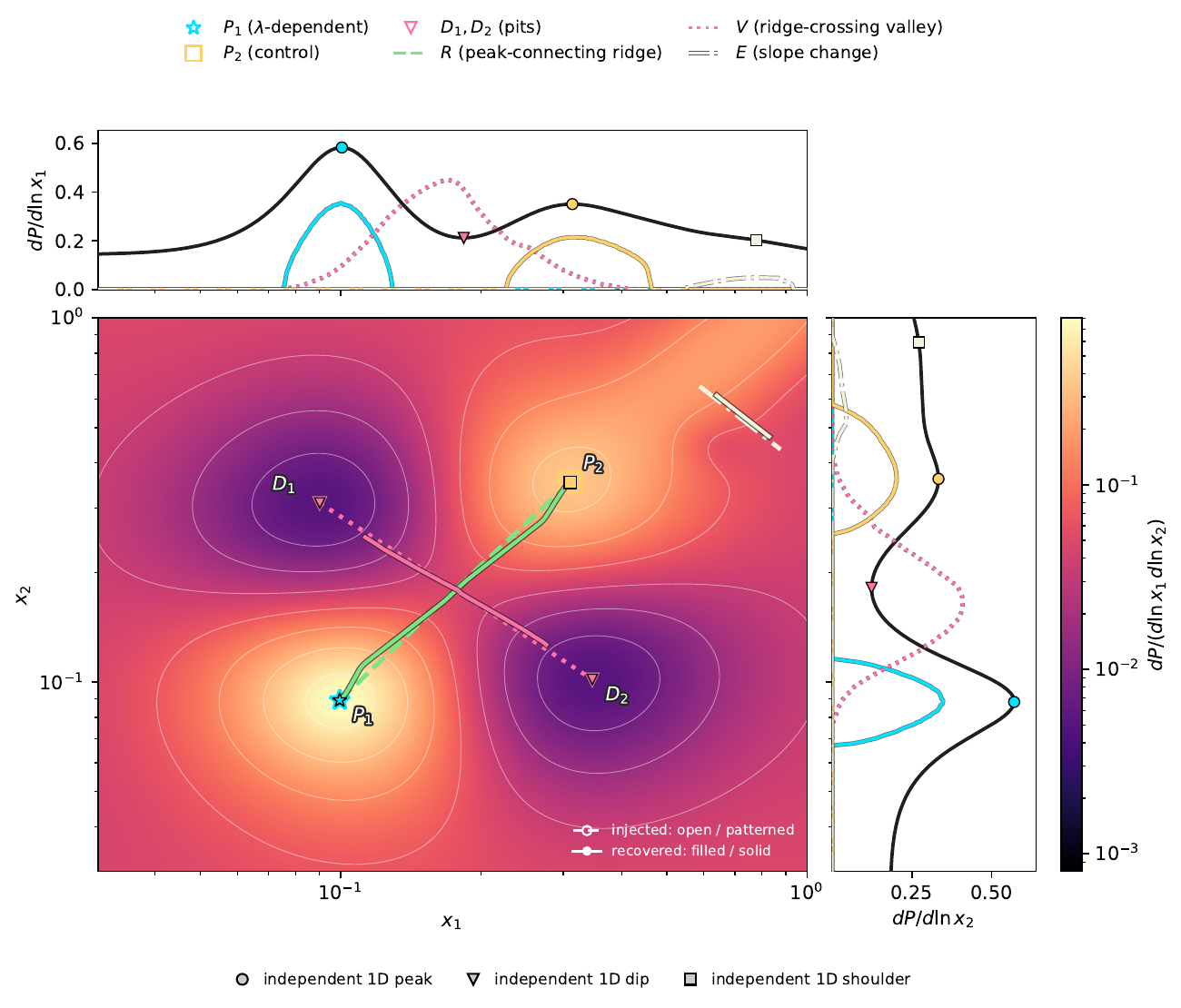}
\caption{
Deterministic recovery of the controlled synthetic density. The central panel compares the injected and reconstructed
peaks, pits, ridge, valley, and shoulder front. Open symbols and
patterned curves indicate the injection; filled symbols and solid
curves indicate the reconstruction. The marginal panels show the
independently recovered one-dimensional features and projections of
the finite two-dimensional measurement regions.
}
\label{fig:controlled_fixed}
\end{figure}

We use the reference two-dimensional smoothing scales $(0,0.025,0.05)$,
automatic persistence selection, the shoulder bound $\alpha_{\max}=2$,
and a $99.5\%$ probability support for the relevant geometric searches.
The two marginals are analysed independently at scales
$(0.05,0.075,0.10)$; these settings were not tuned to the synthetic model.
Figure~\ref{fig:controlled_fixed} illustrates the injected and recovered features in the
joint density and its marginals.
The recovered catalogue exactly matches the injection: two peaks, two
pits, two ridge branches forming one paired ridge event, two valley
branches forming one paired valley event, and one shoulder. No feature
is missed or added, and all are retained at all three two-dimensional
scales. The four extrema lie on their injected grid cells. The independent
marginal analysis likewise recovers the two maxima, intervening minimum,
and high-coordinate shoulder in each marginal. 

\subsection{Posterior ensembles}
\label{sec:controlled_ensemble}

The deterministic test establishes recovery from a known density. We now test
whether the same features remain identifiable across posterior realisations,
whether their uncertainties are recovered, and whether the finder preserves
an injected dependence on an external parameter.\\

\begin{figure}[!t]
\centering
\includegraphics[width=0.45\textwidth]
{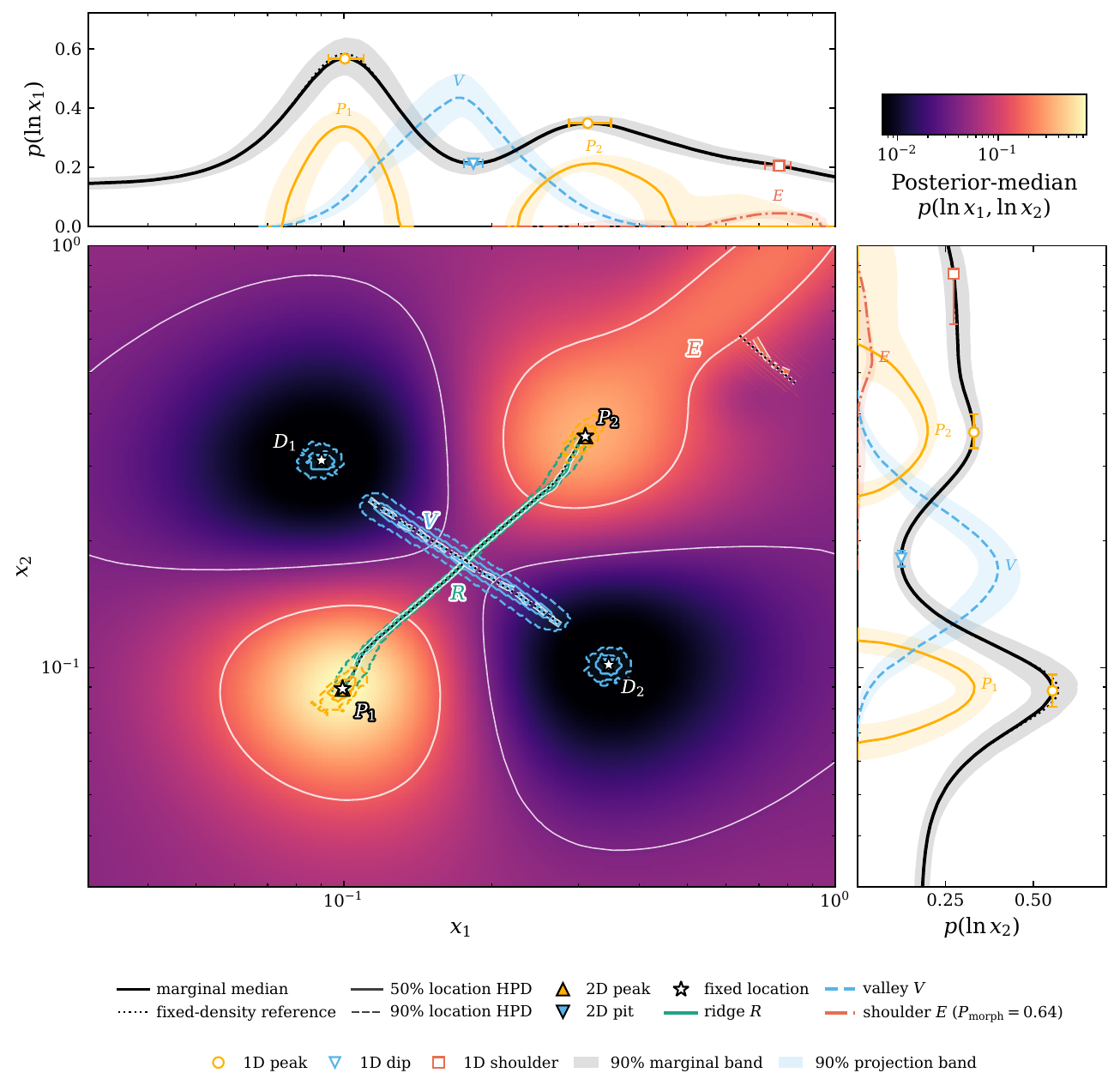}
\caption{
Posterior feature recovery in the controlled ensemble. The central panel
shows the posterior-median density and its $50\%$ and $90\%$ probability
contours. Coloured contours show conditional $50\%$ and $90\%$ location
regions for peaks and pits and location corridors for ridge and valley
branches; faint curves show shoulder fronts across recovered draws.
Black stars and dotted curves indicate the deterministic reference.
The marginal panels show posterior-median densities (black), pointwise
$90\%$ bands (grey), independently recovered 1D feature locations and
intervals, and median projected feature profiles with their uncertainty.
}
\label{fig:controlled_ensemble_geometry}
\end{figure}

\paragraph{Ensemble construction.}
\label{sec:controlled_ensemble_construction}
We generate 1000 equally weighted density maps from the controlled model, in order to mimic the results of a Bayesian posterior.
The full sampling specification is given in
Appendix~\ref{app:ensemble_specification}.
Each realization is paired with a draw $\lambda\sim\mathcal N(0,0.6^2)$, which displaces
$P_1$ along the ridge by $\Delta u_\parallel=0.18\lambda$ and changes its
concentration. The second peak, $P_2$, has no direct $\lambda$-dependence
and acts as a control. 
The finder settings are those
of the deterministic test.

\begin{table}[t]
\caption{
Joint--marginal correspondence in the controlled ensemble: posterior
medians of interval overlap $J$ and projected-weight capture $C$, evaluated
where both measurements are valid. Projections use density for $P_1$, $P_2$
and $E$, and reconstructed missing density for $V$.
}
\label{tab:controlled_ensemble_marginals}
\centering
\small
\setlength{\tabcolsep}{5pt}
\begin{tabular}{@{}lrrrr@{}}
\toprule
Joint feature & \multicolumn{2}{c}{$x_1$ marginal}
              & \multicolumn{2}{c}{$x_2$ marginal} \\
\cmidrule(lr){2-3}\cmidrule(l){4-5}
              & $J$ & $C$ & $J$ & $C$ \\
\midrule
$P_1$ & 0.909 & 1.000 & 0.935 & 0.998 \\
$P_2$ & 0.797 & 0.911 & 0.613 & 0.750 \\
$V$   & 0.273 & 0.534 & 0.385 & 0.692 \\
$E$   & 0.730 & 0.853 & 0.730 & 0.870 \\
\bottomrule
\end{tabular}
\end{table}
\paragraph{Morphology under posterior uncertainty.}
\label{sec:controlled_ensemble_geometry}
Figure~\ref{fig:controlled_ensemble_geometry} shows the reconstructed features with their spatial uncertainty
and the corresponding marginal distributions, together with the underlying truth.

All four critical points are recovered in every draw. Their fixed-density
locations coincide with the posterior medians at the displayed precision
and lie within the posterior location regions; numerical intervals are
provided in
Appendix~\ref{app:ensemble_specification}. Ridge and valley branches are
also recovered with $P_{\rm morph}\geq0.998$, and the fixed-density curves
pass through their posterior corridors. The shoulder is less certain:
$P_{{\rm morph},E}=P_{{\rm loc},E}=0.636$ and
$P_{{\rm reg},E}=0.635$. Thus a feature can be clearly defined on the
reference density while being recovered in only part of the posterior.\\

\begin{figure}[!t]
\centering
\includegraphics[width=0.45\textwidth]
{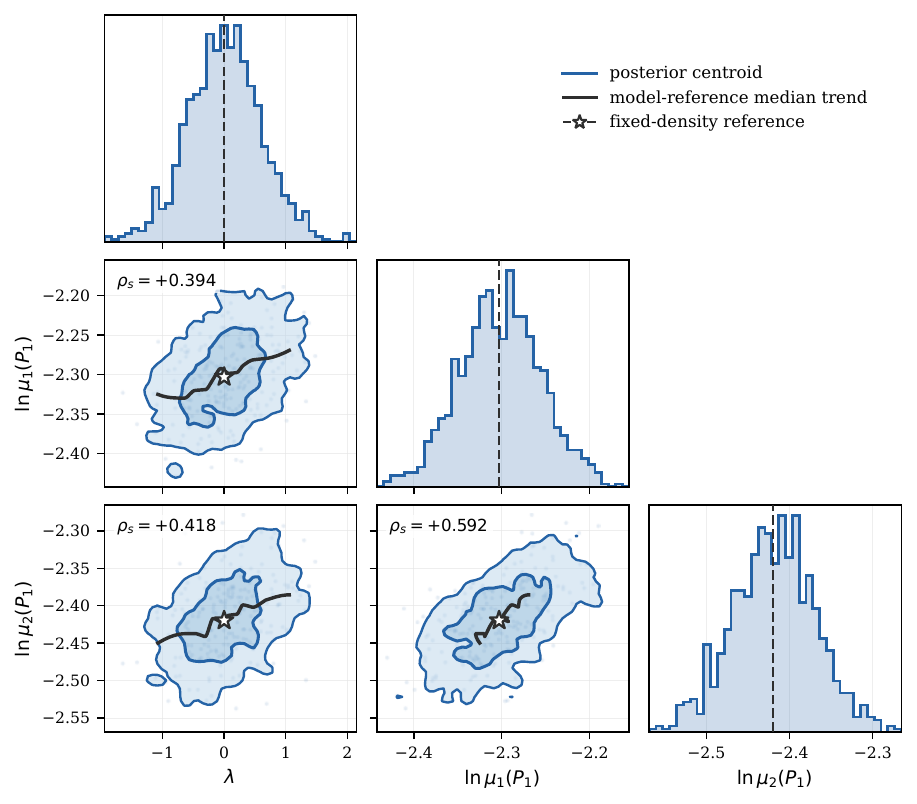}
\caption{
Association of $\lambda$ with the 2D finite-region centroid of $P_1$.
Blue samples and $50\%$ and $90\%$ contours show the recovered posterior.
The solid black curve shows the model-reference median trend in bins of
$\lambda$ (also in the centroid--centroid panel); the star and dashed lines
mark the deterministic reference at $\lambda=0$.
}
\label{fig:controlled_ensemble_lambda}
\end{figure}
\begin{table}[t]
\centering
\caption{
Model-reference and recovered associations between $\lambda$ and the
peak centroids, evaluated on the same posterior draws. The first two
statistics are Spearman correlations with logarithmic centroid coordinates;
the final column is offset-corrected mutual information with the joint
physical centroid.
}
\label{tab:controlled_ensemble_lambda}
\small
\setlength{\tabcolsep}{3pt}
\begin{tabular}{llccc}
\toprule
Feature & Measurement & $\rho_s(\lambda,\ln\mu_1)$ &
$\rho_s(\lambda,\ln\mu_2)$ & $I(\lambda;\boldsymbol{\mu})$ [bit] \\
\midrule
$P_1$ & Model reference & 0.394 & 0.418 & 0.160 \\
 & Recovered & 0.394 & 0.418 & 0.158 \\
\addlinespace[2pt]
$P_2$ & Model reference & -0.021 & -0.015 & 0.000 \\
 & Recovered & -0.021 & -0.015 & 0.000 \\
\bottomrule
\end{tabular}
\end{table}

\paragraph{Correspondence with the marginals.}
We compare independent 1D feature intervals with the projections of
finite 2D measurement regions using interval overlap $J$ and projected-weight
capture $C$ defined in Eqs.~\eqref{eq:method_projection_comparison} and
\eqref{eq:method_capture_fraction}. Table~\ref{tab:controlled_ensemble_marginals} reports their
posterior medians. The valley is particularly illustrative: its relatively
small overlaps ($J=0.273,0.385$) coexist with substantial capture of the
projected deficit ($C=0.534,0.692$). Hence agreement in interval boundaries
and capture of feature weight are distinct properties. The shoulder is
also well captured when measurable ($C=0.853,0.870$).\\

\paragraph{Recovery of the injected parameter dependence.}
\label{sec:controlled_ensemble_lambda}
We finally test the association between $\lambda$ and the finite-region
centroid of $P_1$. The finder
measures this centroid on each density draw; the corresponding model
reference is calculated independently from the known generative density
using the same $\lambda$ and nuisance-parameter draws.
Figure~\ref{fig:controlled_ensemble_lambda} compares the recovered
posterior with the median model-reference trend.

The recovered centroid follows the injected trend in both coordinates.
Table~\ref{tab:controlled_ensemble_lambda} compares the associations using
identical estimators for the recovered and model-reference measurements:
for $P_1$, the Spearman correlations agree at the quoted precision, and
the mutual information is $0.158$ rather than $0.160$ bit. The control
peak $P_2$ exhibits negligible dependence in both measurements.

The main feature identities and the injected parameter association are
stable under several configuration variations; finer smoothing resolution and fixed
persistence cutoffs can additionally admit weaker reference topology. Details are discussed in
Appendix~\ref{app:robustness}.

\section{Applications to gravitational-wave astronomy}
\label{sec:gw_case_study}

In this Section, we apply the framework to existing reconstructions of the compact binary mass spectrum from gravitational-wave observations of the LIGO-Virgo-KAGRA (LVK) catalogs.

We focus on two complementary approaches applied to the GWTC-5.0 catalog~\cite{LIGOScientific:2026wfs}. The first is a "weakly modelled" approach where the population structure is modelled with a free parameter per pixel in the joint mass plane---the \texttt{PixelPop} framework~\cite{Heinzel:2024jlc}. We use the result of the LVK GWTC-5.0 population paper~\cite{LIGOScientific:2026ctl}. 
The corresponding data products are available at Ref.~\cite{ligo_scientific_collaboration_2026_20292639}.

The second approach is a "stongly modelled" one where the two dimensional mass spectrum is modelled with a prescribed functional form whose parameters are inferred from the data. Additionally, in this second case the mass spectrum is inferred alongside the Hubble constant $H_0$ (corresponding to the \emph{spectral-siren} method). A change in the latter can shift the position of the inferred features, which allows studying their mutual dependence as explained in Sec~\ref{sec:parameter_association}. 
Specifically, we use spectral-siren results of the LVK GWTC-5.0 cosmology paper with the \textsc{FullPop-4.0} population model~\cite{LIGOScientific:2026uyd}.
The corresponding data products are available at Ref.~\cite{ligo_scientific_collaboration_2026_20378418}.

\begin{figure}[t]
\centering
\IfFileExists{figures/pixelpop_landscape.pdf}{%
  \includegraphics[width=\linewidth]{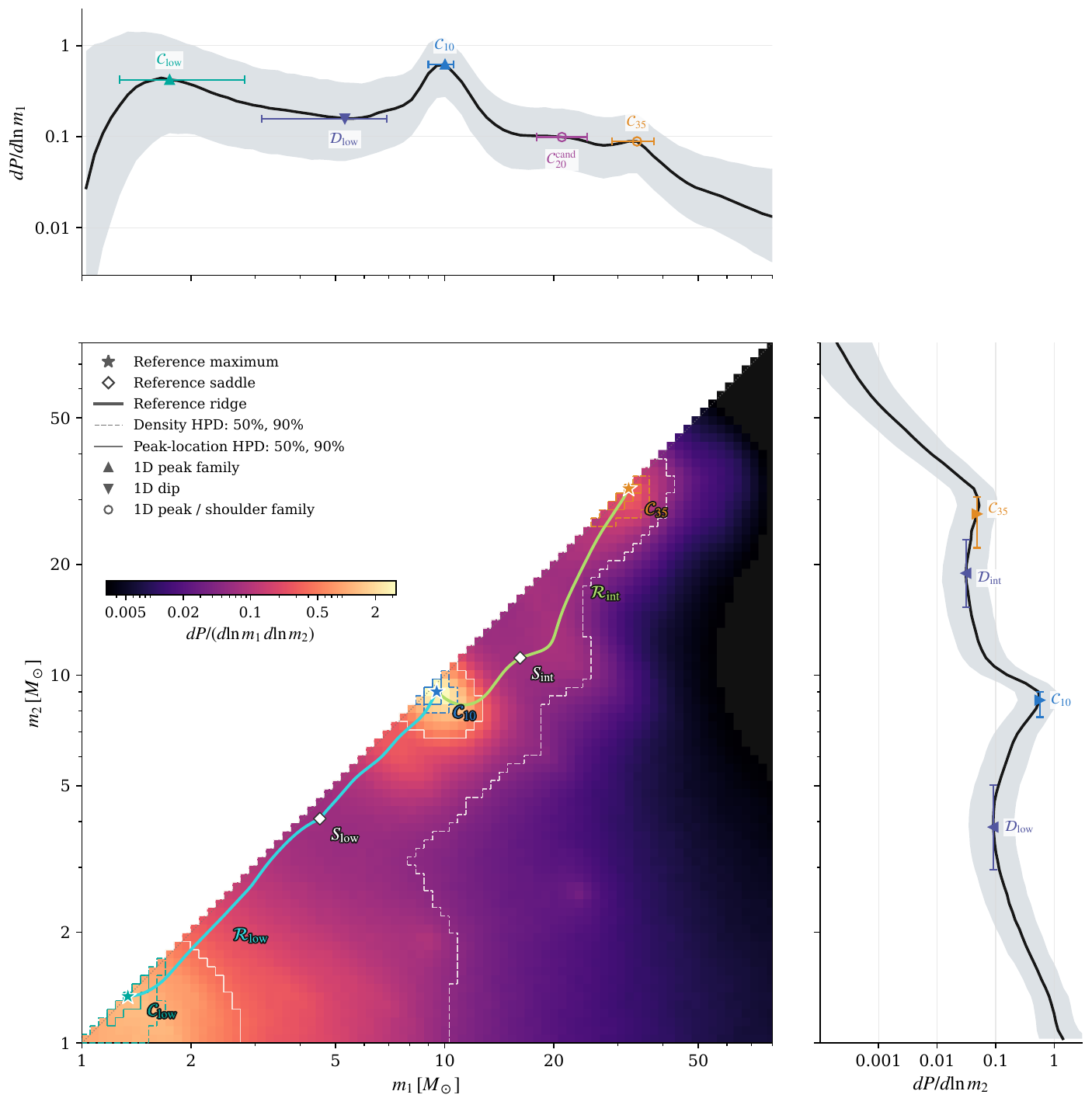}%
}{%
  \fbox{\parbox[c][0.30\textheight][c]{0.92\columnwidth}{\centering
  PixelPop landscape figure: joint density with three reference maxima,
  two saddles and four ridge arms, and aligned independent marginal panels.}}%
}
\caption{Baseline \texttt{PixelPop} mass landscape and independent marginal feature analysis from GWTC-5.0 data~\cite{LIGOScientific:2026ctl}. The central panel shows the normalized posterior-median density $dP/(d\ln m_1\,d\ln m_2)$, with contours enclosing $50\%$ and $90\%$ of its probability. Symbols mark reference maxima and saddles. Thin coloured contours show conditional posterior peak-location regions.
The upper and right panels show median draw-level marginal densities, with pointwise central $90\%$ bands. Feature symbols and horizontal or vertical location intervals summarize the independently analysed marginal peaks, dips, and shoulders.}
\label{fig:pixelpop_landscape}
\end{figure}

\subsection{Weakly modelled reconstruction}
\label{sec:pixelpop_case_study}

\begin{table*}[t]
\centering
\small
\renewcommand{\arraystretch}{1.20}
\caption{Baseline \texttt{PixelPop} feature census from GWTC-5.0 data: stable structure. Joint positions are posterior peak-location medians; all masses are in $M_\odot$. Location and strength intervals are central $90\%$ intervals conditional on recovery and valid measurement, respectively. $P_{\rm loc}$ is the unconditional joint location-recovery probability; $P_{\rm geom}$ additionally requires a valid finite-region measurement. Joint relative prominence is $A_{\rm P}=1-F_{\rm base}/F_{\max}$, where $F_{\rm base}$ is the associated saddle or hierarchy base level. Marginal manifestations summarize the corresponding mass ranges in the independent one-dimensional catalogues; their probabilities describe location recovery, and morphology fractions are conditional on that recovery.}
\label{tab:pixelpop_stable_structure}
\resizebox{\textwidth}{!}{%
\setlength{\tabcolsep}{3pt}%
\begin{tabular}{@{}p{0.10\textwidth}p{0.10\textwidth}p{0.16\textwidth}p{0.10\textwidth}p{0.12\textwidth}p{0.17\textwidth}p{0.17\textwidth}@{}}
\hline
Feature & Type & Joint location & $P_{\rm loc}/$\newline$P_{\rm geom}$ & Joint $A_{\rm P}$ & $m_1$ manifestation & $m_2$ manifestation \\
\hline
$\mathcal C_{\rm low}$ & Boundary concentration
& $m_1=1.41^{+0.24}_{-0.27}$\newline$m_2=1.27^{+0.30}_{-0.24}$
& $99.98\%/$\newline$68.28\%$
& $0.551$\newline$[0.123,0.975]$
& Peak at $1.74^{+1.07}_{-0.48}$;\newline$P_{\rm loc}=98.8\%$;\newline$95.3\%$ peak
& No separate positive marginal family \\

$\mathcal C_{10}$ & Concentration
& $m_1=9.50^{+0.52}_{-0.96}$\newline$m_2=9.01^{+0.49}_{-0.47}$
& $99.90\%/$\newline$90.88\%$
& $0.970$\newline$[0.331,0.991]$
& Peak at $10.02^{+0.55}_{-1.01}$;\newline$P_{\rm loc}=100\%$;\newline$99.8\%$ peak
& Peak at $8.55^{+0.47}_{-0.86}$;\newline$P_{\rm loc}=100\%$;\newline$>99.9\%$ peak \\

$\mathcal C_{35}$ & Boundary concentration
& $m_1=32.15^{+1.75}_{-4.72}$\newline$m_2=30.49^{+3.41}_{-4.48}$
& $99.20\%/$\newline$98.45\%$
& $0.814$\newline$[0.501,0.908]$
& Feature at $33.90^{+3.79}_{-4.98}$;\newline$P_{\rm loc}=79.2\%$;\newline$35.3\%$ peak, $64.7\%$ shoulder
& Peak at $27.42^{+3.07}_{-5.24}$;\newline$P_{\rm loc}=99.98\%$;\newline$95.3\%$ peak \\
\hline
\end{tabular}}
\end{table*}

For the \texttt{PixelPop} reconstruction based on
Ref.~\cite{LIGOScientific:2026ctl}, the input comprises 4000 posterior
draws of $F=dP/(d\ln m_1\,d\ln m_2)$ on a $100\times100$
cell-centred grid over the ordered mass plane $m_2\leq m_1$. 

The baseline two-dimensional analysis uses logarithmic smoothing scales $\sigma_{\ln m}=(0.05,0.10,0.15)$, with the smallest scale close to the typical pixel size. 
The reference hierarchy selects an automatic persistence threshold $T_0=0.04843$, at the geometric midpoint of prominences 0.01513 and 0.15503, separated by a logarithmic gap $G=2.327$. The independent one-dimensional analyses use scales $(0.10,0.15,0.20)$. We vary these assumptions to check the sensitivity of the results; the outcome is discussed in Appendix~\ref{app:gw_sensitivity}. Unless otherwise stated, quoted location and strength intervals are central $90\%$ intervals conditional on recovery and a valid measurement, respectively.

Figure~\ref{fig:pixelpop_landscape} shows the joint landscape with its independently analysed marginals. The central panel displays the normalized posterior-median density, with reference maxima, saddles, and ridge branches superposed.
The upper and right panels show posterior medians and pointwise $90\%$ intervals of the draw-level marginals, together with independently detected peaks, dips, and shoulders.\\

\paragraph{Stable structure.}
The joint catalogue retains three principal concentrations, all recovered in more than $99\%$ of baseline draws,  with posterior-median peak locations
\begin{equation}
\begin{aligned}
\mathcal C_{\rm low}&:\quad (m_1,m_2)=(1.41^{+0.24}_{-0.27},1.27^{+0.30}_{-0.24})\,M_\odot,\\
\mathcal C_{10}&:\quad (m_1,m_2)=(9.50^{+0.52}_{-0.96},9.01^{+0.49}_{-0.47})\,M_\odot,\\
\mathcal C_{35}&:\quad (m_1,m_2)=(32.15^{+1.75}_{-4.72},30.49^{+3.41}_{-4.48})\,M_\odot.
\end{aligned}
\label{eq:pixelpop_stable_locations}
\end{equation}
These corresponds to the main structures discussed in the corresponding analysis~\cite{LIGOScientific:2026ctl}; the framework introduced here shows how to identify their presence, extension and uncertainty in a principled, reproducible way.
 Table~\ref{tab:pixelpop_stable_structure} gives their location uncertainties and relative prominences, separating location recovery, $P_{\rm loc}$, from recovery with a valid finite-region measurement, $P_{\rm geom}$. This distinction is important for $\mathcal C_{\rm low}$: its location is recovered almost invariably, but its finite region is measurable in only $68.3\%$ of draws. 
The three concentrations remain in the catalogue throughout the sensitivity tests, although their recovery probabilities and measured geometry can vary; see Appendix~\ref{app:gw_sensitivity}.

The independent marginal manifestations are not simply copies of the joint maxima. The low-mass concentration is accompanied by a predominantly peak-like primary-mass feature, but no separately identified positive secondary-mass family. $\mathcal C_{10}$ has clear peak counterparts in both marginals. $\mathcal C_{35}$ has a predominantly peak-like secondary-mass counterpart and a primary-mass family that more often takes the form of a shoulder.
This provides additional quantitative insight to the interpretation of marginal features and their relation with the two-dimensional view; in particular, it clarifies the nature of $\mathcal C_{35}$ as a clear two dimensional peak. The corresponding one--dimensional ambiguity appears as an artifact of marginalization~\cite{DeRenzis:2026hcp}.
\\

\paragraph{Tentative structure: ridge connections and the low-mass depletion.}
In addition to the main known structures, the feature finder provides additional insight on tentative emerging features.\footnote{Here \emph{tentative} refers to the joint classification or cross-dimensional interpretation; it does not imply weak marginal evidence in every case.}
In particular, interpretation of the remaining structure requires a combined joint and marginal view. The baseline retains four ridge arms, forming two connections
\begin{equation}
\begin{aligned}
\mathcal C_{\rm low}&\ \longleftarrow\ \mathcal S_{\rm low}\ \longrightarrow\ \mathcal C_{10},\\
\mathcal C_{10}&\ \longleftarrow\ \mathcal S_{\rm int}\ \longrightarrow\ \mathcal C_{35}.
\end{aligned}
\label{eq:pixelpop_ridge_connections}
\end{equation}
The arrows indicate ascent from each saddle to its two connected maxima. 
Their unconditional posterior recovery probabilities are $19.4\%$ and $32.3\%$, respectively, requiring recovery of both arms with the corresponding saddle and peak identities. 

Between $\mathcal C_{\rm low}$ and $\mathcal C_{10}$, the independent marginals contain dips at
\begin{equation}
\begin{aligned}
m_{1,\rm dip}&=5.31^{+1.61}_{-2.18}\,M_\odot,
& P_{\rm loc}&=99.78\%,\\
m_{2,\rm dip}&=3.86^{+1.17}_{-0.90}\,M_\odot,
& P_{\rm loc}&=99.65\%.
\end{aligned}
\label{eq:pixelpop_low_depletion}
\end{equation}
Their median relative depths are 0.659 and 0.805, respectively. The baseline reconstruction therefore supports a depletion in the NS--BH transition between the low-mass and intermediate-mass concentrations in the marginals, while not identifying an above-threshold corresponding under density in the two--dimensional plane.

The low-mass ridge connection provides however an additional way to examine this transition. Its saddle marks the bottleneck at which the two higher-density regions merge as the density threshold is lowered. Conditional on recovery of the complete connection, the posterior saddle location is
\begin{equation}
(m_1,m_2)_{\mathcal S_{\rm low}}^{\rm post}
=
(5.60^{+2.95}_{-1.93},\,3.47^{+1.56}_{-1.08})\,M_\odot.
\label{eq:pixelpop_low_saddle}
\end{equation}
These broad coordinate ranges overlap those highlighted by the marginal dips. The saddle and marginal minima are different objects, but their combined geometry supports interpreting this part of the plane as an under-dense transition between the principal concentrations.\footnote{Note that a ridge can decrease in density toward a bottleneck saddle: its passage through the depleted range does not mean that the depletion is filled, or establish an astrophysical connection between the populations on either side.}\\

\paragraph{Tentative structure: the intermediate-mass transition.}
An additional positive primary-mass family lies at
\begin{equation}
m_1=21.04^{+3.63}_{-3.09}\,M_\odot,
\qquad P_{\rm loc}=66.9\%.
\label{eq:pixelpop_c20}
\end{equation}
We denote it by $\mathcal C_{20}^{\rm cand}$ without assigning it a fourth retained joint concentration. Conditional on recovery, $41.3\%$ of its realizations are peaks and $58.7\%$ are shoulders. Its unconditional probabilities are 0.276 for a peak, 0.393 for a shoulder, and 0.331 for absence. Peak-like realizations have low relative prominence, $A_{\rm P}^{1\rm D}=0.179\,[0.060,0.391]$; shoulder-like realizations have broadly uncertain slope contrast, $\Delta_S=0.796\,[-0.150,1.955]$. No separate positive secondary-mass family provides a counterpart at this scale.

The reference two--dimensional analysis also finds a weak off-diagonal local maximum near $(20.0,12.4)\,M_\odot$, close to the intermediate-mass ridge. Its persistence is however only 0.00731, below $T_0$ and even $T_0/2$, so it does not enter the final peak catalogue.
Additionally, a ridge connects $\mathcal C_{10}$ to $\mathcal C_{35}$ through $\mathcal S_{\rm int}$, whose conditional posterior location is
\begin{equation}
(m_1,m_2)_{\mathcal S_{\rm int}}^{\rm post}
=
(21.04^{+4.97}_{-6.52},\,16.15^{+4.90}_{-6.12})\,M_\odot.
\label{eq:pixelpop_intermediate_saddle}
\end{equation}
Interestingly, the primary-mass median coincides with the marginal candidate's median location. 

This supplies quantitative context for the interpretation of emerging intermediate-mass structure: several weak indicators point to possible structure around the quoted scales, allowing to go beyond a binary choice on the presence or absence of a feature based just on visual inspection of the median density and marginal distributions.
Taken together, the marginal features, the weak local joint maximum, and the saddle-centred ridge connection identify the intermediate-mass range as potentially structured. With the present reconstruction and data, however, the additional local joint maximum remains below the persistence cut, and no additional retained joint valley is identified.

\paragraph{Higher-mass marginal dip.}
The primary-mass catalogue also contains a broad dip near $127\,M_\odot$, recovered in $89.9\%$ of draws. It may be an artifact of an increase in the inferred rate at high mass that can be driven by the PixelPop prior~\cite{Heinzel:2024jlc}, so we do not assign it an astrophysical interpretation.

\subsection{Strongly modelled reconstruction}
\label{sec:fullpop_case_study}

\begin{figure}[t]
\centering
\IfFileExists{figures/fullpop_landscape.pdf}{%
  \includegraphics[width=\linewidth]{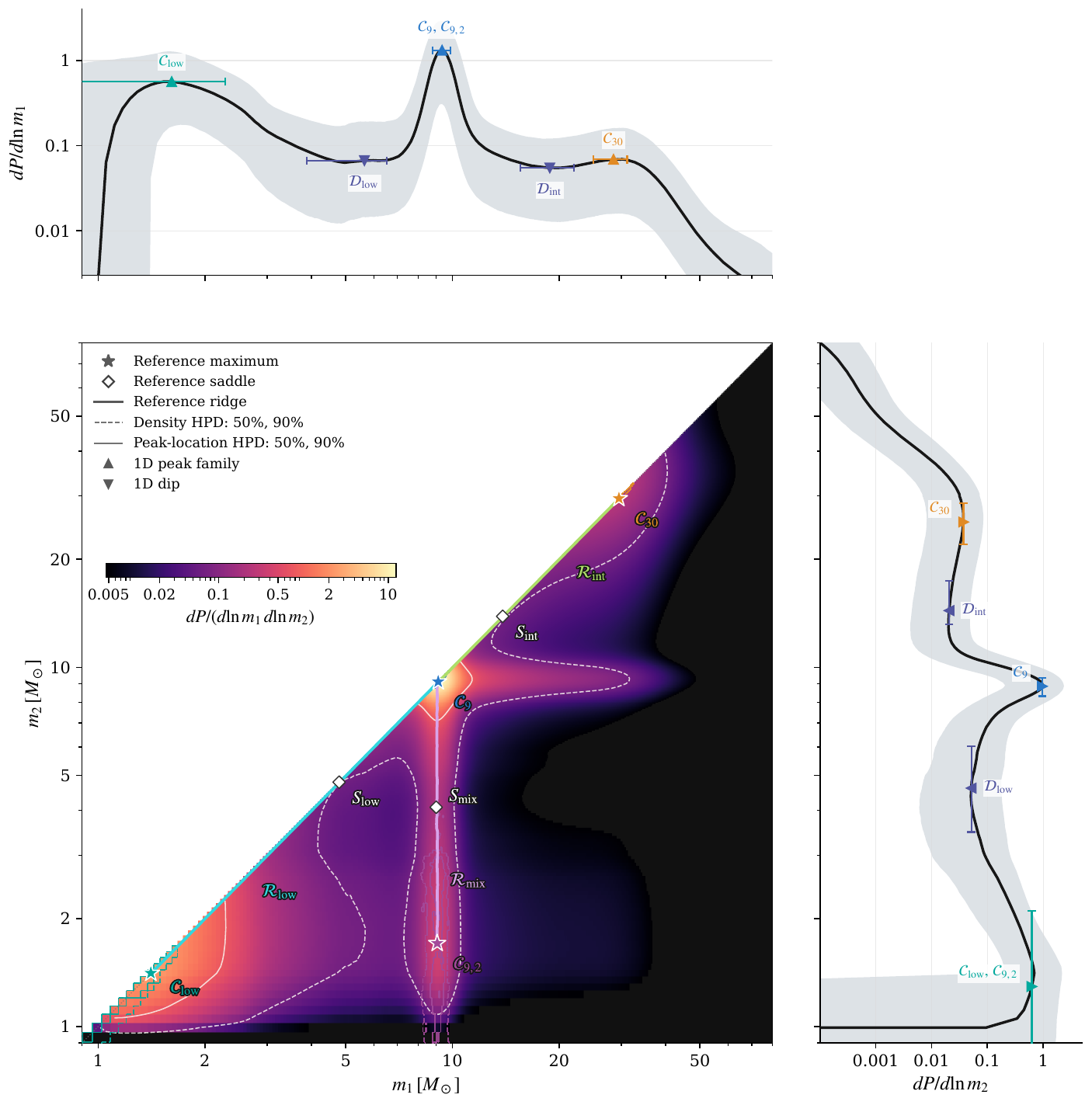}%
}{%
  \fbox{\parbox[c][0.30\textheight][c]{0.92\columnwidth}{\centering
  FullPop landscape: four reference maxima, three saddles and six ridge arms,
  with aligned independent marginal panels. No shoulder fronts or inset.}}%
}
\caption{Baseline GWTC-5.0 cosmology \textsc{FullPop-4.0} mass landscape~\cite{LIGOScientific:2026uyd}, using the same conventions as Fig.~\ref{fig:pixelpop_landscape}. The joint panel shows the normalized posterior-median logarithmic density, probability contours, conditional peak-location regions, and the four reference maxima with three saddle-centred ridge connections. The reference geometry differs from the posterior location medians in Table~\ref{tab:fullpop_concentrations}. The marginal panels show draw-level median densities and pointwise $90\%$ bands, with independent feature-location intervals placed on the curves. Shoulder fronts are omitted.}
\label{fig:fullpop_landscape}
\end{figure}
\begin{table*}[t]
\centering
\small
\renewcommand{\arraystretch}{1.20}
\caption{Baseline \textsc{FullPop-4.0} principal concentrations and their marginal manifestations. Joint locations are posterior medians; all masses are in $M_\odot$. Locations and relative prominences have central $90\%$ intervals conditional on recovery and valid measurement, respectively. $P_{\rm loc}$ and $P_{\rm geom}$ denote unconditional location and finite-region recovery, as in Table~\ref{tab:pixelpop_stable_structure}; $A_{\rm P}=1-F_{\rm base}/F_{\max}$. Marginal locations and probabilities come from the independent 1D catalogues. Shared entries identify overlapping projected manifestations rather than unique associations.}
\label{tab:fullpop_concentrations}
\resizebox{\textwidth}{!}{%
\setlength{\tabcolsep}{3pt}%
\begin{tabular}{@{}
>{\raggedright\arraybackslash}p{0.10\textwidth}
>{\raggedright\arraybackslash}p{0.10\textwidth}
>{\raggedright\arraybackslash}p{0.16\textwidth}
>{\raggedright\arraybackslash}p{0.10\textwidth}
>{\raggedright\arraybackslash}p{0.12\textwidth}
>{\raggedright\arraybackslash}p{0.17\textwidth}
>{\raggedright\arraybackslash}p{0.17\textwidth}@{}}
\hline
Feature & Type & Joint location & $P_{\rm loc}/$\newline$P_{\rm geom}$ & Joint $A_{\rm P}$ & $m_1$ manifestation & $m_2$ manifestation \\
\hline
$\mathcal C_{\rm low}$ & Boundary concentration
& $m_1=1.35^{+0.24}_{-0.48}$\newline$m_2=1.35^{+0.24}_{-0.48}$
& $88.14\%/$\newline$79.24\%$
& $0.985$\newline$[0.862,0.998]$
& Peak at $1.53^{+0.70}_{-0.78}$;\newline$P_{\rm loc}=99.69\%$
& Low-mass family at $1.37^{+0.79}_{-0.47}$;\newline$P_{\rm loc}=99.94\%$ \\

$\mathcal C_{9,2}$ & Unequal-mass concentration
& $m_1=8.99^{+0.72}_{-0.48}$\newline$m_2=1.83^{+0.95}_{-0.95}$
& $74.52\%/$\newline$74.43\%$
& $0.784$\newline$[0.321,0.963]$
& Shared peak at $9.35^{+0.52}_{-0.56}$;\newline$P_{\rm loc}=100\%$
& Same low-mass family as $\mathcal C_{\rm low}$ \\

$\mathcal C_9$ & Boundary concentration
& $m_1=9.23^{+0.48}_{-0.72}$\newline$m_2=9.23^{+0.48}_{-0.72}$
& $99.81\%/$\newline$99.65\%$
& $0.988$\newline$[0.963,0.997]$
& Same primary-mass family as $\mathcal C_{9,2}$
& Peak at $8.85^{+0.50}_{-0.53}$;\newline$P_{\rm loc}=100\%$ \\

$\mathcal C_{30}$ & Boundary concentration
& $m_1=29.28^{+3.44}_{-3.10}$\newline$m_2=29.28^{+3.44}_{-3.10}$
& $72.00\%/$\newline$72.00\%$
& $0.761$\newline$[0.531,0.897]$
& Peak at $28.82^{+2.46}_{-3.85}$;\newline$P_{\rm loc}=80.68\%$
& Peak at $25.49^{+3.53}_{-3.40}$;\newline$P_{\rm loc}=99.90\%$ \\
\hline
\end{tabular}}
\end{table*}

We next apply the feature finder to the parametric \textsc{FullPop-4.0} reconstruction used in the GWTC-5.0 cosmology analysis~\cite{LIGOScientific:2026uyd}.
This case also allows to study the correlation of emerging features with the Hubble constant as quantified by \plandscape{}.
The joint density is modelled with a shared component-mass spectrum and a pairing prescription,
\begin{equation}
\begin{aligned}
\pi(m_1,m_2\mid\Lambda)
&\propto \pi_m(m_1\mid\Lambda)\,\pi_m(m_2\mid\Lambda)\\
&\quad{}\times f(m_1,m_2\mid\Lambda)\,\Theta(m_1-m_2).
\end{aligned}
\label{eq:fullpop_mass_model}
\end{equation}
The common spectrum contains continuum and peak components with smoothing filters, while $f$ controls how component masses combine~\cite{LIGOScientific:2025pvj,LIGOScientific:2025jau}. The factors $\pi_m$ are not themselves the primary- or secondary-mass marginals, which also depend on pairing and the ordered domain.

The input comprises 6836 equally weighted posterior draws of $F=dP/(d\ln m_1\,d\ln m_2)$ on a $250\times250$ nodal grid over $m_2\leq m_1$. 
The baseline uses nominal logarithmic smoothing scales $(0,0.05,0.10)$, and selects the persistence threshold  $T_0=0.05800$.
The independent marginal analyses use scales $(0.10,0.15,0.20)$.
Sensitivity to these choices is examined in Appendix~\ref{app:gw_sensitivity}.

Figure~\ref{fig:fullpop_landscape} uses the same presentation as Figure~\ref{fig:pixelpop_landscape}: the joint density and reference topology are displayed alongside independently analysed draw-level marginals. \\ 

\paragraph{Principal concentrations and shared marginal manifestations.}
The joint catalogue retains the following principal concentrations
\begin{equation}
\begin{aligned}
\mathcal{C}_{\mathrm{low}} &: \quad
(m_1,m_2) =
(1.35^{+0.24}_{-0.48},1.35^{+0.24}_{-0.48})\,M_\odot,\\
\mathcal{C}_{9,2} &: \quad
(m_1,m_2) =
(8.99^{+0.72}_{-0.48},1.83^{+0.95}_{-0.95})\,M_\odot,\\
\mathcal{C}_{9} &: \quad
(m_1,m_2) =
(9.23^{+0.48}_{-0.72},9.23^{+0.48}_{-0.72})\,M_\odot,\\
\mathcal{C}_{30} &: \quad
(m_1,m_2) =
(29.28^{+3.44}_{-3.10},29.28^{+3.44}_{-3.10})\,M_\odot.
\end{aligned}
\end{equation}
Table~\ref{tab:fullpop_concentrations} gives their locations, recovery and relative prominences. $\mathcal C_9$ is the most consistently recovered joint concentration. The others have substantial conditional prominence but lower recovery. 
The unequal-mass maximum however  persists at two baseline scales, while the other maxima persist at all three.

Each marginal contains three principal positive families. $\mathcal C_{9,2}$ shares its primary-mass manifestation with $\mathcal C_9$ and its low-secondary-mass manifestation with $\mathcal C_{\rm low}$. Marginalization therefore combines structures that are distinct in the joint plane. These overlaps are nonexclusive: the marginal catalogues do not provide a unique decomposition into joint concentrations.

The three equal-mass boundary reference maxima $\mathcal{C}_{\rm low}$, $\mathcal{C}_{9}$, $\mathcal{C}_{30}$ correspond to features that are explicitly included in analytic form of the mass spectrum $\pi_m$.
The unequal-mass maximum $\mathcal C_{9,2}$ is instead not prescribed as a separate structure, and not discussed in the corresponding analysis~\cite{LIGOScientific:2026uyd}. Still, it is not an artifact of the algorithm: $\mathcal C_{9,2}$ (and its related ridge, see below) can actually be strongly shaped by the \textsc{FullPop-4.0} assumptions. A shared spectrum with low-mass and $\sim9\,M_\odot$ structure supplies both same-mass and unequal-mass combinations in Eq.~\eqref{eq:fullpop_mass_model}. 
We also note that the presence of this feature and related connections are setting-dependent and can become sub-threshold in some configurations; the main peaks remain instead stable. See Appendix~\ref{app:gw_sensitivity} for details.

\begin{figure*}[t]
    \centering
    \includegraphics[width=0.8\textwidth]{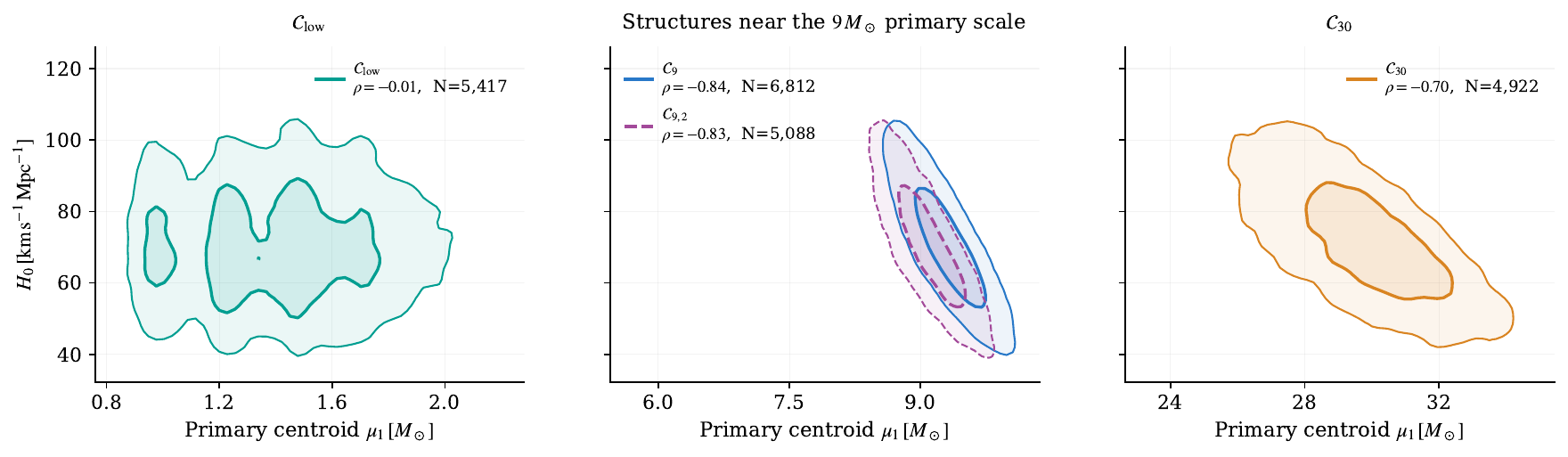}
    \caption{Posterior associations between $H_0$ and the primary-coordinate
    centroid of the four joint concentrations identified in the \textsc{FullPop-4.0} GWTC-5.0 cosmology analysis~\cite{LIGOScientific:2026uyd}. Contours show the $50\%$ and
    $90\%$ highest-density regions of the smoothed posterior samples; annotations
    give the Spearman coefficients. 
    }
    \label{fig:fullpop_h0_correlations}
\end{figure*}
\begin{figure}[t]
    \centering
    \includegraphics[width=0.48\textwidth]{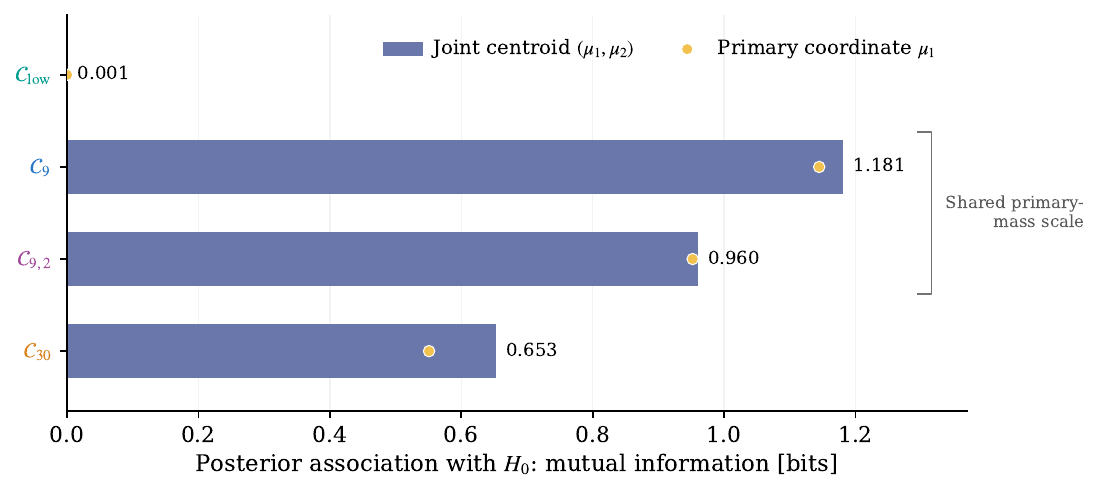}
    \caption{Mutual information between $H_0$ and each identified
    concentration's two-coordinate centroid (bars), in the \textsc{FullPop-4.0} GWTC-5.0 cosmology analysis~\cite{LIGOScientific:2026uyd}. The corresponding
    primary-coordinate information shown by circles. The bracket groups the
    two structures sharing a primary-mass scale. The estimates, expressed in
    bits, describe posterior associations in each feature's valid sample and
    should not be summed.}
    \label{fig:fullpop_h0_information}
\end{figure}

A maximum of the $m_1$-dependent factor can support structure near primary mass $9\,M_\odot$ while the $m_2$-dependent factor traverses its low- and higher-mass features, providing a natural mechanism for the unequal-mass maximum and connecting ridge.
This shows how the independent, agnostic feature-recovery analysis proposed here can help shedding light on the emerging structure induced by specific parametric assumptions.
The feature is geometrically distinct in this reconstruction, but its recovery is conditional on the shared-spectrum and pairing model, and does not establish independent evidence for an additional physical population.
\\

\paragraph{Marginal depletion.}
Both marginals show low-mass and intermediate-mass dips,
\begin{equation}
\begin{aligned}
m_{1,\rm low\ dip}&=5.64^{+0.87}_{-1.74}\,M_\odot,\\
m_{2,\rm low\ dip}&=4.66^{+1.38}_{-1.14}\,M_\odot,\\
m_{1,\rm int\ dip}&=18.88^{+3.21}_{-3.28}\,M_\odot,\\
m_{2,\rm int\ dip}&=14.47^{+3.05}_{-1.23}\,M_\odot.
\end{aligned}
\label{eq:fullpop_marginal_dips}
\end{equation}
The low-mass dips are recovered in approximately $99\%$ of draws and are deep, with median relative depths 0.906 and 0.921. The intermediate primary-mass dip is shallower and less frequently recovered than its secondary-mass counterpart.
Nearly every draw recovering the higher primary-mass positive family also recovers the preceding intermediate dip.\\

\paragraph{Ridge connections.}
Six arms form three paired connections,
\begin{equation}
\begin{aligned}
\mathcal C_{\rm low}&\ \longleftarrow\
\underset{(4.70,\,4.70)}{S_{\rm low}}\
\longrightarrow\ \mathcal C_9,\\
\mathcal C_{9,2}&\ \longleftarrow\
\underset{(8.99,\,4.22)}{S_{\rm mix}}\
\longrightarrow\ \mathcal C_9,\\
\mathcal C_9&\ \longleftarrow\
\underset{(14.01,\,14.01)}{S_{\rm int}}\
\longrightarrow\ \mathcal C_{30}.
\end{aligned}
\label{eq:fullpop_reference_saddles}
\end{equation}
Coordinates beneath the saddle symbols are fixed reference positions in $M_\odot$; arrows indicate ascent toward the maxima. The complete connections are recovered in $59.9\%$, $57.9\%$ and $68.0\%$ of draws, respectively, requiring both arms and the corresponding saddle and peak identities.
The low- and intermediate-mass saddles locate bottlenecks between the principal concentrations, while the mixed connection links the two structures near primary mass $9\,M_\odot$.
As for the $\mathcal C_{9,2}$ concentration, these features can be induced by the assumed parametric form; the strength of the induced structure can be explicitly quantified in the framework proposed here and compared across different parametric assumptions. \\

\paragraph{Logarithmic-slope changes.}
Additional joint shoulder fronts and a tentative primary-mass slope-change family are omitted from the main census and figure. Breaks, peak components and smoothing filters can induce such changes in logarithmic slope, so we do not assign them separate physical interpretations. Our companion study~\cite{DeRenzis:2026hcp} demonstrates this explicitly on simulated data.\\

\paragraph{Association with the Hubble constant}
\label{sec:fullpop_h0_associations}

We finally illustrate the association to a parameter that correlates with feature positions---the Hubble constant $H_0$. For each identified concentration,
we use the density-weighted finite-region centroid
$\boldsymbol{\mu}=(\mu_1,\mu_2)$.
Figure~\ref{fig:fullpop_h0_correlations} shows the posterior associations between
$H_0$ and the primary coordinate $\mu_1$ of each main concentration, together with the  Spearman coefficients. The concentrations near the primary-mass scale of
$9\,M_\odot$ and $\mathcal C_{30}$ exhibit the characteristic negative correlations of spectral-siren anchors, while
the low-mass centroid shows little dependence on $H_0$. 
The two structures
sharing the primary-mass scale near $9\,M_\odot$ are overlaid for comparison.

Figure~\ref{fig:fullpop_h0_information} complements these plots with the
package's mutual-information estimates for the full centroid,
$I(H_0;\boldsymbol{\mu})$, and its primary coordinate alone,
$I(H_0;\mu_1)$.\footnote{Each feature uses the posterior
draws in which its centroid is valid, so the displayed associations are
conditional on feature-specific measurement availability.} 
The values are
non-additive: strong associations of both structures near $9\,M_\odot$ do not
establish independent cosmological contributions or determine their effects
on the uncertainty in $H_0$. This would require a dedicated study.

Both diagnostics identify a clear hierarchy among the principal candidate spectral-siren anchors: $\mathcal C_{9}$ exhibits stronger association with $H_0$ than $\mathcal C_{30}$, with mutual information of approximately 1.18 and 0.65 bits and primary-centroid rank correlations of -0.84 and -0.70, respectively. The low-mass concentration shows instead negligible association through its measured centroid and therefore does not emerge as an appreciable position-based anchor in this posterior.

Together, the two figures demonstrate how \plandscape{} can be used to identify and rank candidate spectral-siren anchors through their correlations and information shared with $H_0$. The features and their finite measurement regions are determined automatically from the reconstructed landscape, under the chosen analysis settings, without manually prescribing specific mass scales or tuning their measurement windows.
Our companion paper~\cite{DeRenzis:2026hcp} applies the same framework to a weakly modelled population reconstruction inferred alongside $H_0$.

\section{Conclusions}

We have presented \plandscape{}, a framework for turning two-dimensional probability densities into catalogues of morphological features, finite measurement regions, and posterior uncertainties. 
The framework brings together ideas from several research fields.
Topological data analysis supplies the concept of persistence and the evolution of
connected density regions, while Morse--Smale descriptions supply
the organisation of extrema and their saddle connections.
Statistical density-ridge methods contribute derivative-based
refinement of curve geometry, and scale-space analysis from computer
vision provides comparisons across smoothing resolutions.
We integrated these constructions with
morphology-specific finite measurement regions and Bayesian
uncertainty propagation in one feature-analysis workflow, turning these geometric objects into quantitative observables whose properties can be followed across density realisations.

We discussed controlled tests that recover the injected morphology and demonstrated how variation in the density propagates into feature recovery, location, and geometry. These tests also reproduce an injected association with an external parameter while leaving a control feature essentially unassociated. 
Separating morphological recovery from localisation and finite-region availability makes it possible to report a feature without implying that all its properties are measurable throughout the posterior, and quantifying such variability.

For gravitational-wave population studies, the framework provides a common quantitative description of structure across different reconstruction methods. The joint and marginal analyses reveal when projection blends distinct concentrations or changes their apparent morphology, while saddle and ridge connections supply context for weaker structure.
We applied the method to state--of--the--art reconstructions of the compact binary mass plane from the latest LVK data. The \texttt{PixelPop} and \textsc{FullPop-4.0} applications display both shared structure and differences shaped by the reconstruction assumptions.
Associations with $H_0$ further show how automatically identified features and measurement regions can quantitatively define a hierarchy of candidate spectral-siren anchors, without prescribing individual mass features or measurement windows by hand. 
This will contribute to the solid interpretation of spectral-siren measurements based on weakly modelled population reconstructions.

A particularly promising application is to compare population-synthesis predictions with weakly modelled reconstructions of the observed population. Predicted and inferred component-mass landscapes could be analysed with consistent measures, domains, and resolutions, and compared through the locations, extents, strengths, and connectivity of their features, together with their uncertainties. This would provide a systematic route from physical population models to testable morphological signatures, extending comparisons beyond visual resemblance or manually selected structures. Such a common representation could help establish which assumptions about binary formation and evolution reproduce the structures inferred from gravitational-wave data.

\section*{Data availability}
\plandscape{} is available open--source at Ref.~\cite{Mancarella_posterior-landscape_morphological_feature_2026}.
Data products used in Sec.~\ref{sec:gw_case_study} are available at~\cite{ligo_scientific_collaboration_2026_20292639,ligo_scientific_collaboration_2026_20378418}

\begin{acknowledgments}
We thank Tom Bertheas for internal LVK review and useful comments on the draft.
This work is supported by the French government under the France 2030 investment plan, as part of the Initiative d'Excellence d'Aix-Marseille Universit\'e -- A*MIDEX AMX-22-CEI-02.
We acknowledge use of computational resources of the MesoCentre of Aix-Marseille University through allocation b435-GWCROSS.
ChatGPT (OpenAI) assisted with developing and debugging the analysis and plotting code under direct direction of the authors, and with drafting and revising portions of the manuscript. All reported numerical results and scientific figures were produced by executing the software locally under the authors' control.
This document has LIGO DCC document number LIGO-P2600571.

{\small This research has made use of data or software obtained from the Gravitational Wave Open Science Center (gwosc.org), a service of the LIGO Scientific Collaboration, the Virgo Collaboration, and KAGRA. This material is based upon work supported by NSF's LIGO Laboratory which is a major facility fully funded by the National Science Foundation, as well as the Science and Technology Facilities Council (STFC) of the United Kingdom, the Max-Planck-Society (MPS), and the State of Niedersachsen/Germany for support of the construction of Advanced LIGO and construction and operation of the GEO600 detector. Additional support for Advanced LIGO was provided by the Australian Research Council. Virgo is funded, through the European Gravitational Observatory (EGO), by the French Centre National de Recherche Scientifique (CNRS), the Italian Istituto Nazionale di Fisica Nucleare (INFN) and the Dutch Nikhef, with contributions by institutions from Belgium, Germany, Greece, Hungary, Ireland, Japan, Monaco, Poland, Portugal, Spain. KAGRA is supported by Ministry of Education, Culture, Sports, Science and Technology (MEXT), Japan Society for the Promotion of Science (JSPS) in Japan; National Research Foundation (NRF) and Ministry of Science and ICT (MSIT) in Korea; Academia Sinica (AS) and National Science and Technology Council (NSTC) in Taiwan.
}

\end{acknowledgments}

\bibliographystyle{apsrev4-2}
\bibliography{refs}

\begin{thebibliography}{41}%
\makeatletter
\providecommand \@ifxundefined [1]{%
 \@ifx{#1\undefined}
}%
\providecommand \@ifnum [1]{%
 \ifnum #1\expandafter \@firstoftwo
 \else \expandafter \@secondoftwo
 \fi
}%
\providecommand \@ifx [1]{%
 \ifx #1\expandafter \@firstoftwo
 \else \expandafter \@secondoftwo
 \fi
}%
\providecommand \natexlab [1]{#1}%
\providecommand \enquote  [1]{``#1''}%
\providecommand \bibnamefont  [1]{#1}%
\providecommand \bibfnamefont [1]{#1}%
\providecommand \citenamefont [1]{#1}%
\providecommand \href@noop [0]{\@secondoftwo}%
\providecommand \href [0]{\begingroup \@sanitize@url \@href}%
\providecommand \@href[1]{\@@startlink{#1}\@@href}%
\providecommand \@@href[1]{\endgroup#1\@@endlink}%
\providecommand \@sanitize@url [0]{\catcode `\\12\catcode `\$12\catcode
  `\&12\catcode `\#12\catcode `\^12\catcode `\_12\catcode `\%12\relax}%
\providecommand \@@startlink[1]{}%
\providecommand \@@endlink[0]{}%
\providecommand \url  [0]{\begingroup\@sanitize@url \@url }%
\providecommand \@url [1]{\endgroup\@href {#1}{\urlprefix }}%
\providecommand \urlprefix  [0]{URL }%
\providecommand \Eprint [0]{\href }%
\providecommand \doibase [0]{https://doi.org/}%
\providecommand \selectlanguage [0]{\@gobble}%
\providecommand \bibinfo  [0]{\@secondoftwo}%
\providecommand \bibfield  [0]{\@secondoftwo}%
\providecommand \translation [1]{[#1]}%
\providecommand \BibitemOpen [0]{}%
\providecommand \bibitemStop [0]{}%
\providecommand \bibitemNoStop [0]{.\EOS\space}%
\providecommand \EOS [0]{\spacefactor3000\relax}%
\providecommand \BibitemShut  [1]{\csname bibitem#1\endcsname}%
\let\auto@bib@innerbib\@empty
\bibitem [{\citenamefont
  {Mancarella}(2026)}]{Mancarella_posterior-landscape_morphological_feature_2026}%
  \BibitemOpen
  \bibfield  {author} {\bibinfo {author} {\bibfnamefont {M.}~\bibnamefont
  {Mancarella}},\ }\href@noop {} {\bibinfo {title} {{posterior-landscape:
  morphological feature finding in two-dimensional probability densities}}},\
  \bibinfo {howpublished}
  {\href{https://github.com/Mik3M4n/posterior-landscape.git}{github.com/Mik3M4n/posterior-landscape.git}}
  (\bibinfo {year} {2026})\BibitemShut {NoStop}%
\bibitem [{\citenamefont {Edelsbrunner}\ \emph {et~al.}(2002)\citenamefont
  {Edelsbrunner}, \citenamefont {Letscher},\ and\ \citenamefont
  {Zomorodian}}]{Edelsbrunner2002Persistence}%
  \BibitemOpen
  \bibfield  {author} {\bibinfo {author} {\bibfnamefont {H.}~\bibnamefont
  {Edelsbrunner}}, \bibinfo {author} {\bibfnamefont {D.}~\bibnamefont
  {Letscher}},\ and\ \bibinfo {author} {\bibfnamefont {A.}~\bibnamefont
  {Zomorodian}},\ }\href {https://doi.org/10.1007/s00454-002-2885-2} {\bibfield
   {journal} {\bibinfo  {journal} {Discrete \& Computational Geometry}\
  }\textbf {\bibinfo {volume} {28}},\ \bibinfo {pages} {511} (\bibinfo {year}
  {2002})}\BibitemShut {NoStop}%
\bibitem [{\citenamefont {Carr}\ \emph {et~al.}(2003)\citenamefont {Carr},
  \citenamefont {Snoeyink},\ and\ \citenamefont {Axen}}]{Carr2003ContourTrees}%
  \BibitemOpen
  \bibfield  {author} {\bibinfo {author} {\bibfnamefont {H.}~\bibnamefont
  {Carr}}, \bibinfo {author} {\bibfnamefont {J.}~\bibnamefont {Snoeyink}},\
  and\ \bibinfo {author} {\bibfnamefont {U.}~\bibnamefont {Axen}},\ }\href
  {https://doi.org/10.1016/S0925-7721(02)00093-7} {\bibfield  {journal}
  {\bibinfo  {journal} {Computational Geometry}\ }\textbf {\bibinfo {volume}
  {24}},\ \bibinfo {pages} {75} (\bibinfo {year} {2003})}\BibitemShut {NoStop}%
\bibitem [{\citenamefont {Edelsbrunner}\ \emph {et~al.}(2003)\citenamefont
  {Edelsbrunner}, \citenamefont {Harer},\ and\ \citenamefont
  {Zomorodian}}]{Edelsbrunner2003MorseSmale}%
  \BibitemOpen
  \bibfield  {author} {\bibinfo {author} {\bibfnamefont {H.}~\bibnamefont
  {Edelsbrunner}}, \bibinfo {author} {\bibfnamefont {J.}~\bibnamefont
  {Harer}},\ and\ \bibinfo {author} {\bibfnamefont {A.}~\bibnamefont
  {Zomorodian}},\ }\href {https://doi.org/10.1007/s00454-003-2926-5} {\bibfield
   {journal} {\bibinfo  {journal} {Discrete \& Computational Geometry}\
  }\textbf {\bibinfo {volume} {30}},\ \bibinfo {pages} {87} (\bibinfo {year}
  {2003})}\BibitemShut {NoStop}%
\bibitem [{\citenamefont {Ozertem}\ and\ \citenamefont
  {Erdogmus}(2011)}]{Ozertem2011PrincipalCurves}%
  \BibitemOpen
  \bibfield  {author} {\bibinfo {author} {\bibfnamefont {U.}~\bibnamefont
  {Ozertem}}\ and\ \bibinfo {author} {\bibfnamefont {D.}~\bibnamefont
  {Erdogmus}},\ }\href {https://jmlr.org/papers/v12/ozertem11a.html} {\bibfield
   {journal} {\bibinfo  {journal} {Journal of Machine Learning Research}\
  }\textbf {\bibinfo {volume} {12}},\ \bibinfo {pages} {1249} (\bibinfo {year}
  {2011})}\BibitemShut {NoStop}%
\bibitem [{\citenamefont {Genovese}\ \emph {et~al.}(2014)\citenamefont
  {Genovese}, \citenamefont {Perone-Pacifico}, \citenamefont {Verdinelli},\
  and\ \citenamefont {Wasserman}}]{Genovese2014Ridges}%
  \BibitemOpen
  \bibfield  {author} {\bibinfo {author} {\bibfnamefont {C.~R.}\ \bibnamefont
  {Genovese}}, \bibinfo {author} {\bibfnamefont {M.}~\bibnamefont
  {Perone-Pacifico}}, \bibinfo {author} {\bibfnamefont {I.}~\bibnamefont
  {Verdinelli}},\ and\ \bibinfo {author} {\bibfnamefont {L.}~\bibnamefont
  {Wasserman}},\ }\href {https://doi.org/10.1214/14-AOS1218} {\bibfield
  {journal} {\bibinfo  {journal} {The Annals of Statistics}\ }\textbf {\bibinfo
  {volume} {42}},\ \bibinfo {pages} {1511} (\bibinfo {year}
  {2014})}\BibitemShut {NoStop}%
\bibitem [{\citenamefont {Witkin}(1983)}]{Witkin1983ScaleSpace}%
  \BibitemOpen
  \bibfield  {author} {\bibinfo {author} {\bibfnamefont {A.~P.}\ \bibnamefont
  {Witkin}},\ }in\ \href
  {https://www.ijcai.org/Proceedings/83-2/Papers/091.pdf} {\emph {\bibinfo
  {booktitle} {Proceedings of the Eighth International Joint Conference on
  Artificial Intelligence ({IJCAI}-83)}}},\ Vol.~\bibinfo {volume} {2}\
  (\bibinfo {address} {Karlsruhe, West Germany},\ \bibinfo {year} {1983})\ pp.\
  \bibinfo {pages} {1019--1022}\BibitemShut {NoStop}%
\bibitem [{\citenamefont {Lindeberg}(1994)}]{Lindeberg1994ScaleSpace}%
  \BibitemOpen
  \bibfield  {author} {\bibinfo {author} {\bibfnamefont {T.}~\bibnamefont
  {Lindeberg}},\ }\href {https://doi.org/10.1007/978-1-4757-6465-9} {\emph
  {\bibinfo {title} {Scale-Space Theory in Computer Vision}}},\ \bibinfo
  {series} {The Kluwer International Series in Engineering and Computer
  Science}, Vol.\ \bibinfo {volume} {256}\ (\bibinfo  {publisher} {Kluwer
  Academic Publishers},\ \bibinfo {address} {Dordrecht},\ \bibinfo {year}
  {1994})\BibitemShut {NoStop}%
\bibitem [{\citenamefont {Favelier}\ \emph {et~al.}(2019)\citenamefont
  {Favelier}, \citenamefont {Faraj}, \citenamefont {Summa},\ and\ \citenamefont
  {Tierny}}]{Favelier2019PersistenceAtlas}%
  \BibitemOpen
  \bibfield  {author} {\bibinfo {author} {\bibfnamefont {G.}~\bibnamefont
  {Favelier}}, \bibinfo {author} {\bibfnamefont {N.}~\bibnamefont {Faraj}},
  \bibinfo {author} {\bibfnamefont {B.}~\bibnamefont {Summa}},\ and\ \bibinfo
  {author} {\bibfnamefont {J.}~\bibnamefont {Tierny}},\ }\href
  {https://doi.org/10.1109/TVCG.2018.2864432} {\bibfield  {journal} {\bibinfo
  {journal} {IEEE Transactions on Visualization and Computer Graphics}\
  }\textbf {\bibinfo {volume} {25}},\ \bibinfo {pages} {1152} (\bibinfo {year}
  {2019})}\BibitemShut {NoStop}%
\bibitem [{\citenamefont {Athawale}\ \emph {et~al.}(2022)\citenamefont
  {Athawale}, \citenamefont {Maljovec}, \citenamefont {Yan}, \citenamefont
  {Johnson}, \citenamefont {Pascucci},\ and\ \citenamefont
  {Wang}}]{Athawale2022MorseEnsembles}%
  \BibitemOpen
  \bibfield  {author} {\bibinfo {author} {\bibfnamefont {T.~M.}\ \bibnamefont
  {Athawale}}, \bibinfo {author} {\bibfnamefont {D.}~\bibnamefont {Maljovec}},
  \bibinfo {author} {\bibfnamefont {L.}~\bibnamefont {Yan}}, \bibinfo {author}
  {\bibfnamefont {C.~R.}\ \bibnamefont {Johnson}}, \bibinfo {author}
  {\bibfnamefont {V.}~\bibnamefont {Pascucci}},\ and\ \bibinfo {author}
  {\bibfnamefont {B.}~\bibnamefont {Wang}},\ }\href
  {https://doi.org/10.1109/TVCG.2020.3022359} {\bibfield  {journal} {\bibinfo
  {journal} {IEEE Transactions on Visualization and Computer Graphics}\
  }\textbf {\bibinfo {volume} {28}},\ \bibinfo {pages} {1955} (\bibinfo {year}
  {2022})}\BibitemShut {NoStop}%
\bibitem [{\citenamefont {Abac}\ \emph {et~al.}(2025)\citenamefont {Abac} \emph
  {et~al.}}]{LIGOScientific:2025hdt}%
  \BibitemOpen
  \bibfield  {author} {\bibinfo {author} {\bibfnamefont {A.~G.}\ \bibnamefont
  {Abac}} \emph {et~al.} (\bibinfo {collaboration} {LIGO Scientific, KAGRA,
  VIRGO}),\ }\href {https://doi.org/10.3847/2041-8213/ae0c06} {\bibfield
  {journal} {\bibinfo  {journal} {Astrophys. J. Lett.}\ }\textbf {\bibinfo
  {volume} {995}},\ \bibinfo {pages} {L18} (\bibinfo {year} {2025})},\ \Eprint
  {https://arxiv.org/abs/2508.18080} {arXiv:2508.18080 [gr-qc]} \BibitemShut
  {NoStop}%
\bibitem [{\citenamefont {Abac}\ \emph
  {et~al.}(2026{\natexlab{a}})\citenamefont {Abac} \emph
  {et~al.}}]{LIGOScientific:2025slb}%
  \BibitemOpen
  \bibfield  {author} {\bibinfo {author} {\bibfnamefont {A.~G.}\ \bibnamefont
  {Abac}} \emph {et~al.} (\bibinfo {collaboration} {LIGO Scientific, VIRGO,
  KAGRA}),\ }\href {https://doi.org/10.3847/2041-8213/ae2c74} {\bibfield
  {journal} {\bibinfo  {journal} {Astrophys. J. Lett.}\ }\textbf {\bibinfo
  {volume} {1004}},\ \bibinfo {pages} {L22} (\bibinfo {year}
  {2026}{\natexlab{a}})},\ \Eprint {https://arxiv.org/abs/2508.18082}
  {arXiv:2508.18082 [gr-qc]} \BibitemShut {NoStop}%
\bibitem [{\citenamefont {Abac}\ \emph
  {et~al.}(2026{\natexlab{b}})\citenamefont {Abac} \emph
  {et~al.}}]{LIGOScientific:2026wfs}%
  \BibitemOpen
  \bibfield  {author} {\bibinfo {author} {\bibfnamefont {A.~G.}\ \bibnamefont
  {Abac}} \emph {et~al.} (\bibinfo {collaboration} {LIGO Scientific, VIRGO,
  KAGRA}),\ }\href@noop {} {\  (\bibinfo {year} {2026}{\natexlab{b}})},\
  \Eprint {https://arxiv.org/abs/2605.27225} {arXiv:2605.27225 [gr-qc]}
  \BibitemShut {NoStop}%
\bibitem [{\citenamefont {Abac}\ \emph
  {et~al.}(2026{\natexlab{c}})\citenamefont {Abac} \emph
  {et~al.}}]{LIGOScientific:2025pvj}%
  \BibitemOpen
  \bibfield  {author} {\bibinfo {author} {\bibfnamefont {A.~G.}\ \bibnamefont
  {Abac}} \emph {et~al.} (\bibinfo {collaboration} {LIGO Scientific, VIRGO,
  Virgo,, KAGRA}),\ }\href {https://doi.org/10.3847/2041-8213/ae771e}
  {\bibfield  {journal} {\bibinfo  {journal} {Astrophys. J. Lett.}\ }\textbf
  {\bibinfo {volume} {1005}},\ \bibinfo {pages} {L51} (\bibinfo {year}
  {2026}{\natexlab{c}})},\ \Eprint {https://arxiv.org/abs/2508.18083}
  {arXiv:2508.18083 [astro-ph.HE]} \BibitemShut {NoStop}%
\bibitem [{\citenamefont {Abac}\ \emph
  {et~al.}(2026{\natexlab{d}})\citenamefont {Abac} \emph
  {et~al.}}]{LIGOScientific:2026ctl}%
  \BibitemOpen
  \bibfield  {author} {\bibinfo {author} {\bibfnamefont {A.~G.}\ \bibnamefont
  {Abac}} \emph {et~al.} (\bibinfo {collaboration} {LIGO Scientific, VIRGO,
  KAGRA}),\ }\href@noop {} {\  (\bibinfo {year} {2026}{\natexlab{d}})},\
  \Eprint {https://arxiv.org/abs/2605.27226} {arXiv:2605.27226 [astro-ph.HE]}
  \BibitemShut {NoStop}%
\bibitem [{\citenamefont {Abac}\ \emph
  {et~al.}(2026{\natexlab{e}})\citenamefont {Abac} \emph
  {et~al.}}]{LIGOScientific:2025jau}%
  \BibitemOpen
  \bibfield  {author} {\bibinfo {author} {\bibfnamefont {A.~G.}\ \bibnamefont
  {Abac}} \emph {et~al.} (\bibinfo {collaboration} {LIGO Scientific, Virgo,
  KAGRA}),\ }\href {https://doi.org/10.3847/2041-8213/ae69dd} {\bibfield
  {journal} {\bibinfo  {journal} {Astrophys. J. Lett.}\ }\textbf {\bibinfo
  {volume} {1007}},\ \bibinfo {pages} {L17} (\bibinfo {year}
  {2026}{\natexlab{e}})},\ \Eprint {https://arxiv.org/abs/2509.04348}
  {arXiv:2509.04348 [astro-ph.CO]} \BibitemShut {NoStop}%
\bibitem [{\citenamefont {Abac}\ \emph
  {et~al.}(2026{\natexlab{f}})\citenamefont {Abac} \emph
  {et~al.}}]{LIGOScientific:2026uyd}%
  \BibitemOpen
  \bibfield  {author} {\bibinfo {author} {\bibfnamefont {A.~G.}\ \bibnamefont
  {Abac}} \emph {et~al.} (\bibinfo {collaboration} {LIGO Scientific, VIRGO,
  KAGRA}),\ }\href@noop {} {\  (\bibinfo {year} {2026}{\natexlab{f}})},\
  \Eprint {https://arxiv.org/abs/2605.27227} {arXiv:2605.27227 [astro-ph.CO]}
  \BibitemShut {NoStop}%
\bibitem [{\citenamefont {Harris}\ \emph {et~al.}(2020)\citenamefont {Harris},
  \citenamefont {Millman}, \citenamefont {van~der Walt}, \citenamefont
  {Gommers}, \citenamefont {Virtanen}, \citenamefont {Cournapeau},
  \citenamefont {Wieser}, \citenamefont {Taylor}, \citenamefont {Berg},
  \citenamefont {Smith}, \citenamefont {Kern}, \citenamefont {Picus},
  \citenamefont {Hoyer}, \citenamefont {van Kerkwijk}, \citenamefont {Brett},
  \citenamefont {Haldane}, \citenamefont {del R{\'i}o}, \citenamefont {Wiebe},
  \citenamefont {Peterson}, \citenamefont {G{\'e}rard-Marchant}, \citenamefont
  {Sheppard}, \citenamefont {Reddy}, \citenamefont {Weckesser}, \citenamefont
  {Abbasi}, \citenamefont {Gohlke},\ and\ \citenamefont
  {Oliphant}}]{Harris2020NumPy}%
  \BibitemOpen
  \bibfield  {author} {\bibinfo {author} {\bibfnamefont {C.~R.}\ \bibnamefont
  {Harris}}, \bibinfo {author} {\bibfnamefont {K.~J.}\ \bibnamefont {Millman}},
  \bibinfo {author} {\bibfnamefont {S.~J.}\ \bibnamefont {van~der Walt}},
  \bibinfo {author} {\bibfnamefont {R.}~\bibnamefont {Gommers}}, \bibinfo
  {author} {\bibfnamefont {P.}~\bibnamefont {Virtanen}}, \bibinfo {author}
  {\bibfnamefont {D.}~\bibnamefont {Cournapeau}}, \bibinfo {author}
  {\bibfnamefont {E.}~\bibnamefont {Wieser}}, \bibinfo {author} {\bibfnamefont
  {J.}~\bibnamefont {Taylor}}, \bibinfo {author} {\bibfnamefont
  {S.}~\bibnamefont {Berg}}, \bibinfo {author} {\bibfnamefont {N.~J.}\
  \bibnamefont {Smith}}, \bibinfo {author} {\bibfnamefont {R.}~\bibnamefont
  {Kern}}, \bibinfo {author} {\bibfnamefont {M.}~\bibnamefont {Picus}},
  \bibinfo {author} {\bibfnamefont {S.}~\bibnamefont {Hoyer}}, \bibinfo
  {author} {\bibfnamefont {M.~H.}\ \bibnamefont {van Kerkwijk}}, \bibinfo
  {author} {\bibfnamefont {M.}~\bibnamefont {Brett}}, \bibinfo {author}
  {\bibfnamefont {A.}~\bibnamefont {Haldane}}, \bibinfo {author} {\bibfnamefont
  {J.~F.}\ \bibnamefont {del R{\'i}o}}, \bibinfo {author} {\bibfnamefont
  {M.}~\bibnamefont {Wiebe}}, \bibinfo {author} {\bibfnamefont
  {P.}~\bibnamefont {Peterson}}, \bibinfo {author} {\bibfnamefont
  {P.}~\bibnamefont {G{\'e}rard-Marchant}}, \bibinfo {author} {\bibfnamefont
  {K.}~\bibnamefont {Sheppard}}, \bibinfo {author} {\bibfnamefont
  {T.}~\bibnamefont {Reddy}}, \bibinfo {author} {\bibfnamefont
  {W.}~\bibnamefont {Weckesser}}, \bibinfo {author} {\bibfnamefont
  {H.}~\bibnamefont {Abbasi}}, \bibinfo {author} {\bibfnamefont
  {C.}~\bibnamefont {Gohlke}},\ and\ \bibinfo {author} {\bibfnamefont {T.~E.}\
  \bibnamefont {Oliphant}},\ }\href {https://doi.org/10.1038/s41586-020-2649-2}
  {\bibfield  {journal} {\bibinfo  {journal} {Nature}\ }\textbf {\bibinfo
  {volume} {585}},\ \bibinfo {pages} {357} (\bibinfo {year}
  {2020})}\BibitemShut {NoStop}%
\bibitem [{\citenamefont {Virtanen}\ \emph {et~al.}(2020)\citenamefont
  {Virtanen}, \citenamefont {Gommers}, \citenamefont {Oliphant}, \citenamefont
  {Haberland}, \citenamefont {Reddy}, \citenamefont {Cournapeau}, \citenamefont
  {Burovski}, \citenamefont {Peterson}, \citenamefont {Weckesser},
  \citenamefont {Bright}, \citenamefont {van~der Walt}, \citenamefont {Brett},
  \citenamefont {Wilson}, \citenamefont {Millman}, \citenamefont {Mayorov},
  \citenamefont {Nelson}, \citenamefont {Jones}, \citenamefont {Kern},
  \citenamefont {Larson}, \citenamefont {Carey}, \citenamefont {Polat},
  \citenamefont {Feng}, \citenamefont {Moore}, \citenamefont {VanderPlas},
  \citenamefont {Laxalde}, \citenamefont {Perktold}, \citenamefont {Cimrman},
  \citenamefont {Henriksen}, \citenamefont {Quintero}, \citenamefont {Harris},
  \citenamefont {Archibald}, \citenamefont {Ribeiro}, \citenamefont
  {Pedregosa}, \citenamefont {van Mulbregt},\ and\ \citenamefont {{SciPy 1.0
  Contributors}}}]{Virtanen2020SciPy}%
  \BibitemOpen
  \bibfield  {author} {\bibinfo {author} {\bibfnamefont {P.}~\bibnamefont
  {Virtanen}}, \bibinfo {author} {\bibfnamefont {R.}~\bibnamefont {Gommers}},
  \bibinfo {author} {\bibfnamefont {T.~E.}\ \bibnamefont {Oliphant}}, \bibinfo
  {author} {\bibfnamefont {M.}~\bibnamefont {Haberland}}, \bibinfo {author}
  {\bibfnamefont {T.}~\bibnamefont {Reddy}}, \bibinfo {author} {\bibfnamefont
  {D.}~\bibnamefont {Cournapeau}}, \bibinfo {author} {\bibfnamefont
  {E.}~\bibnamefont {Burovski}}, \bibinfo {author} {\bibfnamefont
  {P.}~\bibnamefont {Peterson}}, \bibinfo {author} {\bibfnamefont
  {W.}~\bibnamefont {Weckesser}}, \bibinfo {author} {\bibfnamefont
  {J.}~\bibnamefont {Bright}}, \bibinfo {author} {\bibfnamefont {S.~J.}\
  \bibnamefont {van~der Walt}}, \bibinfo {author} {\bibfnamefont
  {M.}~\bibnamefont {Brett}}, \bibinfo {author} {\bibfnamefont
  {J.}~\bibnamefont {Wilson}}, \bibinfo {author} {\bibfnamefont {K.~J.}\
  \bibnamefont {Millman}}, \bibinfo {author} {\bibfnamefont {N.}~\bibnamefont
  {Mayorov}}, \bibinfo {author} {\bibfnamefont {A.~R.~J.}\ \bibnamefont
  {Nelson}}, \bibinfo {author} {\bibfnamefont {E.}~\bibnamefont {Jones}},
  \bibinfo {author} {\bibfnamefont {R.}~\bibnamefont {Kern}}, \bibinfo {author}
  {\bibfnamefont {E.}~\bibnamefont {Larson}}, \bibinfo {author} {\bibfnamefont
  {C.~J.}\ \bibnamefont {Carey}}, \bibinfo {author} {\bibfnamefont
  {{\.I}.}~\bibnamefont {Polat}}, \bibinfo {author} {\bibfnamefont
  {Y.}~\bibnamefont {Feng}}, \bibinfo {author} {\bibfnamefont {E.~W.}\
  \bibnamefont {Moore}}, \bibinfo {author} {\bibfnamefont {J.}~\bibnamefont
  {VanderPlas}}, \bibinfo {author} {\bibfnamefont {D.}~\bibnamefont {Laxalde}},
  \bibinfo {author} {\bibfnamefont {J.}~\bibnamefont {Perktold}}, \bibinfo
  {author} {\bibfnamefont {R.}~\bibnamefont {Cimrman}}, \bibinfo {author}
  {\bibfnamefont {I.}~\bibnamefont {Henriksen}}, \bibinfo {author}
  {\bibfnamefont {E.~A.}\ \bibnamefont {Quintero}}, \bibinfo {author}
  {\bibfnamefont {C.~R.}\ \bibnamefont {Harris}}, \bibinfo {author}
  {\bibfnamefont {A.~M.}\ \bibnamefont {Archibald}}, \bibinfo {author}
  {\bibfnamefont {A.~H.}\ \bibnamefont {Ribeiro}}, \bibinfo {author}
  {\bibfnamefont {F.}~\bibnamefont {Pedregosa}}, \bibinfo {author}
  {\bibfnamefont {P.}~\bibnamefont {van Mulbregt}},\ and\ \bibinfo {author}
  {\bibnamefont {{SciPy 1.0 Contributors}}},\ }\href
  {https://doi.org/10.1038/s41592-019-0686-2} {\bibfield  {journal} {\bibinfo
  {journal} {Nature Methods}\ }\textbf {\bibinfo {volume} {17}},\ \bibinfo
  {pages} {261} (\bibinfo {year} {2020})}\BibitemShut {NoStop}%
\bibitem [{\citenamefont {Lam}\ \emph {et~al.}(2015)\citenamefont {Lam},
  \citenamefont {Pitrou},\ and\ \citenamefont {Seibert}}]{Lam2015Numba}%
  \BibitemOpen
  \bibfield  {author} {\bibinfo {author} {\bibfnamefont {S.~K.}\ \bibnamefont
  {Lam}}, \bibinfo {author} {\bibfnamefont {A.}~\bibnamefont {Pitrou}},\ and\
  \bibinfo {author} {\bibfnamefont {S.}~\bibnamefont {Seibert}},\ }in\ \href
  {https://doi.org/10.1145/2833157.2833162} {\emph {\bibinfo {booktitle}
  {Proceedings of the Second Workshop on the {LLVM} Compiler Infrastructure in
  {HPC}}}},\ \bibinfo {series and number} {{LLVM} '15}\ (\bibinfo  {publisher}
  {Association for Computing Machinery},\ \bibinfo {address} {New York, NY,
  USA},\ \bibinfo {year} {2015})\ pp.\ \bibinfo {pages} {1--6}\BibitemShut
  {NoStop}%
\bibitem [{\citenamefont {Knutsson}\ and\ \citenamefont
  {Westin}(1993)}]{Knutsson1993NormalizedConvolution}%
  \BibitemOpen
  \bibfield  {author} {\bibinfo {author} {\bibfnamefont {H.}~\bibnamefont
  {Knutsson}}\ and\ \bibinfo {author} {\bibfnamefont {C.-F.}\ \bibnamefont
  {Westin}},\ }in\ \href {https://doi.org/10.1109/CVPR.1993.341081} {\emph
  {\bibinfo {booktitle} {Proceedings of the IEEE Computer Society Conference on
  Computer Vision and Pattern Recognition ({CVPR})}}}\ (\bibinfo  {publisher}
  {IEEE Computer Society Press},\ \bibinfo {address} {New York, NY, USA},\
  \bibinfo {year} {1993})\ pp.\ \bibinfo {pages} {515--523}\BibitemShut
  {NoStop}%
\bibitem [{\citenamefont {{Cohen-Steiner}}\ \emph {et~al.}(2007)\citenamefont
  {{Cohen-Steiner}}, \citenamefont {Edelsbrunner},\ and\ \citenamefont
  {Harer}}]{CohenSteiner2007Stability}%
  \BibitemOpen
  \bibfield  {author} {\bibinfo {author} {\bibfnamefont {D.}~\bibnamefont
  {{Cohen-Steiner}}}, \bibinfo {author} {\bibfnamefont {H.}~\bibnamefont
  {Edelsbrunner}},\ and\ \bibinfo {author} {\bibfnamefont {J.}~\bibnamefont
  {Harer}},\ }\href {https://doi.org/10.1007/s00454-006-1276-5} {\bibfield
  {journal} {\bibinfo  {journal} {Discrete \& Computational Geometry}\ }\textbf
  {\bibinfo {volume} {37}},\ \bibinfo {pages} {103} (\bibinfo {year}
  {2007})}\BibitemShut {NoStop}%
\bibitem [{\citenamefont {Crouse}(2016)}]{Crouse2016Assignment}%
  \BibitemOpen
  \bibfield  {author} {\bibinfo {author} {\bibfnamefont {D.~F.}\ \bibnamefont
  {Crouse}},\ }\href {https://doi.org/10.1109/TAES.2016.140952} {\bibfield
  {journal} {\bibinfo  {journal} {IEEE Transactions on Aerospace and Electronic
  Systems}\ }\textbf {\bibinfo {volume} {52}},\ \bibinfo {pages} {1679}
  (\bibinfo {year} {2016})}\BibitemShut {NoStop}%
\bibitem [{\citenamefont {Hyndman}(1996)}]{Hyndman1996HDR}%
  \BibitemOpen
  \bibfield  {author} {\bibinfo {author} {\bibfnamefont {R.~J.}\ \bibnamefont
  {Hyndman}},\ }\href {https://doi.org/10.1080/00031305.1996.10474359}
  {\bibfield  {journal} {\bibinfo  {journal} {The American Statistician}\
  }\textbf {\bibinfo {volume} {50}},\ \bibinfo {pages} {120} (\bibinfo {year}
  {1996})}\BibitemShut {NoStop}%
\bibitem [{\citenamefont {Mann}\ and\ \citenamefont
  {Whitney}(1947)}]{Mann1947U}%
  \BibitemOpen
  \bibfield  {author} {\bibinfo {author} {\bibfnamefont {H.~B.}\ \bibnamefont
  {Mann}}\ and\ \bibinfo {author} {\bibfnamefont {D.~R.}\ \bibnamefont
  {Whitney}},\ }\href {https://doi.org/10.1214/aoms/1177730491} {\bibfield
  {journal} {\bibinfo  {journal} {The Annals of Mathematical Statistics}\
  }\textbf {\bibinfo {volume} {18}},\ \bibinfo {pages} {50} (\bibinfo {year}
  {1947})}\BibitemShut {NoStop}%
\bibitem [{\citenamefont {Cureton}(1956)}]{Cureton1956RankBiserial}%
  \BibitemOpen
  \bibfield  {author} {\bibinfo {author} {\bibfnamefont {E.~E.}\ \bibnamefont
  {Cureton}},\ }\href {https://doi.org/10.1007/BF02289138} {\bibfield
  {journal} {\bibinfo  {journal} {Psychometrika}\ }\textbf {\bibinfo {volume}
  {21}},\ \bibinfo {pages} {287} (\bibinfo {year} {1956})}\BibitemShut
  {NoStop}%
\bibitem [{\citenamefont {Spearman}(1904)}]{Spearman1904Association}%
  \BibitemOpen
  \bibfield  {author} {\bibinfo {author} {\bibfnamefont {C.}~\bibnamefont
  {Spearman}},\ }\href {https://doi.org/10.2307/1412159} {\bibfield  {journal}
  {\bibinfo  {journal} {The American Journal of Psychology}\ }\textbf {\bibinfo
  {volume} {15}},\ \bibinfo {pages} {72} (\bibinfo {year} {1904})}\BibitemShut
  {NoStop}%
\bibitem [{\citenamefont {Shannon}(1948)}]{Shannon1948CommunicationI}%
  \BibitemOpen
  \bibfield  {author} {\bibinfo {author} {\bibfnamefont {C.~E.}\ \bibnamefont
  {Shannon}},\ }\href {https://doi.org/10.1002/j.1538-7305.1948.tb01338.x}
  {\bibfield  {journal} {\bibinfo  {journal} {The Bell System Technical
  Journal}\ }\textbf {\bibinfo {volume} {27}},\ \bibinfo {pages} {379}
  (\bibinfo {year} {1948})}\BibitemShut {NoStop}%
\bibitem [{\citenamefont {Kullback}\ and\ \citenamefont
  {Leibler}(1951)}]{Kullback1951Information}%
  \BibitemOpen
  \bibfield  {author} {\bibinfo {author} {\bibfnamefont {S.}~\bibnamefont
  {Kullback}}\ and\ \bibinfo {author} {\bibfnamefont {R.~A.}\ \bibnamefont
  {Leibler}},\ }\href {https://doi.org/10.1214/aoms/1177729694} {\bibfield
  {journal} {\bibinfo  {journal} {The Annals of Mathematical Statistics}\
  }\textbf {\bibinfo {volume} {22}},\ \bibinfo {pages} {79} (\bibinfo {year}
  {1951})}\BibitemShut {NoStop}%
\bibitem [{\citenamefont {Lindley}(1956)}]{Lindley1956Information}%
  \BibitemOpen
  \bibfield  {author} {\bibinfo {author} {\bibfnamefont {D.~V.}\ \bibnamefont
  {Lindley}},\ }\href {https://doi.org/10.1214/aoms/1177728069} {\bibfield
  {journal} {\bibinfo  {journal} {The Annals of Mathematical Statistics}\
  }\textbf {\bibinfo {volume} {27}},\ \bibinfo {pages} {986} (\bibinfo {year}
  {1956})}\BibitemShut {NoStop}%
\bibitem [{\citenamefont {Wyner}(1978)}]{WYNER197851}%
  \BibitemOpen
  \bibfield  {author} {\bibinfo {author} {\bibfnamefont {A.}~\bibnamefont
  {Wyner}},\ }\href
  {https://doi.org/https://doi.org/10.1016/S0019-9958(78)90026-8} {\bibfield
  {journal} {\bibinfo  {journal} {Information and Control}\ }\textbf {\bibinfo
  {volume} {38}},\ \bibinfo {pages} {51} (\bibinfo {year} {1978})}\BibitemShut
  {NoStop}%
\bibitem [{\citenamefont {Kraskov}\ \emph {et~al.}(2004)\citenamefont
  {Kraskov}, \citenamefont {St{\"o}gbauer},\ and\ \citenamefont
  {Grassberger}}]{Kraskov2004MutualInformation}%
  \BibitemOpen
  \bibfield  {author} {\bibinfo {author} {\bibfnamefont {A.}~\bibnamefont
  {Kraskov}}, \bibinfo {author} {\bibfnamefont {H.}~\bibnamefont
  {St{\"o}gbauer}},\ and\ \bibinfo {author} {\bibfnamefont {P.}~\bibnamefont
  {Grassberger}},\ }\href {https://doi.org/10.1103/PhysRevE.69.066138}
  {\bibfield  {journal} {\bibinfo  {journal} {Physical Review E}\ }\textbf
  {\bibinfo {volume} {69}},\ \bibinfo {eid} {066138} (\bibinfo {year}
  {2004})}\BibitemShut {NoStop}%
\bibitem [{\citenamefont {Frenzel}\ and\ \citenamefont
  {Pompe}(2007)}]{Frenzel2007ConditionalMI}%
  \BibitemOpen
  \bibfield  {author} {\bibinfo {author} {\bibfnamefont {S.}~\bibnamefont
  {Frenzel}}\ and\ \bibinfo {author} {\bibfnamefont {B.}~\bibnamefont
  {Pompe}},\ }\href {https://doi.org/10.1103/PhysRevLett.99.204101} {\bibfield
  {journal} {\bibinfo  {journal} {Physical Review Letters}\ }\textbf {\bibinfo
  {volume} {99}},\ \bibinfo {eid} {204101} (\bibinfo {year}
  {2007})}\BibitemShut {NoStop}%
\bibitem [{\citenamefont {Theiler}\ \emph {et~al.}(1992)\citenamefont
  {Theiler}, \citenamefont {Eubank}, \citenamefont {Longtin}, \citenamefont
  {Galdrikian},\ and\ \citenamefont {Farmer}}]{Theiler1992Surrogate}%
  \BibitemOpen
  \bibfield  {author} {\bibinfo {author} {\bibfnamefont {J.}~\bibnamefont
  {Theiler}}, \bibinfo {author} {\bibfnamefont {S.}~\bibnamefont {Eubank}},
  \bibinfo {author} {\bibfnamefont {A.}~\bibnamefont {Longtin}}, \bibinfo
  {author} {\bibfnamefont {B.}~\bibnamefont {Galdrikian}},\ and\ \bibinfo
  {author} {\bibfnamefont {J.~D.}\ \bibnamefont {Farmer}},\ }\href
  {https://doi.org/10.1016/0167-2789(92)90102-S} {\bibfield  {journal}
  {\bibinfo  {journal} {Physica D: Nonlinear Phenomena}\ }\textbf {\bibinfo
  {volume} {58}},\ \bibinfo {pages} {77} (\bibinfo {year} {1992})}\BibitemShut
  {NoStop}%
\bibitem [{\citenamefont {Heinzel}\ \emph {et~al.}(2025)\citenamefont
  {Heinzel}, \citenamefont {Mould}, \citenamefont {\'Alvarez-L\'opez},\ and\
  \citenamefont {Vitale}}]{Heinzel:2024jlc}%
  \BibitemOpen
  \bibfield  {author} {\bibinfo {author} {\bibfnamefont {J.}~\bibnamefont
  {Heinzel}}, \bibinfo {author} {\bibfnamefont {M.}~\bibnamefont {Mould}},
  \bibinfo {author} {\bibfnamefont {S.}~\bibnamefont {\'Alvarez-L\'opez}},\
  and\ \bibinfo {author} {\bibfnamefont {S.}~\bibnamefont {Vitale}},\ }\href
  {https://doi.org/10.1103/PhysRevD.111.063043} {\bibfield  {journal} {\bibinfo
   {journal} {Phys. Rev. D}\ }\textbf {\bibinfo {volume} {111}},\ \bibinfo
  {pages} {063043} (\bibinfo {year} {2025})},\ \Eprint
  {https://arxiv.org/abs/2406.16813} {arXiv:2406.16813 [astro-ph.HE]}
  \BibitemShut {NoStop}%
\bibitem [{\citenamefont {Collaboration}\ \emph
  {et~al.}(2026{\natexlab{a}})\citenamefont {Collaboration}, \citenamefont
  {Collaboration},\ and\ \citenamefont
  {collaboration}}]{ligo_scientific_collaboration_2026_20292639}%
  \BibitemOpen
  \bibfield  {author} {\bibinfo {author} {\bibfnamefont {L.~S.}\ \bibnamefont
  {Collaboration}}, \bibinfo {author} {\bibfnamefont {T.~V.}\ \bibnamefont
  {Collaboration}},\ and\ \bibinfo {author} {\bibfnamefont {T.~K.}\
  \bibnamefont {collaboration}},\ }\href
  {https://doi.org/10.5281/zenodo.20292639} {10.5281/zenodo.20292639} (\bibinfo
  {year} {2026}{\natexlab{a}})\BibitemShut {NoStop}%
\bibitem [{\citenamefont {Collaboration}\ \emph
  {et~al.}(2026{\natexlab{b}})\citenamefont {Collaboration}, \citenamefont
  {Collaboration},\ and\ \citenamefont
  {Collaboration}}]{ligo_scientific_collaboration_2026_20378418}%
  \BibitemOpen
  \bibfield  {author} {\bibinfo {author} {\bibfnamefont {L.~S.}\ \bibnamefont
  {Collaboration}}, \bibinfo {author} {\bibfnamefont {V.}~\bibnamefont
  {Collaboration}},\ and\ \bibinfo {author} {\bibfnamefont {K.}~\bibnamefont
  {Collaboration}},\ }\href {https://doi.org/10.5281/zenodo.20378418}
  {10.5281/zenodo.20378418} (\bibinfo {year} {2026}{\natexlab{b}})\BibitemShut
  {NoStop}%
\bibitem [{\citenamefont {De~Renzis}\ \emph {et~al.}(2026)\citenamefont
  {De~Renzis}, \citenamefont {Agapito},\ and\ \citenamefont
  {Mancarella}}]{DeRenzis:2026hcp}%
  \BibitemOpen
  \bibfield  {author} {\bibinfo {author} {\bibfnamefont {V.}~\bibnamefont
  {De~Renzis}}, \bibinfo {author} {\bibfnamefont {A.}~\bibnamefont {Agapito}},\
  and\ \bibinfo {author} {\bibfnamefont {M.}~\bibnamefont {Mancarella}},\
  }\href@noop {} {\  (\bibinfo {year} {2026})},\ \Eprint
  {https://arxiv.org/abs/2610.10535} {arXiv:2610.10535 [gr-qc]} \BibitemShut
  {NoStop}%
\bibitem [{\citenamefont {Lloyd}(1982)}]{Lloyd1982Quantization}%
  \BibitemOpen
  \bibfield  {author} {\bibinfo {author} {\bibfnamefont {S.~P.}\ \bibnamefont
  {Lloyd}},\ }\href {https://doi.org/10.1109/TIT.1982.1056489} {\bibfield
  {journal} {\bibinfo  {journal} {IEEE Transactions on Information Theory}\
  }\textbf {\bibinfo {volume} {28}},\ \bibinfo {pages} {129} (\bibinfo {year}
  {1982})}\BibitemShut {NoStop}%
\bibitem [{\citenamefont {Phipson}\ and\ \citenamefont
  {Smyth}(2010)}]{Phipson2010Permutation}%
  \BibitemOpen
  \bibfield  {author} {\bibinfo {author} {\bibfnamefont {B.}~\bibnamefont
  {Phipson}}\ and\ \bibinfo {author} {\bibfnamefont {G.~K.}\ \bibnamefont
  {Smyth}},\ }\href {https://doi.org/10.2202/1544-6115.1585} {\bibfield
  {journal} {\bibinfo  {journal} {Statistical Applications in Genetics and
  Molecular Biology}\ }\textbf {\bibinfo {volume} {9}},\ \bibinfo {eid}
  {Article 39} (\bibinfo {year} {2010})}\BibitemShut {NoStop}%
\bibitem [{\citenamefont {Holm}(1979)}]{H0UpdateHolm1979}%
  \BibitemOpen
  \bibfield  {author} {\bibinfo {author} {\bibfnamefont {S.}~\bibnamefont
  {Holm}},\ }\href@noop {} {\bibfield  {journal} {\bibinfo  {journal}
  {Scandinavian Journal of Statistics}\ }\textbf {\bibinfo {volume} {6}},\
  \bibinfo {pages} {65} (\bibinfo {year} {1979})}\BibitemShut {NoStop}%
\end{thebibliography}%

\FloatBarrier
\clearpage
\appendix

\section{Technical details and supplementary analyses}
\label{app:technical_methods}

This appendix specifies the numerical realisation of
Sec.~\ref{sec:method}. It gives the fixed-scale and multiscale selection
rules, finite-region constructions, posterior correspondence, and
association estimators. The independent one-dimensional finder is described
in full, while the final subsection collects optional flat-region and tail
diagnostics. Unless stated otherwise, numerical values below are defaults
of the version described in the main text.

\subsection{Field representation, connectivity, and persistence selection}
\label{app:technical_fields}
\label{app:technical_topology}

\paragraph{Measure, quadrature, and smoothing.}
The domain mask determines which grid entries contribute to
integration and connectivity. Input fields are normalised in their
declared measure and converted to the selected feature measure.
The spatial quadrature depends explicitly on grid placement.

The default mode retains trapezoidal weights in the selected
density measure and the pointwise transformation in
Eq.~\eqref{eq:method_jacobian}. 

For cell-centred inputs, the declared native coordinate is
$\xi_k=x_k$ or $\xi_k=\log_a x_k$. Interior edges are inferred as
\begin{equation}
 \xi_{k,i+1/2}=\frac{\xi_{k,i}+\xi_{k,i+1}}{2},
\end{equation}
and the two outer edges are extrapolated by half the nearest
centre spacing. All measure widths are derived from these same
physical bins. For example, logarithmic native edges give
\begin{equation}
\begin{split}
 \Delta\log_a x_{k,i}
   &=\xi_{k,i+1/2}-\xi_{k,i-1/2},\\
 \Delta x_{k,i}
   &=a^{\xi_{k,i+1/2}}-a^{\xi_{k,i-1/2}}.
\end{split}
\end{equation}
Linear boundaries are therefore not independently inferred from
physical-coordinate midpoints.

Writing $w^{\rm in}_{k,i}$ and $w^{\rm out}_{k,i}$ for the widths
of a cell in the input and target density measures, conversion uses
\begin{equation}
 F^{\rm out}_{ij}
 =F^{\rm in}_{ij}
 \frac{w^{\rm in}_{1,i}w^{\rm in}_{2,j}}
      {w^{\rm out}_{1,i}w^{\rm out}_{2,j}}.
 \label{eq:method_cell_conversion}
\end{equation}
This preserves the discrete probability of every cell before any
final numerical renormalisation.

For either placement, the normalisation convention is
\begin{equation}
 \sum_{ij}M_{ij}F_{ij}w_{1,i}w_{2,j}=1.
 \label{eq:method_normalisation}
\end{equation}
The selected weights are used consistently for probability
integrals, moments, and marginalisation. In cell-centred mode,
every active mask entry receives its full cell weight, including
diagonal cells of an inclusive ordered mask; no geometric clipping
or fractional boundary weight is inferred. The weights describe
the selected density measure, independently of the coordinates
used for spatial distances.
In analysis coordinates
$g_i$, define the characteristic spacing and the rectangular-domain
diagonal by
$h=[\operatorname{med}(\Delta g_1)\operatorname{med}(\Delta g_2)]^{1/2}$
and $L$, respectively. The Gaussian in
Eq.~\eqref{eq:method_smoothing} has array-coordinate standard deviations
$\sigma_i=s/\operatorname{med}(\Delta g_i)$. The pointwise mask division
corrects the averaging weights, after which $\mathcal Z_s$ restores the
integral in Eq.~\eqref{eq:method_normalisation}.

The highest-density search support is
\begin{equation}
 \mathcal H_\gamma(F_s)=\{\boldsymbol x\in\Omega:F_s(\boldsymbol x)
 \geq c_\gamma\},\qquad
 \int_{\mathcal H_\gamma(F_s)}F_s\,d\nu\geq\gamma,
 \label{eq:method_hpd_support}
\end{equation}
where $c_\gamma$ is selected from the sampled density levels using the same
quadrature. The default $\gamma=0.995$ controls shoulder detection and the
measured part of valleys, but not the peak--pit connectivity sweep. It is
separate from the credible content of posterior location contours.\\

\paragraph{Discrete connectivity and boundary roles.}
Each valid grid node $(i,j)$ is connected to $(i\pm1,j)$,
$(i,j\pm1)$, $(i-1,j-1)$, and $(i+1,j+1)$. This corresponds to the same
diagonal triangulation of every rectangular grid cell. Equal density
values are ordered by grid index, making the pairings deterministic.
Descending and ascending activation implement the join/split constructions
of Sec.~\ref{sec:feature_identification}. During either sweep, a node
points to the previously activated neighbour with the steepest density
increase or decrease per unit analysis-coordinate distance. Paths started
in the components meeting at a saddle therefore terminate at their
associated extrema.

A pit on one boundary remains eligible as a valley endpoint, although
localised-pit region measurements require an interior minimum. A pit at
the intersection of the outer array boundary and an internal mask boundary,
together with its incident valley branches, is excluded. Boundary roles
also enter branch matching, as described below. Global extrema remain
eligible independently of the persistence cutoff and are assigned the
finite stored value
\begin{equation}
 D_s=\max_\Omega F_s-\min_\Omega F_s.
 \label{eq:Ds}
\end{equation}
They are not included in the automatic threshold calibration.\\

\paragraph{Automatic persistence calibration.}
\label{app:technical_persistence}
At each scale, the finite non-essential peak and pit persistences are
pooled. Values below the numerical floor
\begin{equation}
 \Pi_{\rm floor}=\max\{\epsilon_{\rm mach}\max(D_s,1),10^{-4}D_s\}
 \label{eq:method_persistence_floor}
\end{equation}
are discarded. For the remaining sorted values
$\Pi_{(1)}\leq\cdots\leq\Pi_{(n)}$, eligible logarithmic gaps are
\begin{equation}
 G_i=\ln\frac{\Pi_{(i+1)}}{\Pi_{(i)}},\qquad i=1,\ldots,n-2.
 \label{eq:method_persistence_gaps}
\end{equation}
The terminal gap between the two largest values is excluded so that a
single exceptionally strong feature cannot alone set the catalogue
resolution. If $n\geq3$ and the largest eligible gap $G_{i_*}$ satisfies
$G_{i_*}\geq g_{\min}$, then
\begin{equation}
 T_s=\sqrt{\Pi_{(i_*)}\Pi_{(i_*+1)}}.
 \label{eq:method_persistence_threshold_gap}
\end{equation}
The default $g_{\min}=1$ requires a ratio of at least $e$ between the
bounding persistences. If $n\geq3$ but no eligible gap meets that
requirement, the fallback is
\begin{equation}
 T_s=\max\{\Pi_{\rm floor},\tfrac14Q_{0.10}(\{\Pi_{(i)}\})\},
 \label{eq:method_persistence_threshold_fallback}
\end{equation}
where $Q_{0.10}$ is the tenth percentile. With fewer than three eligible
values,
\begin{equation}
 T_s=\Pi_{\rm floor}.
 \label{eq:method_persistence_threshold_smalln}
\end{equation}
A numerical override replaces this calibration. In particular, changing
$g_{\min}$ need not change $T_s$, and raising $g_{\min}$ can activate a
lower fallback cutoff; it is not a monotonic significance parameter.

Non-essential extrema satisfy
\begin{equation}
 \Pi_e\geq T_s.
 \label{eq:method_extremum_selection}
\end{equation}
For a merger with disappearing components $\mathcal D$, the event
persistence is
\begin{equation}
 \Pi_{\rm event}=\max_{c\in\mathcal D}\Pi_c.
 \label{eq:method_event_persistence}
\end{equation}
Each ridge or valley branch requires both a retained extremum endpoint and
\begin{equation}
 \Pi_{\rm event}\geq T_s.
 \label{eq:method_branch_selection}
\end{equation}
This merger quantity is distinct from a branch's longitudinal contrast
$|F_s(e)-F_s(z)|$, which compares that particular endpoint with the saddle.\\

\paragraph{Multiscale correspondence.}
The default two-dimensional scales are
\begin{equation}
 s=(0,0.025,0.05).
 \label{eq:method_default_scales}
\end{equation}
Peaks and pits at the first requested scale supply the initial identities.
Same-type counterparts are assigned one-to-one with distance limit
\begin{equation}
 d_{\rm sc}=\max(3h,0.08L).
 \label{eq:dsc}
\end{equation}
Branches are compared using concatenated saddle and extremum coordinates,
with compatible interior-versus-boundary roles for both endpoints and
\begin{equation}
 d_{\rm sc}^{\rm branch}=\sqrt{2}\,d_{\rm sc}.
 \label{eq:method_branch_scale_distance}
\end{equation}
Peaks, pits, ridges, and valleys require
\begin{equation}
 N_{\rm rec}\geq\left\lceil\frac{2N_{\rm scale}}3\right\rceil,
 \label{eq:method_multiscale_recurrence}
\end{equation}
and retained branches must terminate at extrema satisfying the same
requirement. Shoulder fronts are compared through their first, middle,
and last points, with
\begin{equation}
 d_{\rm sc}^{\rm shoulder}=\sqrt{3}\,d_{\rm sc}.
 \label{eq:method_shoulder_scale_distance}
\end{equation}
When several scales are requested, at least two must contain compatible
fronts. The cluster member with the smallest mean distance to the others
provides the representative shoulder geometry.

\subsection{Curve geometry, shoulders, and finite measurement regions}
\label{app:technical_regions}

\paragraph{Branch refinement and transverse sections.}
Raw saddle--extremum paths are resampled at 160 points. Their interior
points may be moved towards a transverse maximum for a ridge or minimum
for a valley, using the local gradient and Hessian of $F_s$ as in the
refinement described in Sec.~\ref{sec:feature_identification}. Displacements
are limited to $0.75h$, endpoints are fixed, and proposed corrections
outside the domain are rejected. Regularisation over two grid cells then
suppresses residual grid-scale bends.

Finite branch regions use transverse profiles at up to 40 positions along
the curve. The first resolved minimum on either side of a ridge, or maximum
on either side of a valley, supplies a side level $F_\pm(u)$; $F_c(u)$ is
the density on the curve. The crossings
\begin{equation}
 F_{1/2,\pm}(u)=\frac{F_c(u)+F_\pm(u)}2
 \label{eq:method_curve_half_level}
\end{equation}
define the transverse half-level boundaries. The corresponding local
logarithmic contrasts are
\begin{equation}
\begin{split}
 c_{\rm R}(u)&=\ln\frac{F_c(u)}{\sqrt{F_-(u)F_+(u)}},\\
 c_{\rm V}(u)&=\ln\frac{\sqrt{F_-(u)F_+(u)}}{F_c(u)}.
\end{split}
\label{eq:method_transverse_contrast}
\end{equation}
A ridge side without a resolved minimum can use the saddle density as a
fallback; a valley requires resolved maxima on both sides. At least two
valid transverse sections are needed for finite branch geometry.
Branch widths and contrasts are medians over valid sections, with valid
and fallback fractions retained separately. Widths use the physical
coordinates of the transverse boundaries, whereas individual-branch
arclength uses the analysis geometry.

Two same-type arms sharing a saddle define a paired event. At a multiway
merger, endpoints are ordered deterministically and the most widely
separated remaining endpoints are paired successively; an odd remaining
arm is left unpaired. Pairs are defined before multiscale filtering and
survive only when both arms survive. Their measurement window is the
union of the arm strips. Event widths and contrasts use physical-arclength
weights along the joined profiles, and event extents are physical lengths.
If $\overline c$ is the weighted median transverse log contrast,
\begin{equation}
 A_{\rm curve}=1-\exp(-\overline c).
 \label{eq:method_curve_strength_app}
\end{equation}

\paragraph{Point regions and deficit backgrounds.}
Using the half-level in Eq.~\eqref{eq:method_half_level},
the peak region is its connected component in $F_s\geq F_{1/2}$; the pit
region is its component in $F_s\leq F_{1/2}$, using the same connectivity
as the level-set sweep. The base is the paired saddle level. For essential
extrema the nearest adjacent merge level in density is used when available;
otherwise the fallback is zero for a peak and the field maximum for a pit.
The contrasts in Eq.~\eqref{eq:method_point_strength} are reported when
the denominators and contrasts are positive. The full attraction or
drainage basin is stored separately from this finite half-level region.

The missing-density fields used in the common measurements are
\begin{align}
 \Delta F&=(F_{\rm base}-F_s)_+ &&\text{for an interior pit},
 \label{eq:method_pit_deficit}\\
 \Delta F&=(F_{\rm bg}-F_s)_+ &&\text{for a valley}.
 \label{eq:method_valley_deficit}
\end{align}
For a valley, $\ln F_{\rm bg}$ is linearly interpolated between the two
side maxima of each valid transverse profile. 
Boundary pits do not receive
localised deficit centroids or projections even when they remain valid
valley endpoints.\\

\paragraph{Shoulder profiles and selection.}
\label{app:technical_shoulders}
The coordinates and derivatives are those of
Eq.~\eqref{eq:method_shoulder_coordinates}. For derivative estimation,
$\ln F_s(t,r_i)$ is first smoothed along each profile by a Gaussian of
width $0.8$ profile samples. 
Local derivative-fluctuation scales
are estimated from differences,
\begin{equation}
\begin{split}
 \sigma_\alpha&=1.4826\,\operatorname{MAD}(\Delta^2\alpha),\\
 \sigma_\kappa&=1.4826\,\operatorname{MAD}(\Delta\kappa),
\end{split}
\label{eq:method_shoulder_fluctuations}
\end{equation}
where $\operatorname{MAD}$ is the median absolute deviation,
$\Delta^2\alpha$ the second differences of the slope, and $\Delta\kappa$
the first differences of the curvature.

For the ordered landmarks $t_+<t_\alpha<t_-<t_0$, define
\begin{equation}
 \Delta\alpha=\alpha_{\rm pre}-\alpha_{\rm post},
 \label{eq:method_shoulder_contrast_app}
\end{equation}
where the two slopes are secants of $\ln F_s$ over
$[t_+,t_\alpha]$ and $[t_-,t_0]$, respectively. Let
$\mathcal P_\alpha$ be the prominence of the slope maximum and
$\kappa_\pm=\kappa(t_\pm)$. The slope and curvature requirements for a
bounded cross-section are
\begin{equation}
\begin{split}
 \mathcal P_\alpha&\geq\max(5\sigma_\alpha,10^{-3}),\\
 \kappa_+&\geq\max(3\sigma_\kappa,10^{-3}),\\
 -\kappa_-&\geq\max(3\sigma_\kappa,10^{-3}),\\
 \Delta\alpha&\geq\max(3\sigma_\alpha,10^{-3}),\\
 \alpha(t_\alpha)&<\alpha_{\max},
\end{split}
\label{eq:method_shoulder_selection}
\end{equation}
with $\alpha_{\max}=2$ by default. These criteria are separate from the
peak--pit persistence calibration.

Detections are linked through their $t_-$ positions. One missing
intermediate $r$ profile is allowed: for separations of
$\mathrm{gap}=1$ or $2$ sampling steps, linking requires
$|\Delta t_-|/\mathrm{gap}\leq\max(5\Delta t,2.5\Delta r)$.
Here $\Delta t$ and $\Delta r$ are the profile and cross-profile sampling
intervals. A front must span at least
\begin{equation}
 N_{\rm front}\geq\max\{4,\lceil0.06N_r\rceil\}
 \label{eq:method_shoulder_extent}
\end{equation}
contributing profiles. Front portions with a located $t_-$ but incomplete
bounded landmarks contribute to location geometry, not to finite-region
measurements. The longest contiguous bounded segment must independently
meet the same extent requirement. Its onset and terminal curves bound
$\mathcal \rho_f$ within $\mathcal H_\gamma(F_s)$.
The package returns the median bounded-section slope change
$\overline{\Delta\alpha}_f$. Its bounded representation
$A_{{\rm S},f}$ is defined in
Eq.~\eqref{eq:method_shoulder_strength}.
Width is the physical onset-to-terminal separation, and extent is the
physical length of the bounded structure. Support truncation is recorded
separately from physical boundary attachment.\\

\paragraph{Additional region summaries.}
Projected bounds $x_{i,-}$ and $x_{i,+}$ are obtained from the physical
extent of $\mathcal \rho_f$ and define
\begin{equation}
 w_{f,i}=\frac{x_{i,+}-x_{i,-}}{\mu_{f,i}}.
 \label{eq:method_relative_width}
\end{equation}
Point-region shape may also be summarised by an ellipse from uniform
physical-area covariance, with full principal widths equal to four times
the square roots of its eigenvalues. This is a region-shape summary, not a
posterior location contour. Density- or deficit-weighted covariances,
full-basin probability, and finite-region probability remain separate
measurements. Table~\ref{tab:method_catalogue_selection} collects the
feature-specific prescriptions.

\subsection{Posterior correspondence and spatial summaries}
\label{app:technical_posterior}

\paragraph{Matching and conditional measurements.}
The reference catalogue and draw-level extraction follow
Sec.~\ref{sec:posterior_matching}. Point features are represented by their
positions, branches by saddle and extremum endpoints, paired events by a
saddle and two ordered endpoints, and shoulders by three representative
front points. The same-type assignment minimises spatial mismatch, with
the point-distance limit
\begin{equation}
 d_{\rm draw}=\max(4h,0.10L).
 \label{eq:method_draw_distance}
\end{equation}
The corresponding limits are $\sqrt2\,d_{\rm draw}$ for branches and
$\sqrt3\,d_{\rm draw}$ for three-landmark objects. Near-equivalent
alternatives are flagged using default distance margins $h$, $\sqrt2h$,
and $\sqrt3h$, respectively. Ambiguous accepted matches are retained and
their posterior frequency is reported. Unmatched draw candidates do not
extend the reference catalogue, but total draw-level feature counts remain
available.

The availability indicators satisfy
\begin{equation}
 R_{fb}\leq L_{fb}\leq M_{fb}.
 \label{eq:method_nested_supports}
\end{equation}
For any quantity $q_{fb}$ with availability indicator $A^{(q)}_{fb}$,
conditional summaries use
\begin{equation}
 \widetilde\omega^{(q)}_{fb}
 =\frac{\omega_b A^{(q)}_{fb}}
        {\sum_{b'}\omega_{b'} A^{(q)}_{fb'}}.
 \label{eq:method_conditional_weights}
\end{equation}
The required indicator follows the observable: usable point locations,
available curves, valid transverse sections, or valid finite regions.
Weighted medians and central credible intervals use these conditional
weights and retain the corresponding unconditional support.
When the denominator is zero, no conditional distribution is defined;
zero availability is distinct from a measured value of zero.\\

\paragraph{Location distributions and region inclusion.}
Matched point positions are accumulated with posterior weights to estimate
the conditional position distribution. After spatial smoothing and
normalisation by physical-coordinate cell area, the density supplies the
highest-density regions described in Sec.~\ref{sec:posterior_geometry}.
For available curves, sampled positions are rasterised so that each draw
contributes its posterior weight to the curve-location distribution.
The resulting location corridors and point regions are distinct from
finite-window inclusion probabilities.

For the latter, the draw-level indicator is
\begin{equation}
 I_{fb}(\boldsymbol x)=R_{fb}\,
 \mathbf1\{\boldsymbol x\in\mathcal W_{f,b}\}.
 \label{eq:method_region_indicator}
\end{equation}
Its posterior average is Eq.~\eqref{eq:method_region_inclusion}; division
by $P_{{\rm reg},f}$, when positive, conditions on a measurable region.
No fixed posterior credible contour is substituted for the individual
measurement windows in calculating centroids or region probabilities.\\

\paragraph{Separated location configurations.}
The posterior summaries above treat all draw-level realizations matched to the
same reference identity as one feature family. This is appropriate when their
measured locations form a single connected posterior cloud. In some cases,
however, the same tracked identity is recovered in two or more well-separated
regions in mutually exclusive subsets of the posterior draws. Averaging over
such configurations would place the reported centroid between locations that
are not themselves supported and could obscure their different associations
with an external parameter. The package therefore tests whether the valid
draw-level centroids of a tracked feature separate into robust location
configurations and, when they do, reports those configurations separately.
For association observables, valid positive centroid coordinates are
transformed to logarithmic coordinates and whitened. Connected components
of a smoothed $90\%$ highest-density set identify candidate separated
configurations, and weighted $k$-means partitions the draws among them
\citep{Hyndman1996HDR,Lloyd1982Quantization}. A split is accepted only when
five smoothing bandwidths recover the same component count, each component
has at least $2\%$ of the valid posterior weight and 50 draws, and aligned
assignments agree by at least $95\%$. These are mutually exclusive locations
of one tracked identity, not additional features in a single draw.
Co-occurrence is evaluated separately from draw-level availability before
configuration labels are assigned. Accepted configurations are analysed
individually in the continuous-association comparisons.

\subsection{Independent one-dimensional analysis}
\label{app:technical_marginals}

\paragraph{Marginalisation and feature measure.}
Each realisation is expressed as a linear-measure joint density and
renormalised in physical-coordinate quadrature before integration,
\begin{equation}
\begin{split}
 f_{1,b}(x_1)&=\int p_{x,b}(x_1,x_2)\,dx_2,\\
 f_{2,b}(x_2)&=\int p_{x,b}(x_1,x_2)\,dx_1,
\end{split}
\label{eq:method_marginals}
\end{equation}
with both integrals restricted to the valid domain. The selected marginal
feature density is $q_b=f_b$ per $dx$, or $q_b=xf_b$ per $d\ln x$.
Smoothing and matching always use $y=\ln x$, so positive coordinates are
required even when the feature measure is linear. Grids nonuniform in $y$
are interpolated onto a uniform logarithmic grid no coarser than the
finest original log spacing, and each marginal is renormalised.
The workflow currently requires equal-weight posterior draws. Its reference
curve and pointwise credible bands are calculated from individual
marginals, not by marginalising $F_{\rm ref}$.\\

\paragraph{Feature definitions.}
The reference curve is Gaussian-smoothed at each requested width and
renormalised in its selected measure. For the resulting $q_s$:
\begin{itemize}
\item \emph{Peaks} are local maxima. If $q_C$ is the central density and
$\Pi$ its prominence above the base, the half-prominence interval and
relative strength are
\begin{equation}
 q_{1/2}=q_C-\frac\Pi2,\qquad A_{\rm P}^{\rm 1D}=\frac\Pi{q_C}.
 \label{eq:method_marginal_peak}
\end{equation}
\item \emph{Dips} are local minima, found by applying the peak search to
$-q_s$. Their base density is $q_C+\Pi$, giving
\begin{equation}
 q_{1/2}=q_C+\frac\Pi2,\qquad A_{\rm D}^{\rm 1D}=\frac\Pi{q_C+\Pi}.
 \label{eq:method_marginal_dip}
\end{equation}
The bases and interpolated half-prominence crossings use the
one-dimensional peak routines in SciPy \citep{Virtanen2020SciPy}.
Unlike the two-dimensional persistence selection, candidate detection
imposes no additional prominence-amplitude cutoff.
\item \emph{Shoulders} use
\begin{equation}
 \alpha=\frac{d\ln q_s}{dy},\qquad
 \kappa=\frac{d\alpha}{dy}.
 \label{eq:method_marginal_derivatives}
\end{equation}
A slope maximum and its surrounding curvature landmarks identify a
flattening followed by steepening. The position is the following negative
curvature minimum $t_-$, interpreted now along $y$; the bounded interval
uses the same ordered four-landmark construction as in the joint analysis,
from the preceding positive curvature maximum to the subsequent
negative-to-nonnegative crossing. Reference shoulders lacking this bounded
interval are excluded. The derivative search requires density above a
floor, $10^{-5}$ by default. Writing $q_s=x^\beta f_s$, with $\beta=0$
or $1$ for linear or logarithmic measure, $\alpha<\beta$ restricts the
underlying linear density to remain decreasing. Strength is the difference
of the pre- and post-transition secant slopes.
\end{itemize}

\paragraph{Multiscale selection and posterior matching.}
The default widths are
\begin{equation}
 s_{\rm 1D}=(0.050,0.075,0.100),
 \label{eq:method_marginal_scales}
\end{equation}
with a minimum extremum separation of $0.050$, a matching tolerance of
$0.100$ in $y$, and recurrence at two scales. Each first-scale candidate
is tested independently: one coarser-scale detection can support several
first-scale candidates, unlike the one-to-one two-dimensional assignments.

Each posterior marginal repeats the multiscale search. Peaks and shoulders
share a positive-feature family and may realise the same reference
identity; dips match only dips. A retained shoulder near a retained
first-scale peak is removed in favour of the peak. Within each family,
neighbouring reference locations define non-overlapping assignment
intervals with midpoint boundaries in $\ln x$. Candidates propose only
to their nearest compatible identity and must lie inside its interval.
Competing proposals retain the closest candidate, with larger prominence
breaking distance ties; unmatched candidates do not enlarge the reference
catalogue.

Location support, bounded-interval support, and bounded support conditional
on location are reported separately, together with the peak-versus-shoulder
fractions. Medians and central $90\%$ intervals use the 5th, 50th, and 95th
percentiles on the appropriate valid draws. Strengths additionally select
the realised morphology, rather than combining peak contrast and shoulder
slope change. A reference identity with no valid measurements has no
conditional location or strength summary.\\

\paragraph{Measurements and joint counterparts.}
Moving-interval probability and first moments use the unsmoothed
selected-measure marginal and account for partially covered boundary cells.
For peaks and shoulders, the density weights the centroid. For dips, a
background interpolated linearly in $\ln q_s$ between prominence bases
defines a positive deficit relative to the smoothed marginal; its first
moment supplies the depression scale. 

Joint-region projections and the overlap metrics $J,C$ are defined in
Sec.~\ref{sec:marginal_features}. Individual curve branches do not enter
this region-projection comparison; measurable paired ridge and valley
events do. All compatible pairs remain available, while a marginal identity
selects a primary supported joint counterpart with the largest positive
reference overlap, breaking ties by scale proximity. Separated joint
location configurations are compared individually. Draw-level overlap,
capture, and centroid comparisons use the common valid subset. The joint
and marginal intervals remain independently defined; neither is adjusted
to enforce agreement.

\subsection{Association estimators and finite-sample diagnostics}
\label{app:technical_association}

\paragraph{Binary comparisons and valid subsets.}
For present and absent groups of sizes $n_1,n_0$, the rank-biserial effect
used in Sec.~\ref{sec:parameter_association} is
\begin{equation}
 r_{\rm rb}=\frac{2U}{n_1n_0}-1,
 \label{eq:method_binary_association}
\end{equation}
where $U$ is the Mann--Whitney statistic for the present group
\citep{Mann1947U,Cureton1956RankBiserial}. Each group must contain at least
20 draws. Two-dimensional tests compare morphological recovery over all
draws, location availability conditional on recovery, and region
availability conditional on location. Co-occurrence indicators can be
tested similarly. For separated configurations, membership in one is
compared with membership in the others, conditional on a valid parent
region.

Region-based continuous observables include centroid coordinates, enclosed
probability, relative projected widths, strength, extent, integrated
deficit, and bounded/fallback fractions. The two-dimensional objects are
peaks, measurable interior pits, paired curve events, and shoulders;
individual branches and flat regions remain morphological outputs.
Every comparison uses its common valid subset. Within accepted location
configurations, associations are evaluated separately rather than across
their mixture.\\

\paragraph{Nearest-neighbour information estimates.}
Ordinary information uses the first Kraskov--St\"ogbauer--Grassberger
estimator \citep{Kraskov2004MutualInformation}. Scalar arguments are
standardised and multivariate arguments, including the physical centroid
vector, are whitened. A reproducible perturbation of amplitude $10^{-10}$
resolves exact ties. For $N$ valid samples, let $\epsilon_i$ be the distance
to the $k$th joint-space neighbour under the maximum norm. Counts
$n_\lambda(i)$ and $n_Q(i)$ exclude the sample itself and use a strict
radius $\epsilon_i$ in their respective marginal spaces. The estimate in
bits is
\begin{equation}
\begin{split}
 \widehat I(\lambda;Q)=\frac1{\ln2}\Bigg[&\psi(k)+\psi(N)\\
 &-\frac1N\sum_{i=1}^N\big\{\psi[n_\lambda(i)+1]\\
 &\hspace{18mm}+\psi[n_Q(i)+1]\big\}\Bigg],
\end{split}
\label{eq:method_ksg}
\end{equation}
where $\psi$ is the digamma function and $k=5$ by default.
Ordinary information requires $N>\max(k+2,20)$.

Conditional information is estimated directly using the nearest-neighbour
construction of \citet{Frenzel2007ConditionalMI}. For a conditioning variable $Z$,
\begin{equation}
\begin{split}
 \widehat I(\lambda;Q\mid Z)=\frac1{\ln2}\Bigg[\psi(k)
 &+\frac1N\sum_{i=1}^N\big\{\psi[n_Z(i)+1]\\
 &-\psi[n_{\lambda Z}(i)+1]\\
 &-\psi[n_{QZ}(i)+1]\big\}\Bigg].
\end{split}
\label{eq:method_conditional_information}
\end{equation}
The radius is selected in $(\lambda,Q,Z)$ space and the counts in the
indicated subspaces, again excluding the sample itself and using a strict
radius. Each coordinate is standardised separately for this estimator,
which requires $N>\max(k+2,30)$.\\

\paragraph{Empirical null calibration and stability.}
A non-zero association estimate does not by itself establish genuine
dependence. With finitely many posterior draws, mutual-information estimates
can have a non-zero baseline, while apparent differences between binary states
can arise from chance alignment between $\lambda$ and feature recovery.
Because the supplied draws may also be correlated within chains, the
appropriate reference should preserve that chain structure while removing the
specific pairing between the parameter and the feature measurements.

We therefore construct an empirical no-association reference using surrogate
realisations \citep{Theiler1992Surrogate}. In each replicate, the sequence
$\lambda$ is circularly shifted by an independently sampled nonzero offset
within each supplied chain, before selecting the feature-valid draws. The
feature measurements and their validity indicators remain fixed. The shift
preserves the marginal distribution and cyclic ordering of $\lambda$ within
each chain, but changes its draw-by-draw alignment with the feature
measurements. Its interpretation assumes that the supplied chains are
approximately stationary.
For mutual information, the surrogate ensemble estimates the finite-sample
offset expected in the absence of the original alignment. For binary
associations, it provides the reference distribution used to assess how
unusual the observed effect is. We assess the stability of the information
estimates separately by repeating the calculation on subsets of the valid
draws.
With $N_{\rm null}=100$ by default,
ordinary and conditional information are reported as
\begin{equation}
 \widetilde I=\max\{0,\widehat I_{\rm obs}
       -\operatorname{med}_r\widehat I_{{\rm shift},r}\}.
 \label{eq:method_information_offset}
\end{equation}
For conditional-information stability, the 5th and 95th percentiles are
obtained from repeated $80\%$ subsamples, subtracting the full-sample null
median and applying the same truncation. These are descriptive stability
intervals, not posterior credible intervals for the information. The shift
correction is not a significance calibration for conditional independence.

Binary associations use the same surrogate shifts to assess whether the
observed separation of the parameter distributions is larger than expected
from an accidental alignment between $\lambda$ and the binary feature state.
For each shifted replicate, the rank-biserial effect is recomputed, producing
an empirical no-association distribution. We report the two-sided Monte Carlo
probability
\begin{equation}
 p_{\rm MC}=\frac{1+\sum_{r=1}^{N_{\rm null}}
  \mathbf1\{|r_{{\rm rb},r}|\geq|r_{{\rm rb},{\rm obs}}|\}}
  {N_{\rm null}+1}.
 \label{eq:method_binary_null}
\end{equation}
Thus $p_{\rm MC}$ is the fraction of surrogate effects whose magnitude is at
least as large as the observed one, with the plus-one correction appropriate
for randomly sampled null replicates
\citep{Phipson2010Permutation}. A small value indicates that an effect of the
observed magnitude is uncommon after the draw-level alignment has been
broken, whereas a large value means that comparable effects frequently arise
under the surrogate reference.

The sign and magnitude of the association are given by $r_{\rm rb}$;
$p_{\rm MC}$ quantifies only how unusual that magnitude is under the
shift-based no-association construction. It is neither the posterior
probability that no association exists nor a measure of effect size. The
package reports these probabilities separately for each feature channel and
does not apply an automatic multiple-testing correction, because the relevant
family of simultaneous comparisons depends on the scientific application.
When several channels are interpreted jointly, their probabilities should be
adjusted over a prespecified family. For example, the companion application
uses Holm's step-down correction, which controls the family-wise error rate
without requiring independence among the tests
\citep{H0UpdateHolm1979,DeRenzis:2026hcp}.\\

\paragraph{One-dimensional conventions and joint comparisons.}
The independent one-dimensional association routines require at least
200 valid samples for scalar information and use seeded random tie-breaking
perturbations. They report the raw estimate, null median, their difference
without non-negative truncation, and a one-sided Monte Carlo probability
from the fraction of shifted estimates at least as large as observed,
with the same plus-one correction. Chain-preserving shifts in this path
require equal chain lengths. Binary comparisons distinguish located from
absent identities, bounded from unbounded intervals conditional on location,
and peak from shoulder realisations. Continuous quantities include the
finite-band scale, interval probability, relative width, and
morphology-specific strength.

Joint--marginal information comparisons instead use the two-dimensional
estimators on one identical common-valid subset. For marginal centroid
$\mu_{\rm 1D}$, the separate and joint associations of
$\mu_{\rm 1D},\mu_1,\mu_2$ with $\lambda$ are supplemented by
\begin{equation}
\begin{split}
 &I\{\lambda;(\mu_1,\mu_2)\mid\mu_{\rm 1D}\},\\
 &I\{\lambda;\mu_2\mid(\mu_{\rm 1D},\mu_1)\},\\
 &I\{\lambda;\mu_{\rm 1D}\mid(\mu_1,\mu_2)\}.
\end{split}
\label{eq:method_joint_marginal_information}
\end{equation}
These compare dependence retained by the joint location with that retained
by the marginal scale. A binary test separately checks association of
$\lambda$ with membership in the common-valid subset. When both marginals
have primary counterparts in the same joint feature, their four scales are
also grouped for rank-correlation and information summaries; if several
pairs qualify, the one with the largest common-valid sample is selected.
Global-tail observables form a separate family.

\subsection{Optional flat-region and tail analyses}
\label{app:technical_optional}

\paragraph{Plateaus and depression floors.}
These optional diagnostics identify extended, nearly flat regions,
rather than isolated extrema or curves. A plateau has higher median
density than its immediate surroundings; a depression floor has lower
median density. Both are searched only within
$\mathcal H_{0.995}(F_s)$, at the first requested smoothing scale.
They do not use topological persistence or cross-scale recurrence.

To quantify flatness, let
$D_s=\max_\Omega F_s-\min_\Omega F_s>0$ be the density range,
and let $L$ be the diagonal length of the rectangular analysis domain
in the selected coordinates $\boldsymbol g$.
The dimensionless gradient and curvature measures are
\begin{equation}
 \mathcal G=
 \frac{L}{D_s}\|\nabla_{\boldsymbol g}F_s\|,
 \qquad
 \mathcal C=
 \frac{L^2}{D_s}\|H_s\|_{\rm F},
 \label{eq:method_flatness}
\end{equation}
where $H_s$ is the Hessian of $F_s$ with respect to
$\boldsymbol g$, and $\|\cdot\|_{\rm F}$ is its Frobenius norm.
Small $\mathcal G$ indicates a slowly varying density, while small
$\mathcal C$ indicates little variation in its slope.

Let $Q_{0.15}(\mathcal G)$ and $Q_{0.15}(\mathcal C)$ denote the
fifteenth percentiles evaluated over the search support.
Nodes satisfying both flatness conditions are grouped into connected
components. A candidate component must satisfy
\begin{equation}
 \begin{split}
  \mathcal G&<Q_{0.15}(\mathcal G),\\
  \mathcal C&<Q_{0.15}(\mathcal C),\\
  N_{\rm comp}&\geq N_{\min},
 \end{split}
 \label{eq:method_flat_selection}
\end{equation}
where $N_{\rm comp}$ is its number of grid nodes,
$N_{\min}=\max\{9,\lfloor0.0005N_{\rm valid}\rfloor\}$,
and $N_{\rm valid}$ is the number of valid grid nodes in $\Omega$.
The first two inequalities apply to the nodes forming the component;
the third rejects components that are too small.

For each candidate, a surrounding collar two grid-node steps wide
provides the comparison level. Define $\Delta F_{\rm collar}$ as
the median density inside the component minus the median density
in this collar. The component is retained only if
\begin{equation}
 |\Delta F_{\rm collar}|\geq10^{-3}D_s.
 \label{eq:method_flat_contrast}
\end{equation}
Positive contrast identifies a plateau and negative contrast a
depression floor. This signed contrast is distinct from the
non-negative missing-density field $\Delta F$ used for pit and
valley measurements. The connected component itself supplies the
region associated with the diagnostic.
Table~\ref{tab:method_optional_flat_regions} collects the criteria.

\begin{table}[t]
\centering
\caption{
Inclusion criteria and reported contrasts for the optional flat-region
diagnostics. The quantities are defined in the accompanying text.
Flatness conditions apply within $\mathcal H_{0.995}(F_s)$;
the size and collar-contrast tests apply to each connected component.
Both diagnostics use only the first requested smoothing scale.
}
\label{tab:method_optional_flat_regions}
\small
\setlength{\tabcolsep}{3pt}
\renewcommand{\arraystretch}{1.15}
\begin{tabular}{@{}
p{0.22\columnwidth}
p{0.51\columnwidth}
p{0.21\columnwidth}
@{}}
\toprule
\raggedright Diagnostic
&
\raggedright Fixed-scale inclusion
&
\raggedright Reported contrast
\tabularnewline
\midrule

\raggedright Plateau
&
\raggedright
$\begin{aligned}
 \mathcal G&<Q_{0.15}(\mathcal G),\\
 \mathcal C&<Q_{0.15}(\mathcal C),\\
 N_{\rm comp}&\geq N_{\min},\\
 \Delta F_{\rm collar}&\geq10^{-3}D_s.
\end{aligned}$
&
\raggedright
$\Delta F_{\rm collar}>0$.
\tabularnewline[3mm]

\raggedright Depression floor
&
\raggedright
$\begin{aligned}
 \mathcal G&<Q_{0.15}(\mathcal G),\\
 \mathcal C&<Q_{0.15}(\mathcal C),\\
 N_{\rm comp}&\geq N_{\min},\\
 -\Delta F_{\rm collar}&\geq10^{-3}D_s.
\end{aligned}$
&
\raggedright
$\Delta F_{\rm collar}<0$.
\tabularnewline

\bottomrule
\end{tabular}
\end{table}

\paragraph{Global tail scales and ordered domains.}
A tail scale is a cumulative-probability quantile, not the location of a
local morphology. Each marginal supplies the coordinate below which a
chosen probability lies, $0.999$ by default in this summary construction.
It is evaluated separately for each draw without assigning a finite
feature region.

For an ordered domain $x_2\leq x_1$, define
$S_{i,b}(x_*)=\Pr_b(x_i>x_*)$. The probabilities that at least one
coordinate exceeds $x_*$, both do, or the threshold lies between them are
\begin{equation}
\begin{split}
 P_{{\rm any},b}(x_*)&=S_{1,b}(x_*),\\
 P_{{\rm both},b}(x_*)&=S_{2,b}(x_*),\\
 P_{{\rm straddle},b}(x_*)&=\max\{S_{1,b}(x_*)-S_{2,b}(x_*),0\}.
\end{split}
\label{eq:method_ordered_tail_probabilities}
\end{equation}
These are probability contents of a density realisation. Their posterior
variation is uncertainty in the tails, not the recovery probability of a
local feature.

\section{Software inputs and outputs}
\label{app:software_io}

\plandscape{} takes an ensemble of already reconstructed
two-dimensional densities, supplied as an HDF5 or NumPy NPZ array with shape
$(B,N_1,N_2)$, together with two increasing coordinate arrays. A single density
uses $B=1$. A Boolean mask specifies the supported domain, while optional
weights refer to posterior draws. The configuration declares the density
measure, analysis geometry, logarithm base, and whether coordinates are nodes
or finite-volume cell centres. Nodal integration is the default; finite-cell
integration is explicitly selected, with edges inferred in the declared native
linear or logarithmic coordinate. Every active finite cell, including an
ordered-domain diagonal cell, receives its full nominal weight. Optional scalar
samples aligned with the density draws enable parameter-association analysis.

An INI file controls the requested analyses and output location. The command
\texttt{posterior-landscape validate} audits the input and normalization,
while \texttt{posterior-landscape run} executes or resumes the analysis; both
take the INI filename as an argument. The principal products of a combined
analysis are summarized in Table~\ref{tab:software_outputs}. Optional modules
write their corresponding products only when requested.

\begin{table*}[t]
\centering
\small
\renewcommand{\arraystretch}{1.15}
\caption{Principal outputs of \texttt{posterior-landscape}. The consolidated
catalogue and numerical archive provide complementary summary-level and
draw-level access. Additional diagnostic tables and figures are available in
the full output profile.}
\label{tab:software_outputs}
\begin{tabular}{@{}p{0.29\textwidth}p{0.66\textwidth}@{}}
\hline
Product & Contents \\
\hline
\texttt{features.csv}
& Retained reference families, posterior recovery, conditional location
intervals, and class-specific strengths and finite-region measurements. \\
\texttt{results.h5}
& Coordinates and quadrature, reference maps, draw-level presence and validity
flags, locations, centroids, moments, curves, region-inclusion maps, and
projections; paired connections and optional summaries occupy separate groups. \\
Marginal and comparison tables
& Independently identified one-dimensional families and their uncertainties,
with overlaps and capture fractions relative to projected joint regions. \\
Association and tail tables
& Optional rank correlations and mutual information with an aligned scalar,
or global upper-tail scales and probability decompositions. \\
PDF figures
& Landscape overview, independent marginal features, and full-profile
diagnostics for geometry, projections, and associations. \\
Settings, manifests, and log
& Configuration and provenance, package and schema versions, selected
thresholds, normalization diagnostics, and output inventory. \\
\hline
\end{tabular}
\end{table*}

Feature recovery is distinct from measurement availability and from population
probability integrated within a feature region. Locations are summarized
conditional on recovery; finite measurements additionally use their applicable
validity flags. Reference extrema, posterior feature locations, and
finite-region centroids are separate quantities. Likewise, individual ridge
arms and their paired saddle-centred connection are related descriptions rather
than independent features. The saved arrays allow these distinctions to be
retained in subsequent plots and analyses. Association statistics describe
posterior dependence within the available sample and are not additive
contributions to an external parameter's constraint.

The \texttt{essential} profile selects the main reports while retaining the
draw-level HDF5 arrays. The \texttt{full} profile writes all diagnostics and
creates an essential view from the same completed calculation. A one-dimensional
analysis without the joint workflow instead produces its own marginal tables,
figures, manifests, and state files. The original density input should be kept
alongside these products. Installation instructions, configuration examples,
and the detailed output-field reference are maintained in the repository
documentation.

\section{Analytic definition of the synthetic density}
\label{app:synthetic_density}

Here we give the explicit form of the density used for the controlled injection--recovery test in Sec.~\ref{sec:controlled_validation}. 

Explicitly, write $u=\ln x_1$, $v=\ln x_2$, and introduce rotated
coordinates about the centre of the domain,
\begin{equation}
\begin{aligned}
s&=\cos\theta\,(u+1.75)+\sin\theta\,(v+1.75),\\
r&=-\sin\theta\,(u+1.75)+\cos\theta\,(v+1.75),
\end{aligned}
\end{equation}
with $\theta=50^\circ$, $t=(u+v)/2$, and $q=(u-v)/2$.
Define the unnormalised Gaussian
\begin{equation}
G_{a,b}(z,w;\mu,\nu)
=\exp\left[-\frac{(z-\mu)^2}{2a^2}
           -\frac{(w-\nu)^2}{2b^2}\right].
\end{equation}
The density per $du\,dv$ on $\mathcal D=[-3.5,0]^2$ is
\begin{equation}
\begin{aligned}
F_\lambda(u,v)
&=0.92\,\frac{e^{L_\lambda(u,v)}}{Z_\lambda}
 +0.08\,\frac{H(u,v)}{Z_H},\\
H(u,v)&=G_{0.50,0.15}(t,q;-0.25,-0.10),
\end{aligned}
\label{eq:crosspass_density}
\end{equation}
where, with $s_\lambda=-\sqrt{2}(0.62)+0.18\lambda$,
\begin{equation}
\begin{aligned}
L_\lambda={}&
\frac{3(s^2-r^2)}{0.62^2}
 \exp\left[-\frac{s^2+r^2}{2(0.62)^2}\right]\\
&+0.16\tanh\left(\frac{2sr}{1.75^2}\right)\\
&-0.40\,\frac{s}{0.62}
 \exp\left[-\frac{s^2+r^2}{2(0.70)^2}\right]\\
&+0.40\,G_{0.22,0.20}(s,r;s_\lambda,0)\\
&-0.58\,G_{0.08,0.10}(t,q;-0.62,0.15).
\end{aligned}
\label{eq:crosspass_logcore}
\end{equation}
Here $Z_\lambda$ and $Z_H$ normalise $e^{L_\lambda}$ and $H$
separately over $\mathcal D$. The construction combines a rotated
peak--pit pattern, a weak twist and asymmetry, a local perturbation
whose centre depends on $\lambda$, and a leading depression followed
by a broad shelf. 

\section{Robustness to configuration choices}
\label{app:robustness}

To assess how strongly the recovered catalogue depends on configuration
choices, we repeated the controlled tests for a set of variations around the
baseline settings. The goal of this appendix is not to exhaust the full
parameter space of the package, but to verify that the main scientific
conclusions are not driven by a narrow tuning of the defaults.

The variations, listed explicitly in
Table~\ref{tab:controlled_robustness_configurations}, probe the aspects of
the method most likely to affect the recovered catalogue: two-dimensional
resolution, persistence selection, the probability support used for extended
geometry, the independent one-dimensional smoothing scale, and numerical
grid resolution.
The baseline
configuration is the one used in the main text. Each variation is compared
with that baseline in terms of the recovered feature families and, for the
posterior ensemble, the associated support values and summary measurements.

Table~\ref{tab:controlled_robustness_configurations} summarizes the
configuration variations used in the stress tests. Each run changes one
analysis choice relative to the baseline, leaving all other settings fixed.
The deterministic and posterior-ensemble tests use the same variations where
applicable; grid resolution is varied only for the deterministic density,
whereas the non-zero first smoothing scale is used only for the posterior
ensemble.

\begin{table*}[t]
\centering
\caption{
Configuration variations used in the controlled robustness tests.
Only the setting shown in each row is changed relative to the baseline.
The baseline uses two-dimensional scales $(0,0.025,0.05)$, automatic
persistence selection, probability support $\gamma=0.995$, one-dimensional
scales $(0.050,0.075,0.100)$, and a $321\times321$ grid.
Here $T_0$ denotes the automatic persistence threshold obtained at the
baseline first scale for the corresponding dataset. A numerical persistence
threshold is held fixed across smoothing scales, whereas the baseline
automatic procedure recalibrates the threshold at each scale.
}
\label{tab:controlled_robustness_configurations}
\small
\setlength{\tabcolsep}{5pt}
\begin{tabular}{llll}
\toprule
Configuration &
Changed setting &
Value &
Applied to \\
\midrule

Baseline
& --- 
& reference settings
& fixed + ensemble \\

Fine 2D scales
& 2D smoothing scales
& $(0,0.0125,0.025)$
& fixed + ensemble \\

Coarse 2D scales
& 2D smoothing scales
& $(0,0.05,0.10)$
& fixed + ensemble \\

Half persistence cutoff
& persistence threshold
& $0.5\,T_0$
& fixed + ensemble \\

Double persistence cutoff
& persistence threshold
& $2\,T_0$
& fixed + ensemble \\

Reduced support
& probability support $\gamma$
& $0.990$
& fixed + ensemble \\

Extended support
& probability support $\gamma$
& $0.999$
& fixed + ensemble \\

Fine 1D scales
& marginal smoothing scales
& $(0.025,0.0375,0.050)$
& fixed + ensemble \\

Coarse 1D scales
& marginal smoothing scales
& $(0.100,0.150,0.200)$
& fixed + ensemble \\

Non-zero first scale
& 2D smoothing scales
& $(0.0125,0.025,0.05)$
& ensemble only \\

Coarse grid
& grid resolution
& $161\times161$
& fixed only \\

Fine grid
& grid resolution
& $641\times641$
& fixed only \\

\bottomrule
\end{tabular}
\end{table*}

The deterministic recovery is robust to all tested configuration changes.
Every fixed-density analysis recovers the injected two-dimensional catalogue,
including variations of the smoothing scales, persistence selection,
probability support, and numerical grid resolution. Once posterior
draw-to-draw variation is introduced, the same large-scale feature identities
remain stable, but additional weak topology can appear in the reference
catalogue for finer two-dimensional smoothing or when the scale-dependent
automatic persistence selection is replaced by a fixed cutoff. The additional
critical points have substantially smaller prominence than the injected
critical points, providing a useful diagnostic of their secondary character.
The independent one-dimensional representation of the broad shoulder is also
more sensitive to marginal smoothing than its two-dimensional counterpart.

\subsection{Fixed-density stress tests}
\label{app:robustness_fixed}

\begin{figure*}[t]
    \centering
    \includegraphics[width=0.75\textwidth]{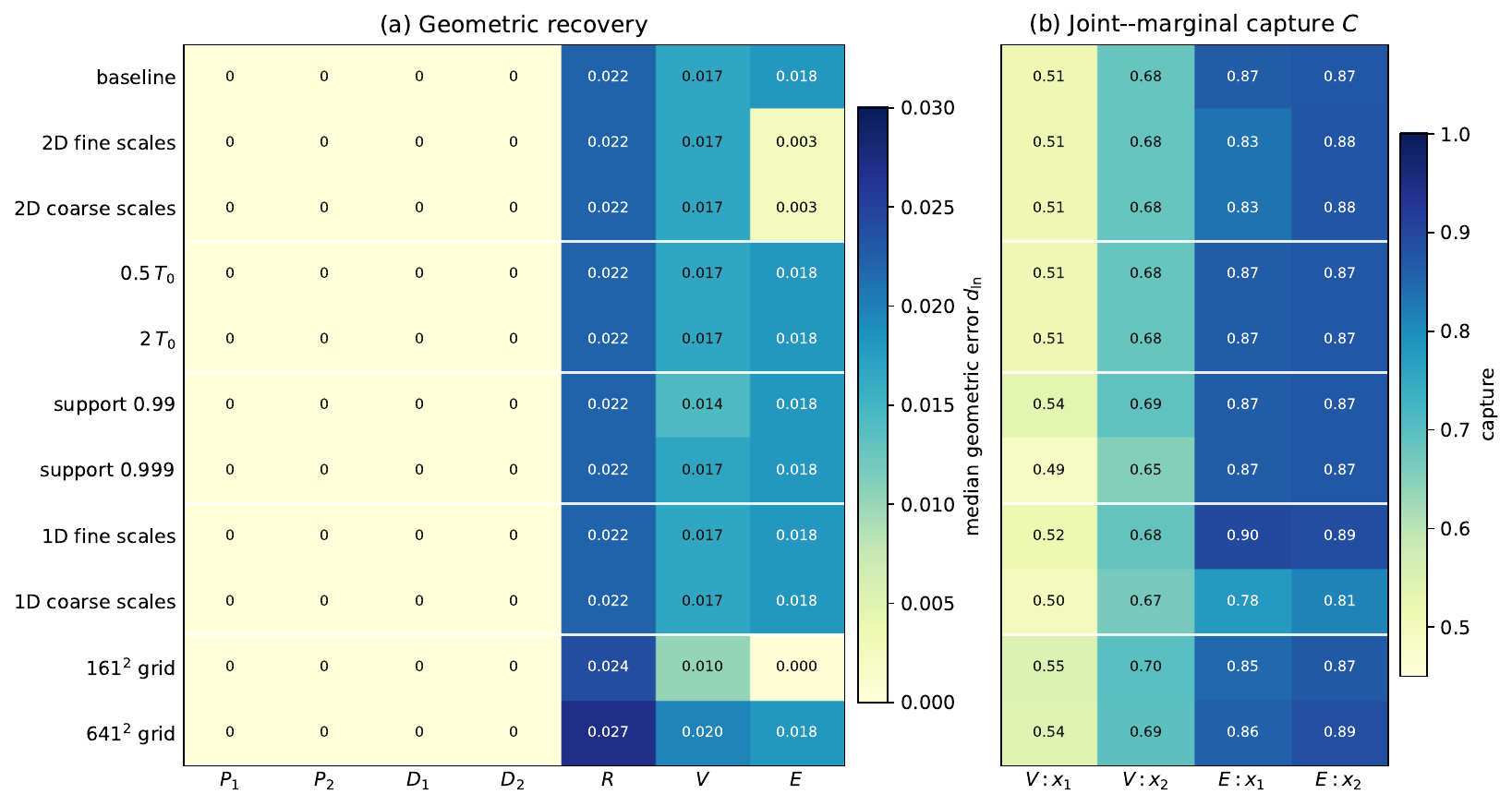}
\caption{
Robustness of the controlled fixed-density recovery under configuration
variations. Panel~(a) gives the geometric discrepancy from the injected
critical points and extended features. All four critical points remain on
their injected grid locations, and the ridge, valley, and shoulder retain
small geometric errors throughout. Panel~(b) gives the joint--marginal
capture $C$ for the valley and shoulder projections. No tested
fixed-density configuration adds or removes a two-dimensional feature. 
}
    \label{fig:appendix_fixed_robustness}
\end{figure*}

Figure~\ref{fig:appendix_fixed_robustness} summarizes the deterministic
stress tests. All configurations recover exactly the injected
two-dimensional catalogue: the two peaks, two pits, ridge, valley, and
shoulder remain present, with no additional feature family introduced.

Panel~(a) quantifies the corresponding geometric recovery. The four critical
points remain on their injected grid locations in every configuration, while
the median logarithmic discrepancies of the extended features remain small.
The largest value among the displayed ridge, valley, and shoulder errors is
$0.027$. The shoulder locus is somewhat more sensitive to the smoothing
scale and grid discretization than the critical points, but without changing
its identity or producing additional topology.

Panel~(b) shows the corresponding joint--marginal capture. These quantities
vary modestly with probability support and with the independent
one-dimensional smoothing scales, as expected because they depend on the
finite-region boundaries and on independently inferred marginal intervals.
They do not indicate a change in the underlying two-dimensional catalogue.

The deterministic experiment therefore provides a useful control: when the
underlying density itself is fixed, neither the tested numerical resolution
nor the analysis settings generate spurious catalogue structure. 

\subsection{Posterior-ensemble robustness}
\label{app:robustness_ensemble}

\begin{figure*}[t]
    \centering
    \includegraphics[width=0.8\textwidth]{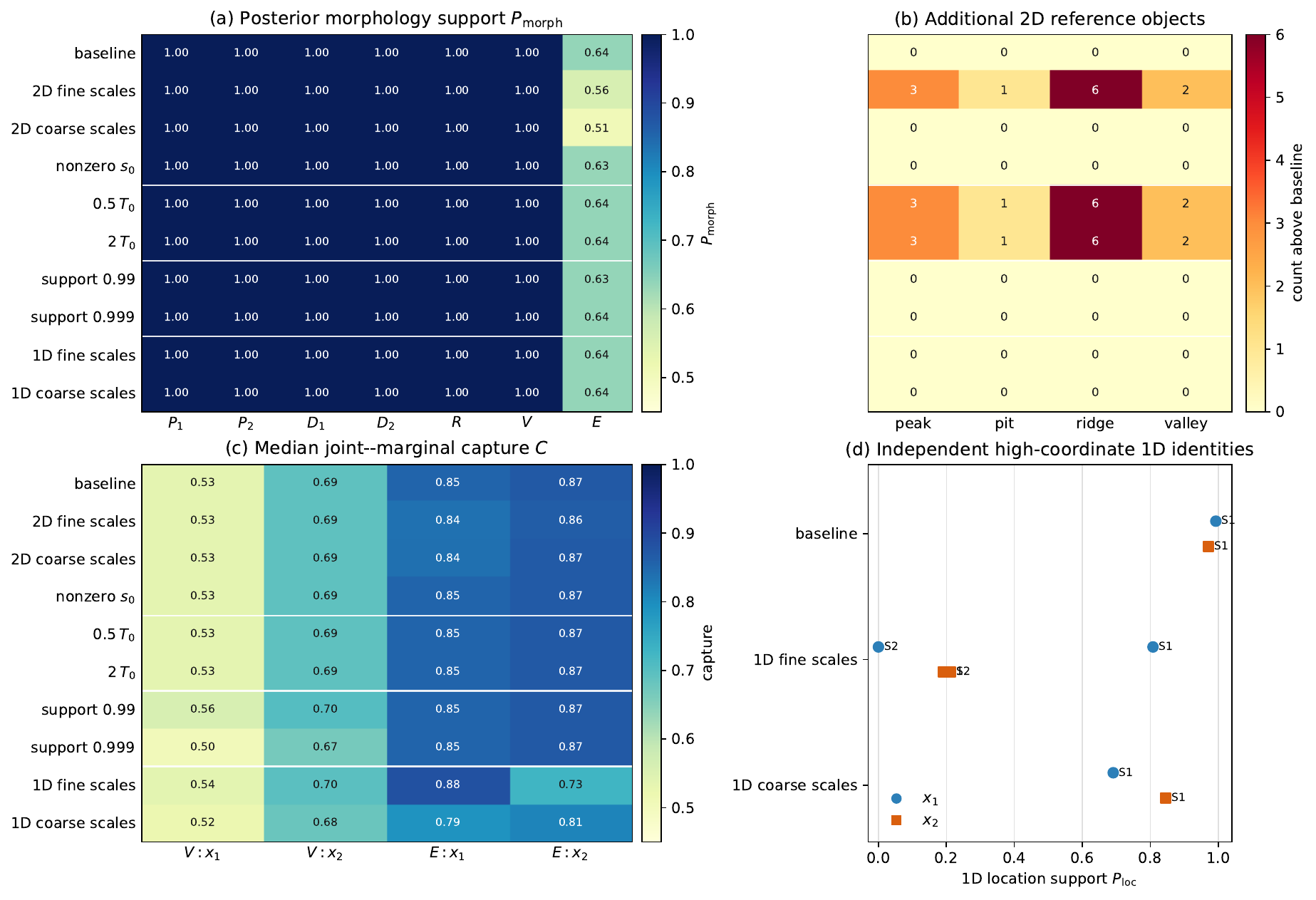}
\caption{
Robustness of the posterior-ensemble analysis. Panel~(a) gives the morphology
support $P_{\rm morph}$ of the baseline two-dimensional identities. The
peaks, pits, ridge, and valley remain essentially fully supported throughout;
the shoulder varies between approximately $0.51$ and $0.64$. Panel~(b)
counts reference objects absent from the baseline catalogue. Additional
objects occur only for the fine two-dimensional scales and the two fixed
persistence-threshold runs; each contains three additional peaks, one pit,
six ridge branches, and two valley branches. The four added critical points
have substantially lower prominence than the baseline critical points, while
the additional branches arise from the resulting change in topological
connectivity. Panel~(c) shows the median joint--marginal capture $C$ for the
valley and shoulder. Panel~(d) shows the location support of the
high-coordinate one-dimensional positive-feature identities for the
reference, fine, and coarse marginal smoothing scales. The latter sensitivity
concerns the marginal representation of the shoulder rather than its
two-dimensional recovery.
}
    \label{fig:appendix_ensemble_robustness}
\end{figure*}

Figure~\ref{fig:appendix_ensemble_robustness} shows the corresponding
robustness analysis for the posterior ensemble. Here the issue is not only
whether a feature is present in the reference catalogue, but also whether its
posterior support and summary measurements remain stable across
configurations.

The principal two-dimensional morphology remains stable across the posterior
stress tests. As shown in panel~(a) of
Fig.~\ref{fig:appendix_ensemble_robustness}, the baseline peaks, pits, ridge,
and valley retain essentially unit morphology support throughout. The
shoulder remains the less certain feature already identified in the reference
analysis, with $P_{\rm morph}$ varying from approximately $0.51$ to $0.64$
as the two-dimensional smoothing scales are changed.

The principal sensitivity appears instead in the number of secondary
reference objects. Panel~(b) shows that additional two-dimensional objects
occur only for the fine-scale analysis and for the two runs in which the
scale-dependent automatic persistence selection is replaced by fixed
thresholds $0.5T_0$ and $2T_0$. Each of these catalogues contains three
additional peaks and one additional pit, together with six ridge branches
and two valley branches generated by the expanded critical-point
connectivity. No additional reference object occurs for the coarse scales,
the non-zero first scale, either probability-support variation, or either
one-dimensional smoothing variation.

The four additional critical points are weak relative to the injected
catalogue. Their reference prominences lie between approximately
$2.9\times10^{-3}$ and $1.1\times10^{-2}$, whereas the smallest prominence
among the baseline critical points is approximately $4.2\times10^{-2}$.
Thus the new critical points occupy a clearly lower-prominence part of the
catalogue. The associated branches should not all be described as
low-prominence, since some connect a newly resolved weak extremum to an
existing strong extremum. Rather, they are consequences of introducing the
additional weak critical points into the topological graph.

In this
controlled example, prominence and persistence therefore provide useful
indicators for distinguishing the dominant morphology from weak topology
revealed only under some analysis resolutions. The automatic
scale-dependent persistence selection used in the baseline acts precisely in
this direction; the experiment does not motivate a universal numerical
prominence cutoff.

The independent one-dimensional analysis shows a different kind of
sensitivity. Finer marginal smoothing resolves the broad high-coordinate
transition into several competing positive-feature identities, while coarse
smoothing reduces its location support and its capture of the projected
two-dimensional shoulder. This affects the marginal decomposition rather
than the two-dimensional morphology: changing the one-dimensional scales
does not alter the recovered two-dimensional shoulder, whose morphology
support remains at its baseline value. This distinction is expected because
the shoulder is injected as a structure of the joint density, whereas its
one-dimensional manifestations arise only after marginalization.\\

\paragraph{Interpretation of the additional families.}
The additional ensemble objects are absent from the known injected
morphology and from every deterministic stress test. Their appearance
therefore does not indicate instability of the main recovered features.
Instead, posterior variation may create weak secondary structure in the
reference field that can enter the catalogue at finer effective resolution
or under a different persistence-selection rule. The clear separation in
prominence between the four added critical points and the baseline critical
points shows why persistence-based selection is useful: it provides a
principled way to control how far into this weak topological structure the
catalogue is allowed to extend. The six additional ridge branches and two
valley branches largely reflect the connectivity induced by those four new
critical points and should not be interpreted as eight independent new
physical structures.\\

\paragraph{Interpretation of the one-dimensional sensitivity.}
The one-dimensional fine-scale runs show that the broad high-coordinate
structure is not represented by a unique marginal decomposition. This is not
surprising for a shoulder-like transition, whose definition relies on local
slope and curvature structure rather than on an isolated extremum. The
two-dimensional shoulder remains stable because it follows the morphology of
the full injected density. The one-dimensional catalogue is correspondingly
more sensitive to how that broad transition is smoothed and partitioned after
projection.\\

\paragraph{Summary.}
For the deterministic density, the injected morphology is recovered under
all tested numerical and analysis variations. With posterior uncertainty,
the dominant two-dimensional feature identities remain stable, while weak
non-injected critical points can become resolved under finer effective
smoothing resolution or altered persistence selection. Their substantially
lower prominence provides a natural diagnostic of their secondary status.
The remaining sensitivity is chiefly in the one-dimensional representation
of the broad shoulder, rather than in its two-dimensional recovery.

\section{Posterior-like ensemble specification}
\label{app:ensemble_specification}

The ensemble used in Section~\ref{sec:controlled_ensemble} contains 1000
equally weighted density draws. Table~\ref{tab:controlled_ensemble_draws}
gives the prescribed sampling distribution. All quantities are drawn
independently. This deliberate simplification isolates the designed
$P_1$--$\lambda$ dependence from nuisance-parameter correlations while still
allowing the nuisance draws to broaden the recovered feature posteriors.

\begin{table*}[t]
\centering
\caption{
Sampling specification for the posterior-like ensemble. Here $c$, $\theta$,
$w$, and $A$ are the centre, orientation, width, and contrast of the coherent
ridge--valley landscape. The quantities $\Delta t_E$ and $f_E$ perturb the
position and strength of the slope-change feature. For log-normal draws, the
table gives the standard deviation of the logarithm relative to the fiducial
value.
}
\label{tab:controlled_ensemble_draws}
\small
\begin{tabular}{lcc}
\toprule
Quantity &
Fiducial value &
Sampling distribution \\
\midrule
$\lambda$
& $0$
& $\mathcal{N}(0,0.60^2)$ \\

Landscape centre $c$
& $-1.75$
& $\mathcal{N}(-1.75,0.025^2)$ \\

Orientation $\theta$
& $50^\circ$
& $\mathcal{N}(50^\circ,(2^\circ)^2)$ \\

Landscape width $w$
& $0.62$
& $\ln(w/0.62)\sim\mathcal{N}(0,0.05^2)$ \\

Landscape contrast $A$
& $3.0$
& $\ln(A/3.0)\sim\mathcal{N}(0,0.08^2)$ \\

Shoulder displacement $\Delta t_E$
& $0$
& $\mathcal{N}(0,0.04^2)$ \\

Injected shoulder amplitude factor $f_E$
& $1$
& $\ln f_E\sim\mathcal{N}(0,0.20^2)$ \\
\bottomrule
\end{tabular}
\end{table*}

Every density draw uses the same $321\times321$ grid, uniform in
$(\ln x_1,\ln x_2)$, and is normalized with respect to
$d\ln x_1\,d\ln x_2$. All draws receive equal posterior weight.

\paragraph{Supplementary critical-point locations.}
Table~\ref{tab:controlled_ensemble_locations} reports the coordinate-wise
posterior locations corresponding to the joint location regions shown in
Fig.~\ref{fig:controlled_ensemble_geometry}. 

\begin{table}[t]
\centering
\caption{
Posterior locations of the critical points in the controlled ensemble.
Each location is the posterior median followed by the coordinate-wise central
$90\%$ credible interval, conditional on location availability.
}
\label{tab:controlled_ensemble_locations}
\small
\setlength{\tabcolsep}{3pt}
\begin{tabular}{lccc}
\toprule
Feature & $P_{\mathrm{morph}}$ & $x_1$ location & $x_2$ location \\
\midrule
$P_1$ & $1.000$ & $0.0995\;[0.0921,\,0.1086]$ & $0.0892\;[0.0808,\,0.0973]$ \\
$P_2$ & $1.000$ & $0.3103\;[0.2874,\,0.3350]$ & $0.3538\;[0.3241,\,0.3904]$ \\
$D_1$ & $1.000$ & $0.0902\;[0.0826,\,0.0973]$ & $0.3103\;[0.2874,\,0.3313]$ \\
$D_2$ & $1.000$ & $0.3461\;[0.3206,\,0.3778]$ & $0.1017\;[0.0942,\,0.1098]$ \\
\bottomrule
\end{tabular}
\end{table}

\section{Sensitivity of the GW results to analysis settings}
\label{app:gw_sensitivity}

In this Appendix we discuss the sensitivity of the GW analysis to variations around the baseline settings.

\subsection{Weakly modelled approach}

We start from the \texttt{PixelPop} results of Sec.~\ref{sec:pixelpop_case_study}. Table~\ref{tab:pixelpop_sensitivity} summarizes all completed sensitivity runs, each varying one choice at a time. All configurations retain the same three principal joint maxima, with overlapping central $90\%$ location intervals in both coordinates, but their recovery probabilities and measured regions can vary. Halving the persistence threshold $T_0$ reveals no fourth retained maximum and increases the median number of draw-level maxima from 20 to 29. Doubling it reduces this count to 13 and lowers $\mathcal C_{35}$ recovery to $85.9\%$. Stable catalogue identity therefore does not imply invariant posterior support.\\

\begin{table*}[t]
\centering
\small
\renewcommand{\arraystretch}{1.20}
\caption{Sensitivity of the GWTC-5.0 \texttt{PixelPop} catalogue to variations around the baseline configuration. Only the stated setting changes from the baseline, which uses $T_0=0.04843$, 2D scales $(0.05,0.10,0.15)$, marginal scales $(0.10,0.15,0.20)$, and probability-support mass 0.995. Automatic thresholds are reselected for the 2D smoothing variations. Ridge probabilities describe recovery of a complete two-arm connection, including the corresponding saddle and peak identities. Its baseline recovery is $19.4\%$ for the low-mass pair and $32.3\%$ for the intermediate-mass pair. Changes to 2D settings leave independent marginal results unchanged; marginal-only changes leave the joint results unchanged. Absence of a marginal reference family does not imply zero posterior probability for every possible feature in that mass range.}
\label{tab:pixelpop_sensitivity}
\begin{tabular}{@{}p{0.20\textwidth}p{0.25\textwidth}p{0.48\textwidth}@{}}
\hline
Variation & Changed setting & Main result \\
\hline
Half persistence & $T=0.02421425\simeq T_0/2$ & Same three peaks and both ridge pairs; pair recovery $17.8\%$ (low) and $30.0\%$ (intermediate); more draw-level fragmentation. \\
Double persistence & $T=0.09685700\simeq2T_0$ & Same three peaks and both ridge pairs; $\mathcal C_{35}$ recovery $85.9\%$; pair recovery $20.9\%$ and $32.8\%$. \\
Fine 2D smoothing & $(0.025,0.05,0.075)$; $T=0.06133$ & Three peaks and both ridge pairs, recovered in $5.7\%$ and $11.2\%$ of draws; additional boundary pit and valley arm; no fourth retained maximum. \\
Coarse 2D smoothing & $(0.10,0.20,0.30)$; $T=0.02401$ & Low-mass ridge pair retained ($41.9\%$); intermediate pair absent from the final catalogue; two shoulders recovered in $23.1\%$ and $56.3\%$ of draws. \\
Zero first 2D scale & $(0,0.05,0.10)$; $T=0.08711$ & Intermediate ridge pair retained ($9.45\%$); low pair absent from the final catalogue; additional boundary pit and valley arm. \\
Probability support & Mass 0.990 or 0.999 & Retained catalogue and stored joint numerical measurements unchanged. \\
Fine marginal smoothing & $(0.05,0.075,0.10)$ & $\mathcal C_{20}^{\rm cand}$ recovery $97.8\%$, predominantly peak-like but weak in contrast; both low-mass dips and the intermediate $m_2$ dip recovered in every draw. \\
Coarse marginal smoothing & $(0.20,0.30,0.40)$ & No separate positive $m_1$ families near 20 or $35\,M_\odot$, and no intermediate $m_2$ dip; both low-mass marginal dips remain. \\
\hline
\end{tabular}
\end{table*}

\paragraph{Principal concentrations and ridge connectivity.}
The three principal concentrations are present under every tested choice. Their location intervals overlap, even when finer smoothing produces substantially more draw-level extrema or reduces the fraction of draws with valid finite peak regions. This distinguishes the stability of the principal mass scales from the detailed shape of each measured concentration.

The ridge connections are more sensitive. Both paired connections survive the half- and double-threshold runs, but their recovery probabilities do not vary monotonically with the persistence cut. Recovery requires the corresponding saddle to connect the same two peak families; changing the cut changes the simplified hierarchy against which that connection is matched. Removing or retaining additional draw-level extrema can therefore alter connection recovery even when the principal peaks remain.

Fine 2D smoothing retains both paired ridges but lowers their recovery to $5.7\%$ and $11.2\%$. Coarse smoothing retains only the low-mass pair, whereas the zero-first configuration retains only the intermediate-mass pair in the final catalogue. The latter is recovered in $9.45\%$ of draws. Thus the reference ridges provide useful geometric context, but their precise posterior connectivity is less stable than the locations of the three concentrations.\\

\paragraph{Marginal depletions and the intermediate-mass candidate.}
The low-mass depletion is more stable in the independent marginals than its complete ridge representation. Under coarse marginal smoothing, both low-mass dips remain identifiable, with recovery probabilities $82.0\%$ in $m_1$ and $95.6\%$ in $m_2$. Under fine marginal smoothing both are recovered in every draw. The intermediate secondary-mass dip is also recovered in every draw at the finer scales, but is absent from the coarse marginal catalogue.

The primary-mass candidate near $20\,M_\odot$ remains resolution-dependent. Fine marginal smoothing raises its location recovery to $97.8\%$ and its conditional peak fraction to $73.8\%$, while its peak-like prominence remains only $0.202\,[0.047,0.475]$, with a central $90\%$ interval. Coarse marginal smoothing retains neither this family nor the separate positive primary-mass family associated with $\mathcal C_{35}$. No tested configuration introduces a fourth retained joint maximum near $20\,M_\odot$.

An audit of the baseline reference hierarchy before the persistence cut clarifies this distinction. At the baseline smoothing scale, it contains a weak off-diagonal local maximum at approximately $(19.96,12.39)\,M_\odot$ with persistence 0.00731, below both $T_0$ and $T_0/2$. Local pits in the NS--BH transition have still smaller persistences, below $2\times10^{-4}$. These are properties of the single reference field, rather than posterior feature-recovery probabilities. They show why marginal structure and saddle-centred ridge context can coexist with the absence of an additional above-threshold joint maximum or valley corridor.\\

\paragraph{Boundary structures and probability support.}
Fine and zero-first 2D smoothing additionally retain a boundary pit and a valley arm absent from the baseline. The pit is frequently recovered, in $99.85\%$ and $99.73\%$ of draws, respectively, but valid deficit-weighted measurements are available in only $7.2\%$ and $10.8\%$. The associated valley arms are recovered in $99.63\%$ and $99.48\%$ of draws; valid finite geometry is available in $34.35\%$ and $39.20\%$. Their reference curves follow the lowest secondary-mass grid edge near $m_2=1.03\,M_\odot$, rather than the NS--BH transition. Their appearance therefore does not establish a joint valley counterpart to either marginal depletion.

Coarse 2D smoothing instead yields two broad shoulders, recovered in $23.1\%$ and $56.3\%$ of draws. These extra features depend on the smoothing choice and do not join the baseline census. Changing the probability-support mass from 0.995 to 0.990 or 0.999 leaves the retained joint catalogue and its stored numerical measurements unchanged. The combined analysis therefore separates stable joint concentrations, strongly recovered marginal depletions, and weaker local or scale-dependent structure, while using ridges to locate the intervening saddle-centred transitions.

\subsection{Strongly modelled approach}

Here we discuss results for the \textsc{FullPop-4.0} results of Sec.~\ref{sec:fullpop_case_study}.

Table~\ref{tab:fullpop_sensitivity} summarizes nine variations around the
\textsc{FullPop-4.0} baseline of Sec.~\ref{sec:fullpop_case_study}, using
the same 6836 posterior draws and $250\times250$ nodal grid. The three
diagonal concentrations, $\mathcal C_{\rm low}$, $\mathcal C_9$ and
$\mathcal C_{30}$, remain in every catalogue, with central $90\%$
location intervals overlapping those of the baseline in both coordinates.
The unequal-mass concentration $\mathcal C_{9,2}$ and its mixed ridge
connection are more sensitive to the selection settings.

\begin{table*}[t]
\centering
\small
\renewcommand{\arraystretch}{1.20}
\caption{Sensitivity of the GWTC-5.0 \textsc{FullPop-4.0} catalogue.
Only the stated setting changes from the baseline, which uses
$T_0=0.05800$, nominal 2D scales $(0,0.05,0.10)$, marginal scales
$(0.10,0.15,0.20)$, and probability-support mass 0.995. Automatic
thresholds are recalibrated at each scale; numerical thresholds instead
apply a common cutoff across scales. Peak recovery denotes location
support; ridge recovery requires the complete two-arm connection and
its endpoint identities. Changes to 2D settings leave the independent
marginal summaries unchanged, and marginal-only changes leave the
joint summaries unchanged.}
\label{tab:fullpop_sensitivity}
\begin{tabular}{@{}p{0.20\textwidth}p{0.25\textwidth}p{0.48\textwidth}@{}}
\hline
Variation & Changed setting & Main result \\
\hline
Half persistence & $T=0.02899755\simeq T_0/2$
& Three diagonal peaks and two ridge pairs; $\mathcal C_{9,2}$ and its mixed connection are absent. $\mathcal C_{30}$ recovery is $88.9\%$; intermediate ridge-pair recovery is $80.8\%$. \\
Double persistence & $T=0.11599020\simeq2T_0$
& Same three peaks and two ridge pairs; $\mathcal C_{30}$ recovery falls to $40.8\%$, and intermediate ridge-pair recovery to $40.0\%$. \\
Fine 2D smoothing & $(0,0.025,0.05)$
& All four peaks and three ridge pairs remain, with unchanged peak and ridge posterior measurements. $\mathcal C_{9,2}$ recurs at all three scales rather than two. \\
Coarse 2D smoothing & $(0,0.10,0.20)$
& Three diagonal peaks and two ridge pairs; $\mathcal C_{9,2}$ and its mixed connection are absent. \\
Nonzero first 2D scale & $(0.025,0.05,0.10)$; $T_0=0.02644$
& Four peaks and three ridge pairs remain; $\mathcal C_{30}$ recovery rises to $92.0\%$. An outer-edge pit and valley arm appear, without valid deficit-weighted region measurements. \\
Probability support & Mass 0.990 or 0.999
& Peak and ridge measurements are unchanged. The shoulder census changes from four families to three or nine, respectively. \\
Fine marginal smoothing & $(0.05,0.075,0.10)$
& Retains the baseline marginal mass families, including the higher-mass peaks and intermediate dips; some peak/shoulder classifications change. \\
Coarse marginal smoothing & $(0.20,0.30,0.40)$
& Each marginal retains only the low-mass and near-$9\,M_\odot$ positive families and their intervening dip. Higher-mass positive and dip families are no longer retained separately. \\
\hline
\end{tabular}
\end{table*}

\paragraph{Principal concentrations and ridge connectivity.}
The low-mass and intermediate ridge pairs remain under every tested
choice, while the mixed connection follows the setting dependence of
$\mathcal C_{9,2}$. Its disappearance even in the half-threshold run
reflects the multiscale selection: replacing automatic, scale-specific
thresholds with a common numerical cutoff changes which features recur
at the required two of three scales. The final catalogues therefore
need not be nested as the first-scale threshold is lowered. Finer 2D
scales retain the unequal-mass peak at all three scales, while coarser
scales remove it from the retained catalogue.

Using a nonzero first scale changes both the reference field and its
automatically selected threshold. Recovery can consequently increase
substantially even though the retained peak locations remain compatible
with the baseline. The tests distinguish stable principal mass scales
from setting-dependent recovery probabilities and detailed connectivity.\\

\paragraph{Independent marginals.}
The near-$9\,M_\odot$ positive family is recovered in every draw in
both marginals under both scale variations. The low-mass depletion also
survives coarse marginal smoothing, with dip recovery of $90.8\%$ in
$m_1$ and $95.2\%$ in $m_2$. The higher-mass positive families and
intermediate dips remain at finer scales but cease to be separately
catalogued at coarser scales. Their marginal representation is therefore
more resolution-dependent than the low-mass depletion.\\

\paragraph{Shoulders and boundary structures.}
Shoulder families depend on both smoothing and probability support,
even when peak and ridge measurements are unchanged. This supports
their separate treatment outside the principal census. The additional
pit and valley arm in the nonzero-first run lie at the lowest
secondary-mass grid edge, $m_2=0.4\,M_\odot$. Neither provides valid
deficit-weighted region measurements, and no paired valley event is
retained. They therefore do not establish a localized joint valley
counterpart to the marginal depletions.

\end{document}